%% file: rrtf.tex
\documentclass[11pt,a4paper]{article}

\usepackage{hep-title}
\preprint{FERMILAB-PUB-24-0005-T-V,IQuS@UW-21-072,IFJPAN-IV-2024-4}

\usepackage{epsfig}
\usepackage[usenames,dvipsnames]{color}

\usepackage{amssymb}
\usepackage{amsmath}
\usepackage[utf8]{inputenc}
\usepackage{graphicx}
\usepackage{float}
\usepackage{color}
\usepackage{xcolor}
\usepackage{rotating}
\usepackage{bigints}
\usepackage{bbold}
\usepackage{enumerate}
\usepackage[hidelinks]{hyperref}
\usepackage{cleveref}
\usepackage{float}
\usepackage{url}
\usepackage{tablefootnote}
\usepackage{eucal}
\usepackage{setspace}
\usepackage{lineno}
\usepackage{xspace}
\usepackage{authblk}
\usepackage[comma,square,numbers,sort&compress]{natbib}

\newcommand{\beq}{\begin{equation}}
\newcommand{\eeq}{\end{equation}}
\newcommand{\bea}{\begin{eqnarray}}
\newcommand{\eea}{\end{eqnarray}}
\newcommand{\eg}{\textit{e.g.}}

\newcommand{\ie}{\textit{i.e.}}
\newcommand{\dd}{\mathrm{d}}

\newcommand{\pT}{p_{\rm T}}
\newcommand{\jpsi}{J/\psi}
\newcommand{\psip}{\psi(2{\rm S})}
\newcommand{\YiS}{\ensuremath{\Upsilon(1\mathrm{S})}}
\newcommand{\YiiS}{\ensuremath{\Upsilon(2\mathrm{S})}}
\newcommand{\YiiiS}{\ensuremath{\Upsilon(3\mathrm{S})}}
\newcommand{\raa}{R_{\mathrm{AA}}}
\newcommand{\sqrtsNN}{\sqrt{s_{\mathrm{NN}}}}
\newcommand{\ingredients}[1]{\vspace{3mm}
{\bf Model ingredients}\vspace{-2mm}\begin{itemize}\setlength{\itemindent}{0em}\setlength\itemsep{-0.2em}\renewcommand{\labelitemi}{-} #1 \end{itemize}}
\newcommand{\ingredient}[1]{\item {\em #1}}

\begin{document}

\title{Comparative Study of Quarkonium Transport in Hot QCD Matter}

\author[1]{A.~Andronic\thanks{Editor}}
\author[2]{P.B.~Gossiaux$^*$}
\author[3]{P.~Petreczky$^*$}
\author[4]{R.~Rapp$^*$} 
\author[5]{M.~Strickland$^*$}
\author[6]{J.P.~Blaizot}
\author[7]{N.~Brambilla}
\author[8,9]{P.~Braun-Munzinger}
\author[10]{B.~Chen}
\author[11]{S.~Delorme}
\author[12]{X.~Du}
\author[13,12]{M. A.~Escobedo}
\author[12]{E. G.~Ferreiro}
\author[14]{A.~Jaiswal}
\author[15]{A.~Rothkopf}
\author[8]{T.~Song}
\author[9]{J.~Stachel}
\author[16]{P.~Vander Griend}
\author[17]{R.~Vogt}
\author[4]{B.~Wu} 
\author[2]{J.~Zhao}
\author[18]{X.~Yao}

\affil[1]{Institut f\"ur Kernphysik, Universit\"at M\"unster, Germany}
\affil[2]{SUBATECH, IMT Atlantique, Universit\'e de Nantes, CNRS-IN2P3, Nantes, France}
\affil[3]{Physics Department, Brookhaven National Laboratory, Upton, USA}
\affil[4]{Cyclotron Institute and Department of Physics and Astronomy, Texas A\&M University, College Station, USA}
\affil[5]{Kent State University, USA}
\affil[6]{CEA Saclay, France}
\affil[7]{Technical University of Munich, TUM School of Natural Sciences, Munich, Germany}
\affil[8]{Research Division and EMMI, GSI Helmholtzzentrum f\"ur Schwerionenforschung, Darmstadt, Germany}
\affil[9]{Physikalisches Institut, Ruprecht-Karls-Universit\"at Heidelberg, Heidelberg, Germany}
\affil[10]{Tianjin University, China}
\affil[11]{IFJ-PAN, Krakow, Poland}
\affil[12]{IGFAE, University of Santiago de Compostela, Spain}
\affil[13]{Universitat de Barcelona i Institut de Ciències del Cosmos, Martí i Franquès 1, 08028 Barcelona, Spain}
\affil[14]{National Institute of Science Education and Research, An OCC of Homi Bhabha National Institute, Jatni-752050, India}
\affil[15]{University of Stavanger, Norway}
\affil[16]{University of Kentucky and Fermilab, USA}
\affil[17]{LLNL and UC Davis, USA}
\affil[18]{InQubator for Quantum Simulation, Department of Physics, University of Washington, Seattle, WA 98195, USA}


\maketitle 

\begin{abstract}
This document summarizes the efforts of the EMMI Rapid Reaction Task Force on ``Suppression and (re)generation of quarkonium in heavy-ion collisions at the LHC'', centered around their 2019 and 2022 meetings. It provides a review of existing experimental results and theoretical approaches, including lattice QCD calculations and semiclassical and quantum approaches for the dynamical evolution of quarkonia in the quark-gluon plasma as probed in high-energy heavy-ion collisions. The key ingredients of the transport models are itemized to facilitate comparisons of calculated quantities such as reaction rates, binding energies, and nuclear modification factors. A diagnostic assessment of the various results is attempted and coupled with an outlook for the future.
\end{abstract}

\section{Introduction}

High-energy nucleus-nucleus collisions, such as those at the Large Hadron Collider (LHC) and Relativistic Heavy-Ion Collider (RHIC), provide the unique opportunity to create in the laboratory the conditions of the microsecond-old Universe, in particular the color-deconfined state called the Quark-Gluon Plasma (QGP). The microscopic understanding of the remarkable properties of the QGP and its hadronization remains a key challenge in nuclear physics. 

The quarkonium families, charmonia and bottomonia, play a critical role in this objective and have long been prominent observables used to probe the fundamental color force in the hot QCD medium. Theoretically, the vacuum heavy-quark (HQ) potential provides a well-calibrated starting point for the study of quarkonium interactions in medium, see Refs.~\cite{Rapp:2008tf,BraunMunzinger:2009ih,Kluberg:2009wc,Mocsy:2013syh,Liu:2015izf} for recent reviews.
In particular, the string term in the HQ potential 
characterizes the long-range nonperturbative part of the force and is associated with the confining property of QCD. It is expected to play a central role in the transition from hadronic to partonic degrees of freedom, and may well be responsible for the strongly coupled properties of the QGP as evidenced by its transport properties, up to temperatures of 2-3 times the pseudo-critical temperature, $T_{\rm pc}$, of the QCD crossover transition~\cite{Liu:2016ysz}. 

Similar to the way quarkonium spectroscopy provides information about the QCD force in vacuum, a systematic investigation of the in-medium force must involve the study of different quarkonium states in matter, \eg, as they subsequently dissolve with increasing temperature. The complexity in describing the in-medium properties of quarkonia and their implementation into transport calculations in ultra-relativistic heavy-ion collisions (URIHCs) prevents their use as a straightforward thermometer of the medium produced in these reactions. 
On the contrary, using available information on the space-time and temperature evolution in URHICs from other sources (\eg, hydrodynamic simulations or electromagnetic radiation), one can utilize quarkonium observables to deduce their in-medium properties and infer the fundamental interactions in QCD matter. 
In the vacuum, the 1S ground-state bottomonia ($\YiS$ and $\eta_b$) are small enough in size to be sensitive to the color-Coulomb $1/r$ part of the HQ potential. However,  excited bottomonia and all charmonia are predominantly bound by the non-perturbative string term, which is characterized by a linearly increasing potential $\sim \sigma r$ with a ``string tension" $\sigma \simeq$ 1 GeV/fm (and/or residual mesonic forces). 
Thus, charmonia and excited bottomonia are excellent probes of the in-medium confining force, as originally envisioned for the $\jpsi$ meson~\cite{Matsui:1986dk}.

In the cooling of the expanding fireball, quarkonia can also be (re-)generated through recombination of individual heavy quarks and antiquarks diffusing through the medium. In particular, quarkonium formation can also occur through quarks and antiquarks from different 
initial pairs. This mechanism~\cite{BraunMunzinger:2000px,Thews:2000rj,Grandchamp:2001pf}
has been shown to be critical for understanding the rise of $\jpsi$ production from RHIC to the LHC where (re)generation seems to constitute the main part of the yield observed in central Pb--Pb collisions~\cite{Rapp:2017chc}.
The data are also compatible with production of $\jpsi$ exclusively through statistical hadronization at the QCD crossover phase boundary~\cite{Andronic:2017pug}.
Precise measurements of the $c\bar{c}$ production cross section and the extraction of the charm-quark diffusion coefficient~\cite{Rapp:2018qla} are key objectives for the experimental program with heavy ions at the LHC in Runs 3 and 4 (2021-2029)~\cite{Citron:2018lsq} and will be important for drawing more definite conclusions. 
The measured elliptic flow, characterized by the flow coefficient $v_2$, is significant for the $\jpsi$ mesons; it is consistent with transport model predictions at relatively low transverse momentum ($\pT$) and requires additional ingredients (such as space-momentum correlations of diffusing charm quarks) at higher $\pT$~\cite{He:2021zej}. The recent measurement of a rather large $v_2$ of $\jpsi$ mesons in high-multiplicity p-Pb events~\cite{Acharya:2017tfn} came as a surprise and cannot be reproduced by transport model calculations~\cite{Du:2018wsj}, suggesting that initial-state effects could be the origin of the azimuthal correlations.

Regarding bottomonia, the current understanding suggests that (re)generation is less important for $\YiS$, but possibly figures as a major component in the strongly suppressed yield of excited states~\cite{Du:2017qkv,Brambilla:2023hkw}.
However, the interplay of the reaction rates and in-medium binding energies, and thus the strength of the underlying HQ potential (\ie, its in-medium screening) remains a matter of debate~\cite{Zhou:2014hwa,Krouppa:2017jlg,Hoelck:2016tqf,Du:2019tjf} and is also amenable to an interpretation based on comover interactions modeled with cross sections extracted from p-Pb data.
It is therefore of great importance to obtain additional information about the typical time at which quarkonia are produced, in particular through $\pT$ spectra and elliptic flow, which contain information about the fireball collectivity imprinted on the quarkonia at the time of their decoupling.
No significant elliptic flow has been observed experimentally for bottomonia~\cite{Acharya:2019hlv,Sirunyan:2020qec}.

On the theoretical side, the basic objects are the quarkonium spectral functions which encode the information on the quarkonium binding energies 
and the (inelastic) reaction rates as well as melting temperatures.
Steadily-improving constraints on the determination of the quarkonium spectral functions are available from thermal lattice QCD (lQCD), see Refs.~\cite{Mocsy:2013syh,He:2022ywp} for reviews.
The information from the spectral functions can then be utilized in heavy-ion phenomenology via transport models. The latter provide the connection between first-principles information from lQCD and experiment that greatly benefits the extraction of robust information on the in-medium QCD force and its emergent transport properties, most notably the inelastic reaction rates of quarkonia. 
Thus far most transport models are based on rate equations and/or semiclassical Boltzmann equations~\cite{Rapp:2017chc}. In recent years transport approaches utilizing open-quantum system (OQS) frameworks have been developed in a Schr\"odinger-Langevin~\cite{Blaizot:2015hya,Katz:2015qja,Kajimoto:2017rel}
or density-matrix~\cite{Akamatsu:2014qsa,Brambilla:2016wgg,Brambilla:2022ynh}
formulation, see also the review in Ref.~\cite{Akamatsu:2020ypb}. These approaches can test the classical approximation underlying the Boltzmann and rate equation treatments and possibly quantify the corrections. Quantum effects may be particularly relevant at high $\pT$ in connection with finite formation times of quarkonia which are enhanced by the Lorentz time dilation in the moving frame; schematic treatments of this effect in semiclassical approaches suggest that varying formation times can lead to observable differences for high-momentum charmonia and bottomonia yields~\cite{Zhao:2010nk,Hoelck:2016tqf,Du:2017qkv,Aronson:2017ymv,Krouppa:2017jlg}. 

Finally, the implementation of phase-space distributions of heavy quarks diffusing through the QGP into quarkonium transport is being investigated by a few groups~\cite{Song:2012at,Yao:2017fuc,Du:2022uvj},
which, as mentioned above, has direct impact on the magnitude and $\pT$ dependence of (re)generation processes~\cite{Grandchamp:2002wp,Zhao:2007hh}.  In particular, the role of non-perturbative effects in the HQ interactions in the QGP (which are believed to be essential to explain the large elliptic flow observed for D mesons)~\cite{Rapp:2018qla}
needs to be accounted for; the associated large scattering widths are likely to require quarkonium transport implementations beyond semiclassical (or perturbative) approximations, which reiterates the motivation for quantum treatments of recombination processes.

The larger experimental data samples in Runs 3 and 4 at the LHC, combined with improved detector performance and measurement techniques, will lead to significant improvements over the current measurements, with extended kinematic coverage (in $\pT$) and the possibility to reach currently unobserved quarkonium states~\cite{Citron:2018lsq}.
But even at present, the uncertainties in the quarkonium data are in some cases significantly smaller than model uncertainties. Improvements are needed in the conceptual aspects discussed above, as well as on the input parameters of the models. Clearly, the modeling of quarkonium production has reached a stage where the complexity of the problem, in connection with the precision reached by experiment, can no longer be handled by individual group efforts.   
A broadly-vetted consensus on the future path can only be achieved in a setting such as the one offered by a Rapid Reaction Task Force (RRTF) within the ExtreMe Matter Institute (EMMI) at GSI Darmstadt, which allows for much more in-depth discussions and working sessions, with the main protagonists from around the world in one room.
We considered it critical to initiate a broad initiative now, to enable the theoretical progress that is needed for the interpretation of Runs 3 and 4 data at the LHC.

Based on the physics case outlined above, the most important issues where consensus and progress is most urgent were identified to be: i) the identification and model comparisons of transport parameters; ii) the controlled implementation of constraints from lattice QCD, and iii) the significance of quantum transport treatments. More concretely, these were addressed guided by the following questions:

1) To what extent are the currently employed transport approaches (mostly carried out in semiclassical approximations) consistent in their treatment of quarkonium dissociation and regeneration?

2) What are the equilibrium limits of the transport approaches and how do the former compare to the results of the statistical hadronization model?

3) What is the significance of the effects on quantum transport of the quarkonium wave packets, and what is needed to develop quantum transport into a realistic phenomenology?

4) How can the abundant information from lattice QCD (quarkonium correlation functions, heavy-quark free energies and susceptibilities, and the open heavy-flavor sector) be systematically implemented into transport approaches?

5) What are the ultimate model uncertainties, and will those allow conclusions regarding the fundamental question of the existence of hadronic correlations in a deconfined medium?

It is clear that the major experimental effort placed on measuring quarkonium with increasing precision needs to be matched by the effort on the theory front.
Our RRTF was intended to stimulate a coordinated pooling of resources and lead the way towards a ``unified'' extraction of the fundamental QCD heavy-quark potential.
We also reiterate that the modeling of heavy-quark diffusion in quarkonium production in a deconfined medium has important implications for the extraction of the diffusion coefficients of heavy quarks in QGP~\cite{Rapp:2018qla,He:2022ywp}.

Our article is structured as follows: we start out with a brief overview of the experimental status of quarkonium measurements at RHIC and the LHC in Sec.~\ref{sec_exp} 
and then collect some basic theoretical concepts for both the description of heavy quarkonia in medium as well as their implementation into transport models for heavy-ion collisions in Sec.~\ref{sec_theo}. 
In Sec.~\ref{sec_models} a rather detailed yet compact account of all transport approaches employed in the present effort is given; this is organized into a total of 19 categories (addressing model components and ingredients) that each research group was asked to elaborate on within their approach. 
In Sec.~\ref{sec_hw} we compare and discuss the numerical results of targeted calculations that all transport models were asked to carry out as far as their approach allows.  
We summarize and conclude our effort in Sec.~\ref{sec_concl}.

\section{Experimental overview}
\label{sec_exp}

Quarkonium yields and the nuclear modification factor $\raa$ have been measured in heavy-ion collisions (including p/d-A collisions, which we will not discuss here) at the LHC and RHIC as a function of centrality and transverse momentum for both the charm and the bottom sector, both for ground states and for radially-excited states (the $\psip$ meson has not been measured yet in Au-Au collisions at RHIC).
We briefly outline the status of quarkonium measurements in nucleus-nucleus collisions, with emphasis on the RHIC and LHC data.

\subsection{Charmonium}

After the studies of charmonium production at the SPS ($\sqrtsNN = 17.3$ GeV)~\cite{NA50:2004sgj,Arnaldi:2008zz} remained somewhat inconclusive, there was a question as to whether or not the observed suppression pattern of charmonium states is to be related to QGP effects.  Since then, measurements at RHIC ($\sqrtsNN=200$ GeV) revealed a strong suppression of $\jpsi$ production in Au-Au collisions compared to pp~\cite{PHENIX:2006gsi,PHENIX:2011img,STAR:2013eve,STAR:2019fge}, of a similar magnitude as measured at SPS and stronger at forward rapidity than at midrapidity.

\begin{figure}[htb]
\includegraphics[width=0.46\linewidth]{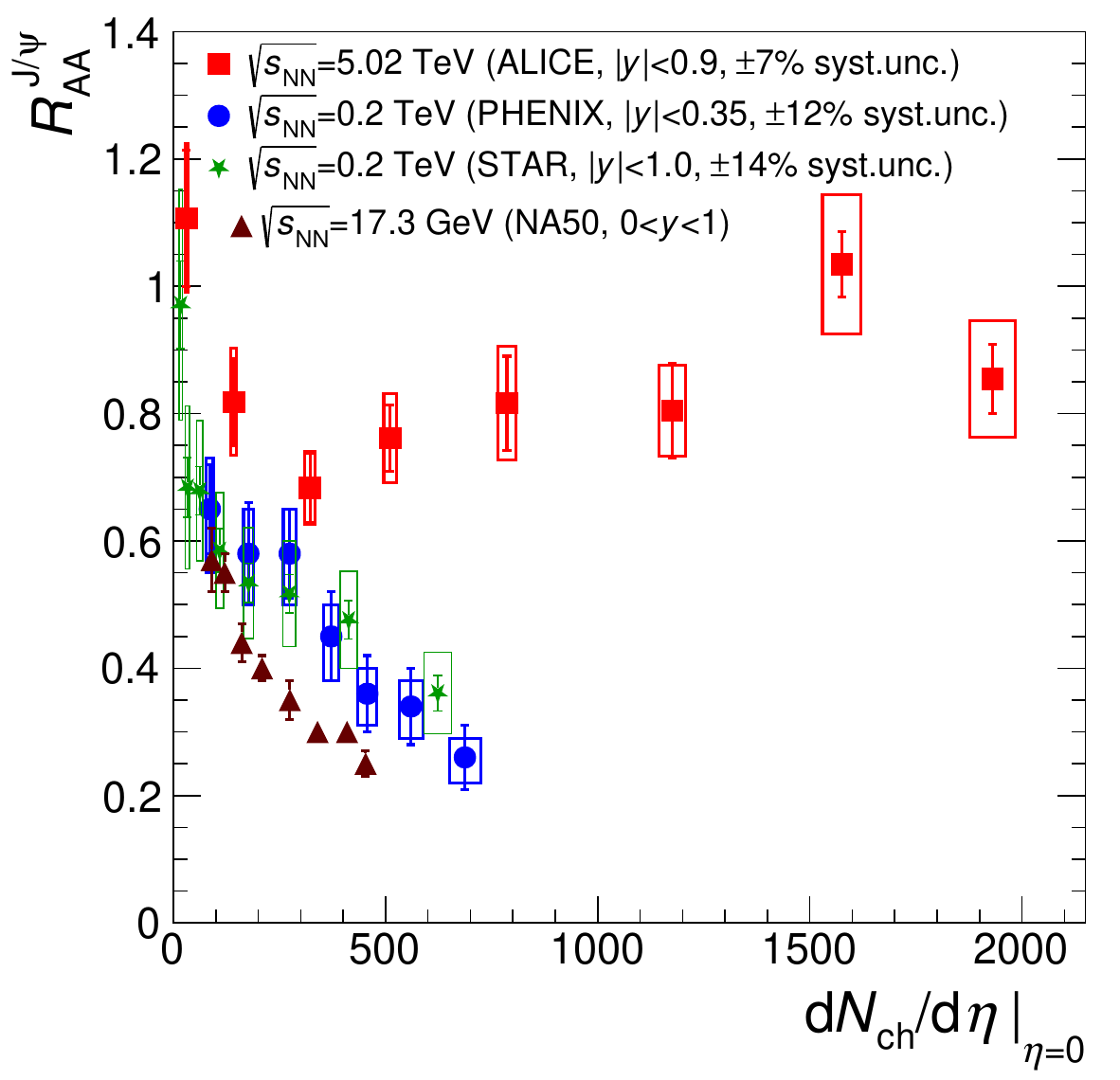}
\includegraphics[width=0.46\linewidth]{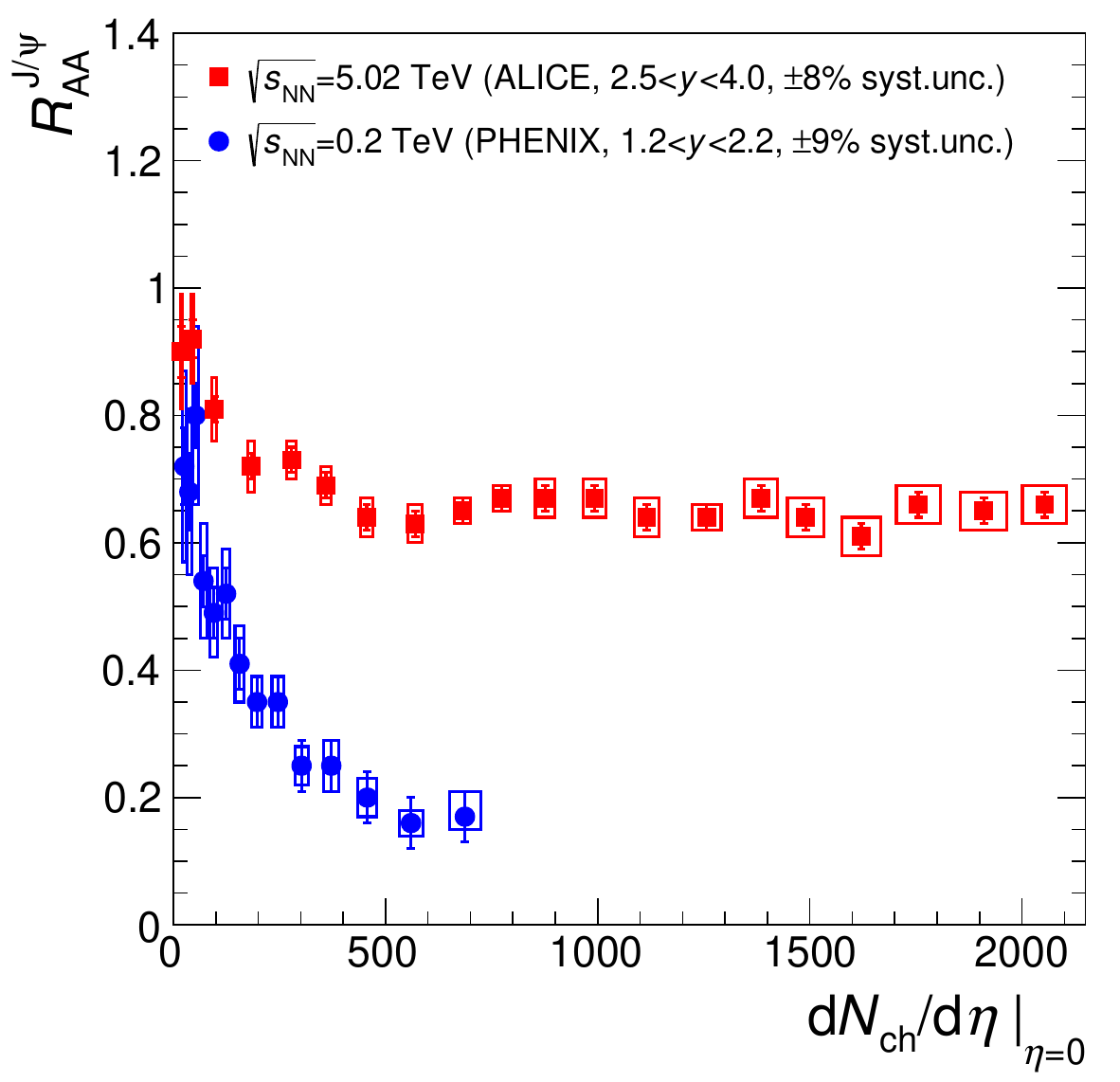}
\caption{The dependence of $\raa$ of $\jpsi$ mesons on the charged particle multiplicity (at midrapidity) for Pb--Pb at 5.02 TeV and Au-Au collisions at 200 GeV, measured at midrapidity (left) and forward rapidity (right). In the left-hand plot, at midrapidity, the data at the SPS from the NA50 collaboration \cite{NA50:2004sgj} (as shown in \cite{ALICE:2022wpn}) are included.}
\label{fig:exp-jpsi-nch}
\end{figure}

In U-U collisions at RHIC, the production yield of $\jpsi$ was measured to be higher than in Au-Au collisions for the same number pf participating nucleons, $N_{\rm part}$ values, in central collisions~\cite{PHENIX:2015rkp}.
Another observation at RHIC was that the $\jpsi$ suppression is similar at $\sqrtsNN=39$ and 62.4 GeV compared to 200 GeV \cite{PHENIX:2012xtg,STAR:2016utm}.
A strong suppression was observed in Cu-Au collisions \cite{PHENIX:2014yic}, while no suppression was measured in Cu-Cu \cite{STAR:2013eve}, albeit with large uncertainties.

\begin{figure}[hbt]
\includegraphics[width=0.495\linewidth]{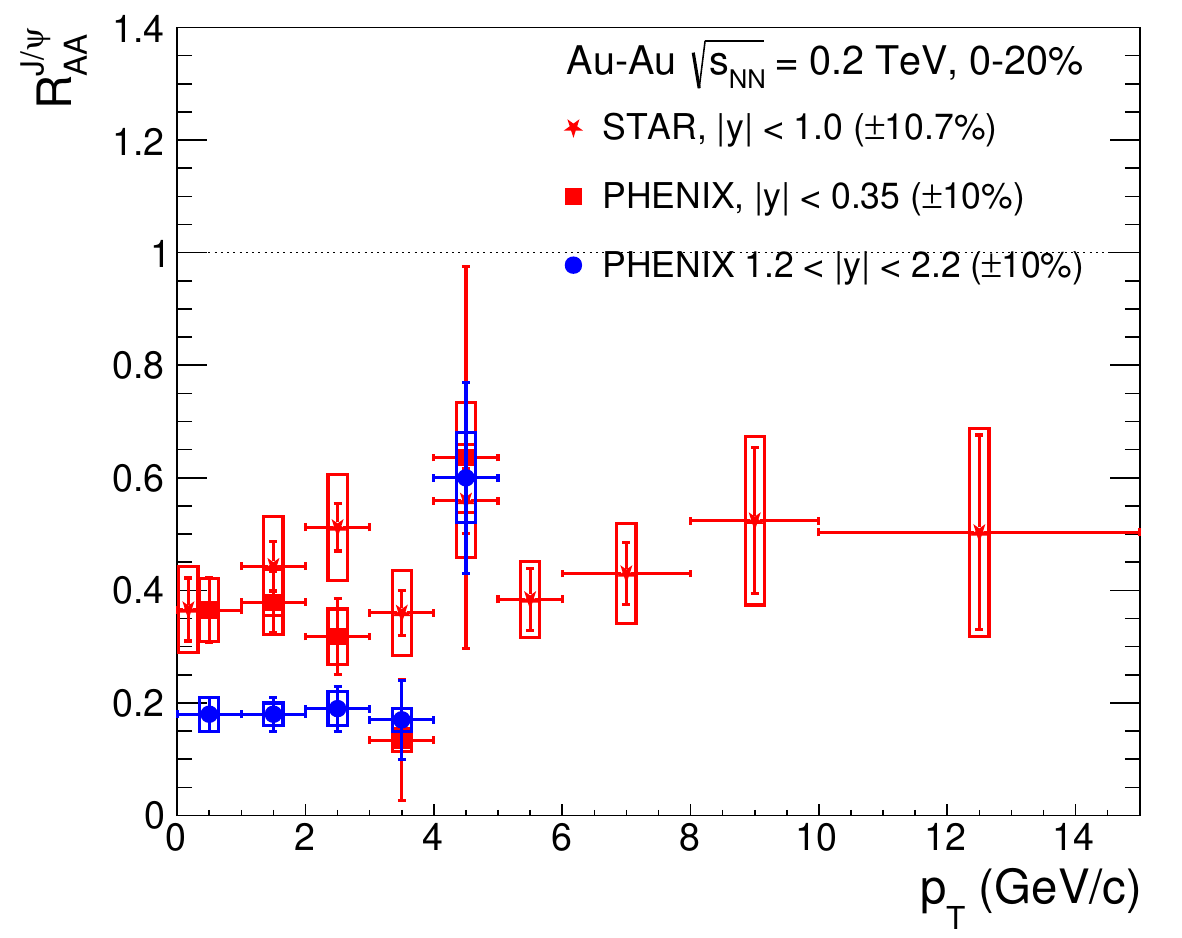}
\includegraphics[width=0.495\linewidth]{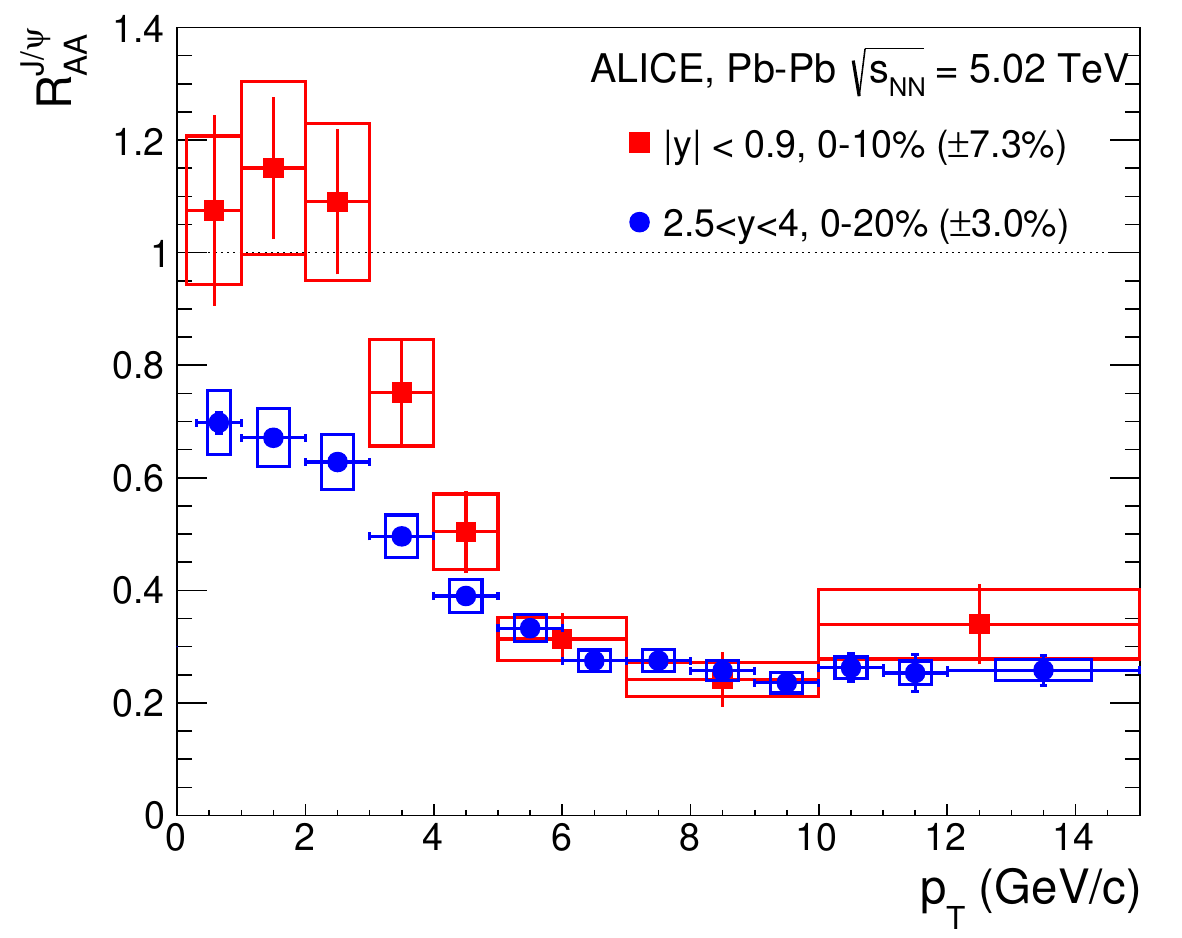}
\caption{The $\pT$ dependence of $\raa$ of $\jpsi$ mesons in central Au-Au (left panel) and Pb-Pb (right panel) collisions.}
\label{fig:exp-jpsi-pt1}
\end{figure}

A clearly different behavior was measured in Pb-Pb collisions at the LHC, where a significantly reduced suppression was observed (larger $R_{AA}$) \cite{ALICE:2012jsl,ALICE:2013osk}. The current LHC $\jpsi$ data \cite{ALICE:2023gco} are compared to the RHIC data in Fig.~\ref{fig:exp-jpsi-nch} as a function of centrality, expressed as the charged-particle pseudorapidity density $\dd N_{\mathrm{ch}}/\dd \eta$ (at $\eta=0$), a proxy for the energy density of the fireball. As at RHIC, the $\raa$ values are larger at midrapidity than at forward rapidity.

\begin{figure}[htb]
\includegraphics[width=0.495\linewidth]{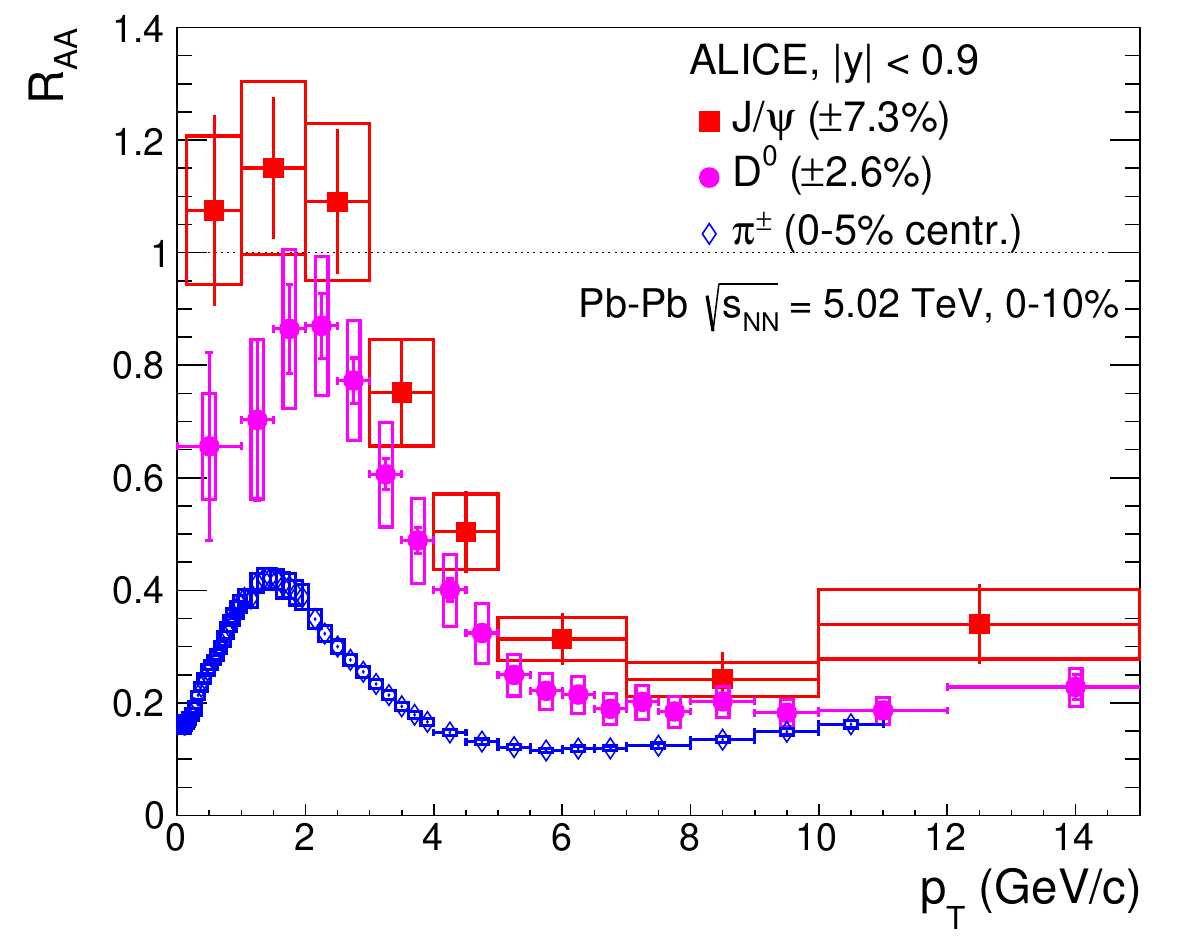}
\includegraphics[width=0.495\linewidth]{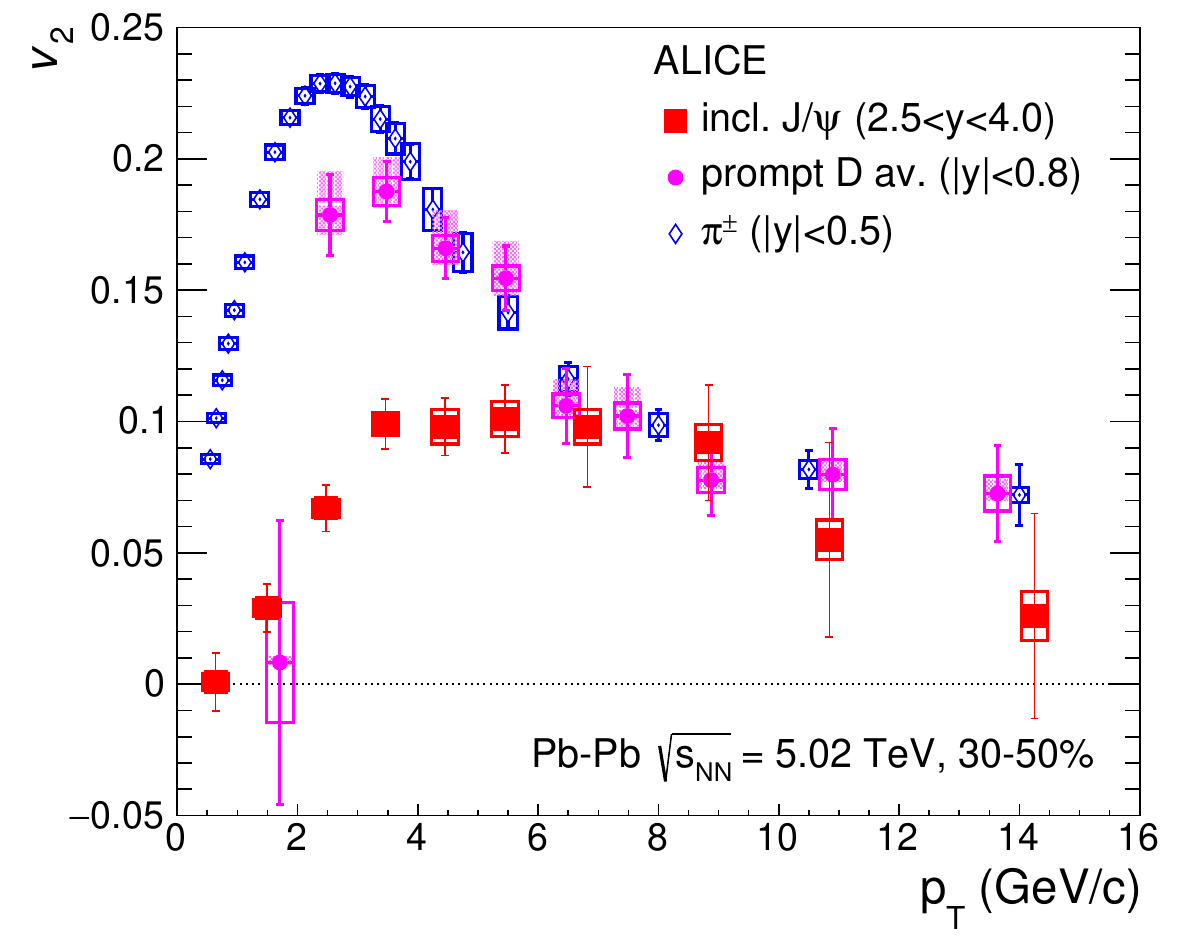}
\caption{The $\pT$ dependence of $\raa$ (left panel, central collisions) and $v_2$ (right panel, 30-50\% centrality) of $\jpsi$ mesons in comparison to D mesons and pions in  Pb--Pb collisions at 5.02 TeV.}
\label{fig:exp-jpsi-pt2}
\end{figure}

Another marked difference between the LHC and RHIC data can be observed in Fig.~\ref{fig:exp-jpsi-pt1}. While the RHIC data exhibit no significant $p_T$ dependence, the LHC data show significantly larger $\raa$ values at low $\pT$, even with a hint at exceeding unity, at midrapidity.
The $\jpsi$ $\raa$ data are compared to those of the D mesons and pions for central Pb-Pb collisions in the the left panel of Fig.~\ref{fig:exp-jpsi-pt2}. The observed ordering is a fingerprint of the different production mechanisms of the charm  and light quarks and suggests as well a (re)generation mechanism for the $\jpsi$ mesons in QGP and/or at hadronization (QCD crossover phase transition).

While the elliptic flow of $\jpsi$ was found to be compatible with zero at RHIC (albeit with large uncertainties)~\cite{STAR:2012jzy}, it is large at the LHC \cite{ALICE:2017quq,CMS:2023mtk}. This is shown in the right panel of Fig.~\ref{fig:exp-jpsi-pt2}, where the $\jpsi$ $v_2$ data are compared to those of D mesons and pions for mid-central collisions.

\begin{figure}[hbt]
\centerline{\includegraphics[width=.54\textwidth]{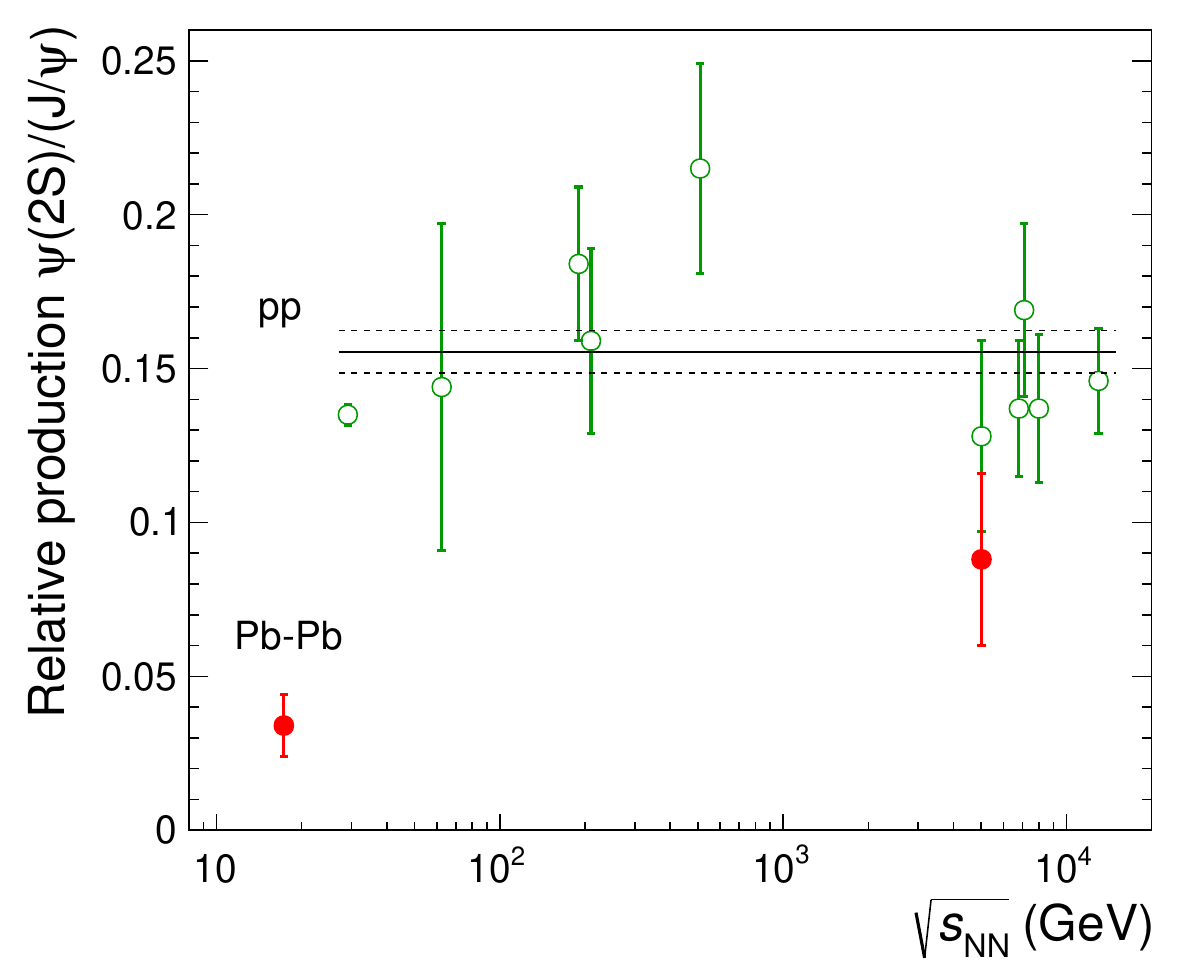}}
\caption{The production ratio of $\psip$ to $\jpsi$ as a function of collision energy measured in pp (pA) collisions (circles) and central Pb-Pb collisions (filled circles).}
\label{fig:exp-psip2j}
\end{figure}

The relative production of $\psip$ and $\jpsi$ mesons as a function of collision energy is shown in Fig.~\ref{fig:exp-psip2j}. The data points for pp collisions are from experiments at the SPS~\cite{Abreu:1998rx}, HERA (pA)~\cite{Abt:2006va}, RHIC~\cite{Adare:2016psx,PHENIX:2019ihw}, and the LHC~\cite{Aaij:2011jh,Aaij:2012ag,Acharya:2017hjh}.
The average value of the pp measurements is represented by the black horizontal line with the 1$\sigma$-uncertainty represented by the dashed lines. The point for central Pb--Pb collisions at SPS energy is from the NA50 experiment~\cite{Alessandro:2006ju} and the point at the LHC is from ALICE~\cite{ALICE:2022jeh}.
  
\subsection{Bottomonium}
\begin{figure}[htb]
\includegraphics[width=0.495\linewidth]{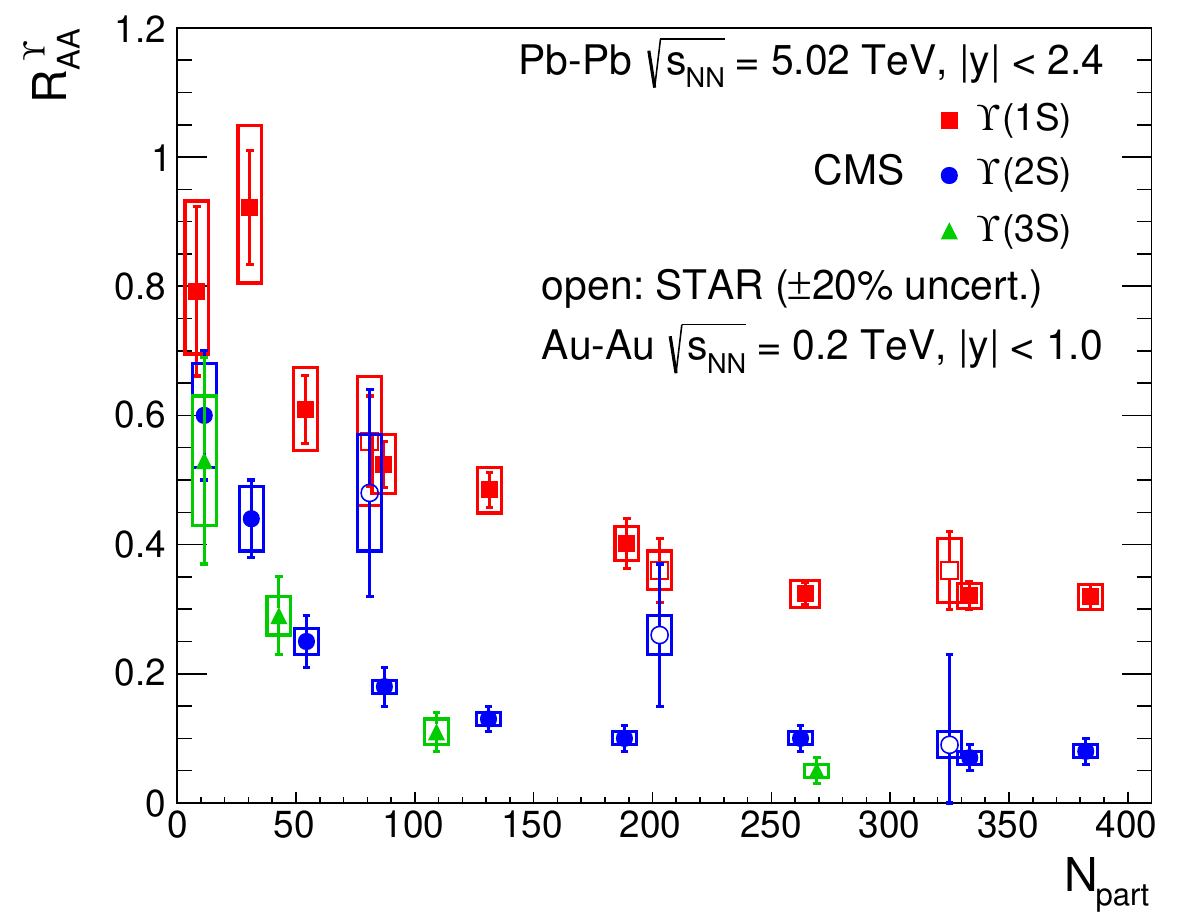}
\includegraphics[width=0.495\linewidth]{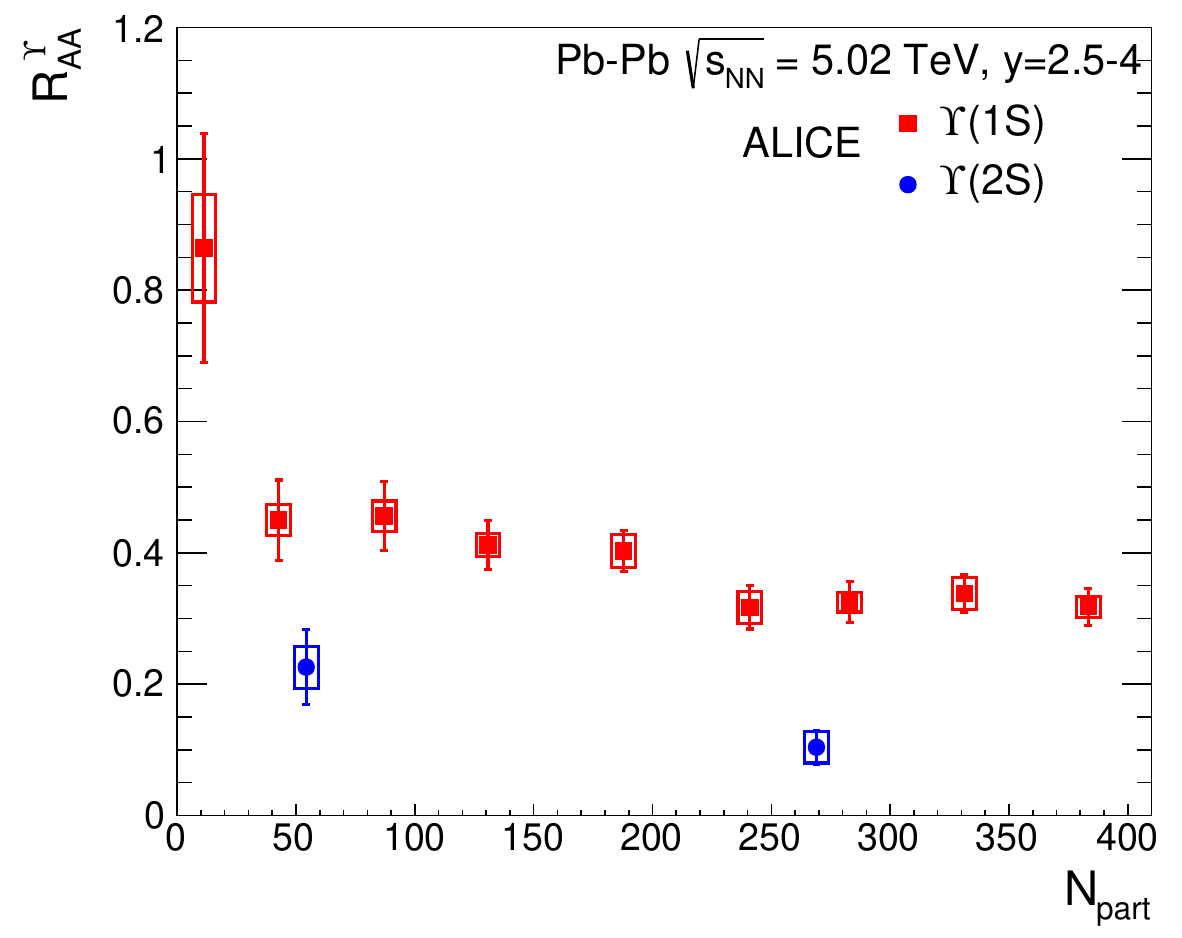}
\caption{The dependence of $\raa$ of $\Upsilon$ mesons on $N_{\mathrm{part}}$ for Pb--Pb at 5.02 TeV and Au-Au collisions at 200 GeV, measured at midrapidity (left) and forward rapidity (right).}
\label{fig:exp-y-npart}
\end{figure}

The early observation by the CMS collaboration at the LHC of the suppression of $\Upsilon$ mesons \cite{CMS:2012gvv}, which was clearly stronger for the radially-excited states, was followed by the current precise measurements of $\raa$ for the $\Upsilon$ states by CMS \cite{CMS:2017ycw,CMS:2018zza,CMS:2023lfu}
and ATLAS \cite{ATLAS:2022exb} at midrapidity 
and ALICE at forward rapidity \cite{ALICE:2020wwx}.
A similar $\YiS$ and $\YiiS$ suppression pattern (and magnitude, for $\YiS$) was measured at RHIC by STAR \cite{STAR:2022rpk}.
A summary of the data is presented in Fig.~\ref{fig:exp-y-npart}. 

The  suppression shows a similar magnitude at forward rapidity and midrapidity.
It is gradually stronger going from the 1S to the 2S and 3S states; this significant pattern is denoted as ``sequential suppression'', with the picture of the melting of the excited states in QGP and their "missing" feed-down to the 1S state. 

\begin{figure}[hbt]
\centerline{\includegraphics[width=.55\textwidth]{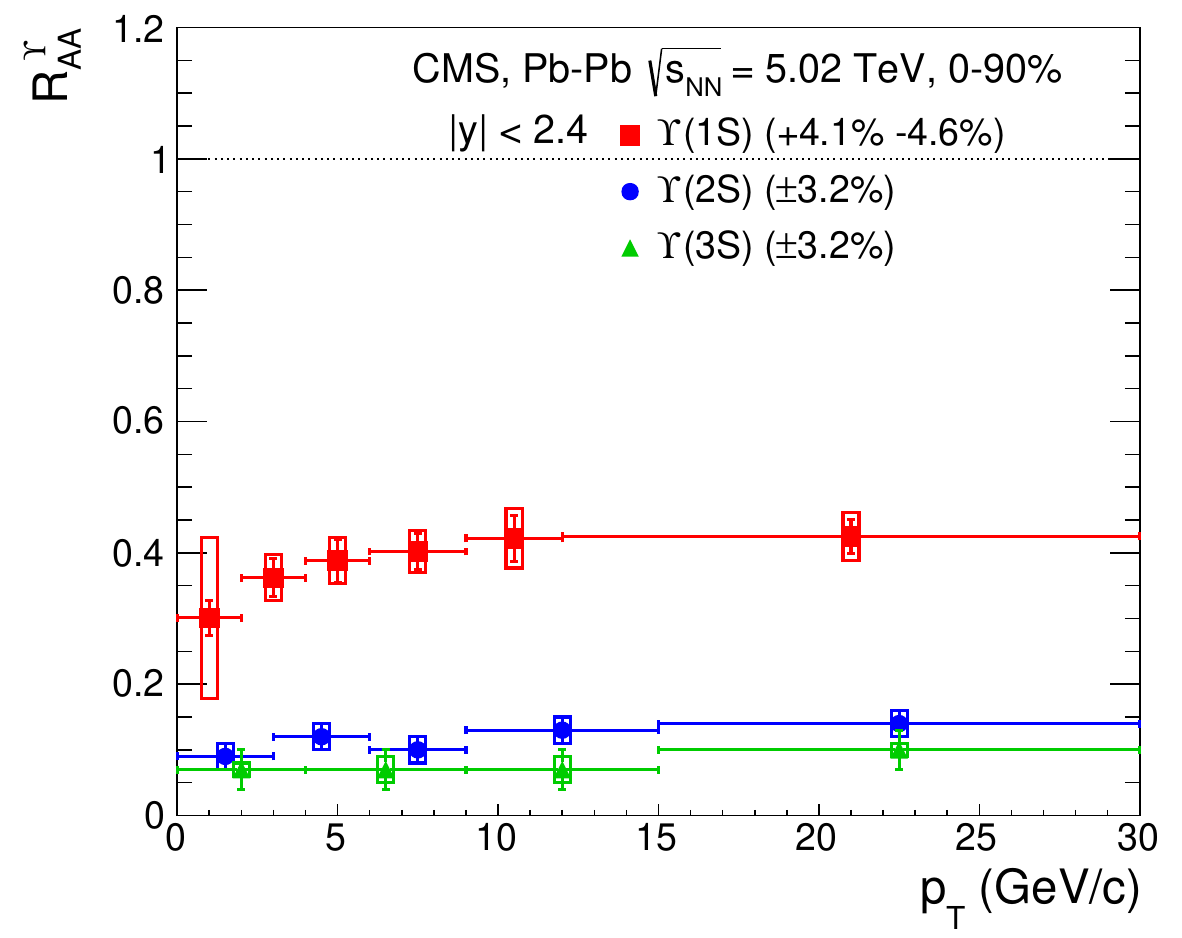}}
\caption{The dependence of $R_{AA}$ of $\Upsilon$ mesons on $\pT$ for 0-90\% Pb--Pb collisions at 5.02 TeV, at midrapidity.}
\label{fig:exp-y-pt}
\end{figure}

The data are shown as a function of $\pT$ in Fig.~\ref{fig:exp-y-pt} for midrapidity in 0-90\% Pb-Pb collisions. No prominent features were observed, except a small increase of $\raa$ vs. $\pT$ for the $\YiS$ state.
Elliptic flow of $\Upsilon$ mesons was measured to be small, both at forward rapidity~\cite{Acharya:2019hlv} and at midrapidity~\cite{Sirunyan:2020qec}, compatible with zero.


\section{Theoretical background}
\label{sec_theo}
One of the advantages of quarkonium physics in QCD matter is the availability of first-principle lQCD computations for a variety of finite-temperature quantities, such as quarkonium correlation functions, heavy-quark free energies and susceptibilities. This is particularly relevant for a strongly coupled medium where non-perturbative effects are expected to be prevalent and perturbative calculations must be interpreted with care. However, since lQCD computations are carried out in Euclidean space-time, the information is usually not readily applicable for use in transport simulations of heavy-ion collisions. 

In Sec.~\ref{ssec_theo-meth}, we
discuss some of the methods that have been used to assess in-medium quarkonium properties and their implementation into transport models, and in Sec.~\ref{ssec_lqcd} we give a more detailed account of pertinent lQCD results. 

\subsection{Theoretical Methods}
\label{ssec_theo-meth}
A key concept in facilitating the connection between lQCD and effective-model calculations are in-medium spectral functions. Quite generally, the latter encode information on the degrees of freedom in the medium (through the presence or absence of well-defined peaks), their in-medium masses and (if applicable) binding energies and the elastic and inelastic reaction rates represented by their widths. Both binding energies and reaction rates are related to the in-medium interaction between heavy quarks and between the heavy quarks and the thermal partons of the medium, respectively. This connection also highlights the importance of utilizing a reasonably realistic model of the thermal medium in which the heavy quarks and quarkonia are embedded.

In practice one, therefore, needs effective descriptions of the in-medium quarkonium dynamics. In the present context, this brings in another asset of the HQ sector, namely the use of 1/$m_Q$ expansion schemes, such as HQ effective theory (which amounts to neglecting the antiparticle components of the Dirac spinors), non-relativistic QCD (NRQCD), and potential NRQCD (pNRQCD). The latter, in particular, is based on neglecting the energy transfer in HQ scatterings, which allows one to convert the 4D Bethe-Salpeter equation into 3D Lippmann-Schwinger type equations (or, in coordinate space, a Schr\"odinger equation), rendering the problem much more tractable.
One further needs appropriate many-body approaches to implement the in-medium physics. Among the ones being used are: (i) perturbative Lagrangian-based hard-thermal loop (HTL) perturbative approaches, (ii) the dynamical quasi-particle model (DQPM) which utilizes a running coupling and accounts for parton spectral functions which can describe the lQCD equation of state (EoS) down to temperatures close to $T_{\rm pc}$~\cite{Moreau:2019vhw}, and (iii) the thermodynamic $T$-matrix, which is a non-perturbative Hamiltonian-based approach that solves the 1-and 2-body correlation functions self-consistently based on a unified 2-body potential for both heavy and light-parton  interactions~\cite{Riek:2010fk,Liu:2017qah}.

To make contact with experiment, the equilibrium physics of quarkonia has to be implemented in transport descriptions. This has traditionally been done in semiclassical models, including the comover interaction model (CIM)~\cite{Capella:2005cn}, simulations of the Boltzmann equation~\cite{Spieles:1999kp,Zhang:2002ug,Yan:2006ve} and kinetic rate equations~\cite{Thews:2000rj,Grandchamp:2003uw,Song:2011xi}) based on dissociation rates in QGP and/or hadronic matter as well as suitably computed equilibrium limits to account for regeneration processes. Constraints from lQCD have been considered by, \eg, using the results for the HQ free or internal energies as underlying potentials to compute in-medium binding energies~\cite{Zhao:2010nk,Liu:2010ej,Krouppa:2017jlg}, and to test the masses and widths of charmonium spectral functions via pertinent Euclidean correlator ratios~\cite{Zhao:2010nk}.

Due to the small energy scale (or long time scale) provided by the in-medium quarkonium binding energies, especially at temperatures in the vicinity of the quarkonium ``dissociation temperature", several efforts are underway to develop quantum transport equations based on the framework of open-quantum systems. 
Most of them utilize a classification of regimes where the binding energy, relative to the temperature, is either small (where quantum effects are expected to be relevant) or large (where semiclassical approximations are reliable); these are further combined with scale hierarchies from HQ effective field theories. Information from lQCD can be implemented through transport coefficients, such as  the HQ diffusion coefficient, albeit without explicit 3-momentum dependencies (which will be challenging to obtain from lQCD). The question of regeneration, especially in the case of multiple quark pairs, has not been addressed yet at a level suitable for phenomenological applications. The equilibration of a single pair has been studied in Ref.~\cite{Miura:2019ssi} and in semiclassical limits for multiple pairs in Refs.~\cite{Blaizot:2017ypk,Villar:2022sbv}.
All of these considerations will figure in the model descriptions given in Sec.~\ref{sec_models}.

\subsection{Lattice QCD Results}
\label{ssec_lqcd}
Lattice-QCD calculations can provide first-principles input into theoretical modeling of quarkonium production in heavy-ion collisions.
Many quantities of interest, like in-medium quarkonium masses and widths, or transport coefficients, are encoded in the spectral functions, defined as the imaginary part of the retarded meson correlation
functions~\cite{Bazavov:2009us,Mocsy:2013syh}. 
For example, the in-medium widths of quarkonia are closely related to the reaction rates used in transport models. If the widths are reasonably small, quarkonium states can be identified by peaks in the spectral functions. As temperature increases, the peaks become broader and ultimately can no longer be used to identify quarkonium states. For example, if 
the width of the peak is much larger than the energy splitting between different quarkonium states it is no longer possible to extract well defined quarkonium states. Obtaining the spectral functions from lQCD is challenging because the latter is formulated in Euclidean time, and the correlation functions are given in terms of integrals over spectral functions. Temporal correlation functions are related to spectral functions via a Laplace transformation, while spatial correlation
functions are related to spectral functions via a double integral transformation~\cite{Mocsy:2013syh}.
Lattice QCD calculations can also be combined with EFT approaches. For example, information about quarkonium spectral function can be obtained using a lattice formulation of NRQCD~\cite{Aarts:2010ek,Aarts:2011sm,Kim:2014iga,Larsen:2019zqv,Larsen:2019bwy,Larsen:2020rjk}.
In this way one avoids large discretization effects due to HQ masses An additional benefit arises from the fact that meson
correlators in NRQCD do not obey periodic boundary conditions, 
which effectively implies that information on meson
correlators can be obtained from doubling the temporal extent in the Euclidean time direction.
Heavy-Quark Effective Theory (HQET) can be used for lQCD calculations of the HQ
diffusion coefficient~\cite{Caron-Huot:2009ncn,Bouttefeux:2020ycy,Laine:2021uzs}.

Most lattice studies of quarkonium spectral functions use point meson operators, \ie,  meson operators
with the quark and antiquark field located at the same spatial point. It turns out that temporal meson correlation functions with point meson operators have limited sensitivity to the in-medium properties of quarkonia~\cite{Rapp:2008tf,Mocsy:2013syh}. This is due to large contributions from the continuum part
of the spectral function to the correlators of point meson operators, as well as the rather small temporal lattice extent at high temperatures~\cite{Mocsy:2007yj,Petreczky:2008px}. 
Therefore, no conclusive results on the in-medium properties of the quarkonium states could be obtained from the temporal correlation functions of point meson operators. There only seems to be a  consensus that the 1S bottomonium state can survive in the QGP for $T>400$ MeV, with a small mass shift~\cite{Aarts:2011sm,Kim:2014iga,Kim:2018yhk}.
The study of spatial meson correlation functions is not limited to small separations, rendering them
more sensitive to the in-medium properties of quarkonia~\cite{Karsch:2012na,Bazavov:2014cta,Petreczky:2021zmz}. In particular, indications were found that 1S charmonia states may dissolve for temperatures of 200-300\,MeV~\cite{Karsch:2012na,Bazavov:2014cta}, 
while 1S bottomonium states will dissolve for temperatures above 500\,MeV \cite{Petreczky:2021zmz}. The latter finding is consistent with the analysis of bottomonium spectral function in lattice NRQCD.

Correlators of extended meson operators, \ie, meson operators with quark and antiquark fields separated by some spatial distance, are more sensitive to the in-medium modification of quarkonia, since the contribution of the continuum part to the spectral function is reduced.
Using NRQCD with extended operators it was possible to analyze in-medium masses and widths of different bottomonium states~\cite{Larsen:2019zqv,Larsen:2019bwy}. Interestingly, it was found that the in-medium mass shift of all bottomonium states is small and compatible with zero within estimated errors. The in-medium width of different bottomonium states was found to increase with  temperature, and that the magnitude of the width follows a hierarchy in the sizes of the different states~\cite{Larsen:2019zqv,Larsen:2019bwy}.

The in-medium modification of $Q\bar{Q}$ interactions in QGP has been traditionally studied in terms of the free energy and singlet free energy of a static $Q \bar Q$ pair. The latter quantity can be defined
in Coulomb gauge. State-of-the-art calculations in 2+1 flavor QCD with physical quark masses suggest that color screening in the free energy sets in at distances 
$r \simeq 0.3/T$ \cite{Bazavov:2018wmo}.
Previous studies of the $Q\bar Q$ free energy for two~\cite{Kaczmarek:2005ui} and three~\cite{Petreczky:2004pz} degenerate quark flavors with unphysical masses have been used as input potentials in some phenomenological models.

The $Q\bar Q$ free energy characterizes the interactions at time scales much larger than the inverse
temperature. For quarkonia physics, it is more relevant to consider a complex potential
defined in terms of Wilson loops~\cite{Rothkopf:2011db}. The first calculation of the complex
potential along these lines with 2+1 flavor QCD with unphysical quark masses found~\cite{Burnier:2014ssa} that the real part of the potential is screened. A parametrization of these results using a generalized Gauss law model~\cite{Lafferty:2019jpr} has also been used in some phenomenological
models. The corresponding lattice calculations are performed on $N_{\tau}=12$
lattices and with limited statistics. Another lattice study that also uses $N_{\tau}=12$ lattices, but with much larger statistics, extracted a different result~\cite{Bala:2021fkm}. Here, simple but physically
motivated parametrizations of the spectral functions were used to obtain the real and imaginary parts of the potential: the real part of the potential turns out not to be screened in general~\cite{Bala:2021fkm}. The only way to obtain a screened potential from the lattice results of Ref.~\cite{Bala:2021fkm} is to use a perturbative HTL inspired representation of the spectral function \cite{Bala:2021fkm}. However, it turns out that, although HTL results for the equation of state of in good agreement with lattice calculations at high temperatures \cite{Haque:2014rua,Haque:2020eyj}, HTL calculations are in disagreement with the lattice results on the Wilson line correlators even
for very high temperatures~\cite{Bala:2021fkm}. More recently, the analysis of Wilson line correlators from Ref.~\cite{Bala:2021fkm} has been repeated using lattices with much larger temporal extent, $N_{\tau}=20-36$, and it was
again found that the real part of the potential is not screened~\cite{Bazavov:2023dci}. Very recently, a microscopic calculation of the Wilson line correlators has been carried out in the thermodynamic $T$-matrix approach~\cite{ZhanduoTang:2023ewm}; it was found that, with a refined input potential which also shows little screening at small and intermediate distances, the lattice results can be reasonably well reproduced, but with very large quarkonium widths.

The HQ momentum diffusion coefficient, $\kappa$, is also used as an input to a few phenomenological models of quarkonia production. It can be calculated on the lattice using HQET, as mentioned above, in terms of a chromo-electric correlation function with fundamental Wilson lines~\cite{Caron-Huot:2009ncn}. 
Most lQCD calculations of $\kappa$ have been performed in quenched QCD~\cite{Banerjee:2011ra,Francis:2015daa,Brambilla:2020siz,Altenkort:2020fgs,Banerjee:2022gen,Brambilla:2022xbd}. Very recently the first lattice calculation of the HQ diffusion constant in 2+1 flavor QCD appeared~\cite{Altenkort:2023oms}, which includes the HQ  mass dependence of $\kappa$ and the related HQ
spatial diffusion coefficient, ${\cal D}_s$~\cite{Altenkort:2023eav}.
Within the pNRQCD approach one can consider a chromo-electric correlator with adjoint Wilson lines. This correlation function defines another transport coefficient
$\kappa_{\rm adj}$, which enters the pNRQCD-based open quantum system approach of quarkonium production~\cite{Brambilla:2017zei}. To leading order in perturbation theory,  $\kappa=\kappa_{\rm adj}$,  but this is not true in general. Unfortunately, no lattice QCD calculations of $\kappa_{\rm adj}$ are available so far.

As will be discussed below, phenomenological models of quarkonium production in heavy-ion collisions require knowledge of  the in-medium HQ masses. 
Generalized charm and bottom susceptibilities can be useful for constraining these masses. Lattice calculations of these quantities exist both in quenched~\cite{Petreczky:2008px} and full QCD~\cite{Petreczky:2009cr,Bazavov:2014yba,Bellwied:2015lba,Mukherjee:2015mxc,Bazavov:2023xzm}.
It was inferred that the in-medium charm- and bottom-quark masses decrease with increasing temperature~\cite{Petreczky:2008px,Petreczky:2009cr,Altenkort:2023eav}.

\section{Model Descriptions}
\label{sec_models}
In this section we provide a synopsis of each of the phenomenological approaches to describe heavy-quarkonium transport in URHICs that has been included in the present effort.  
By summarizing each approach, guided by the same list of key ingredients, assumptions and inputs, we attempt to lay out the strengths and areas of future improvement in the various model calculations.

\input{models/dukemit.tex}

\input{models/munichksu.tex}

\input{models/nantes.tex}

\input{models/phsd.tex}

\input{models/saclay.tex}

\input{models/comover.tex}

\input{models/shm.tex}

\input{models/tamu.tex}

\input{models/tsinghua.tex}

\input{models/vogt.tex}


\section{Comparisons of Model Ingredients and Interpretation of Results}
\label{sec_hw}
In this section we will confront the theoretical inputs and results of the transport models to quantitatively analyze key components of in-medium quarkonium kinetics and how they manifest
themselves in phenomenologically-relevant outcomes. We start by collecting the temperature evolution 
of the medium expansion models in Sec.~\ref{ssec_med-evo} for the case of central Pb-Pb~($\sqrtsNN$ = 5 TeV) collisions and then turn to the arguably most fundamental transport parameter, \ie, the reaction rate, for various charmonia and bottomonia as a function of temperature in Sec.~\ref{ssec_rate}. To aid in the interpretation of these results, we inspect the inputs for binding energies and HQ masses in Sec.~\ref{ssec_Eb}, the resulting
3-momentum dependence of the rates in Sec.~\ref{ssec_mom-dep} and the spatial dependence of the imaginary part of the potentials (as applicable) in Sec.~\ref{ssec_ImV}. We then turn to more phenomenologically-oriented studies, by testing the medium evolution models with a prescribed temperature dependent reaction rate to compute an $\raa$ in Sec.~\ref{ssec_fix-rate}, and by studying the impact of formation time effects in early quarkonium evolution on their suppression factor in Sec.~\ref{ssec_tauf}.

\subsection{Medium Evolution}
\label{ssec_med-evo}
\begin{figure}[htb]
\begin{center}
\includegraphics[width=0.6\linewidth]{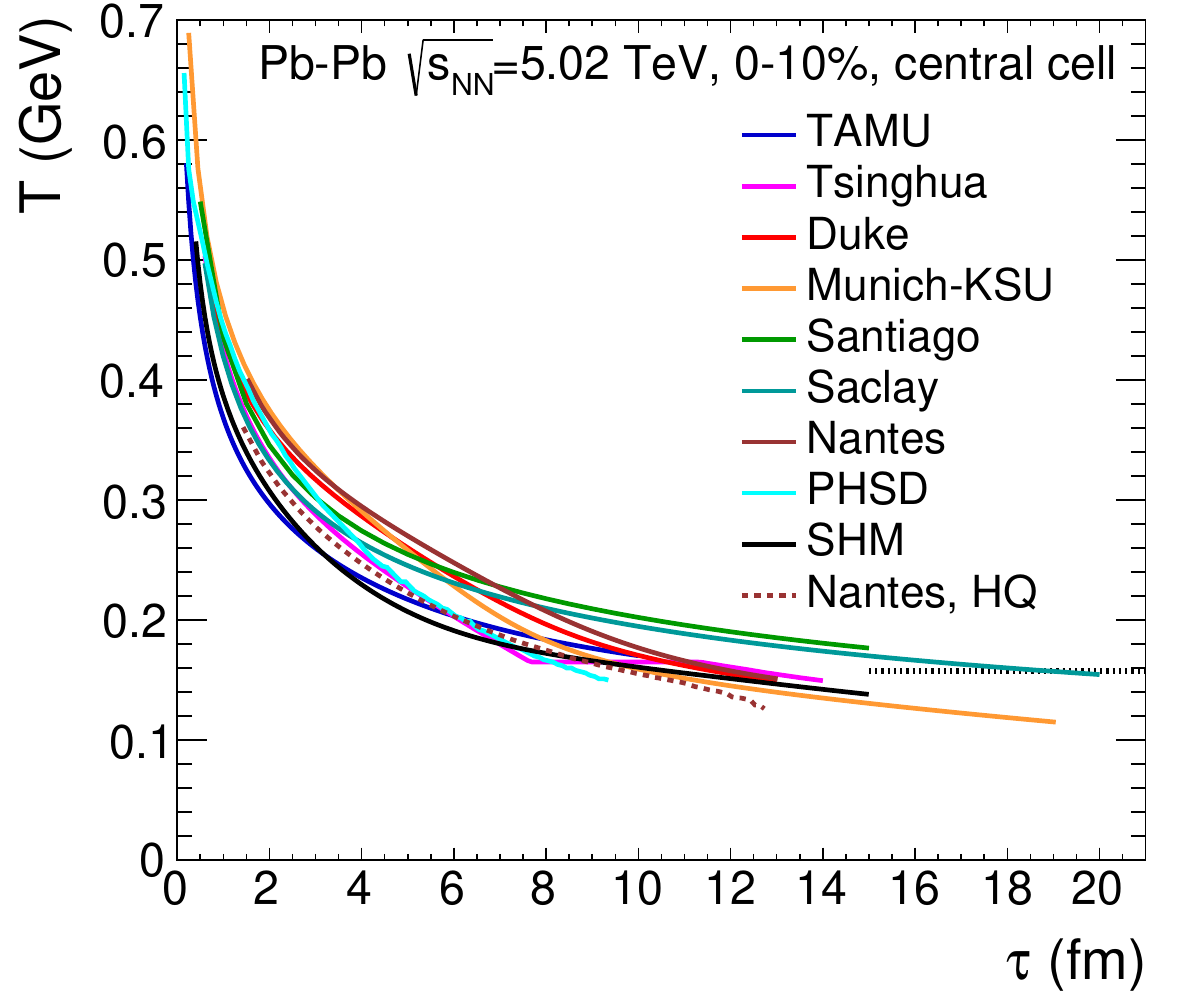}
\end{center}
\vspace{-7mm}
\caption{Temperature evolution of the central cell in 0-10\% central Pb-Pb collisions at 5.02 TeV as a function of proper time in the fireball expansion models employed in the current work. For the Nantes approach the local (average) $T$ as ``seen'' by the a heavy-quark pair is plotted in addition (dashed line). The SHM uses MUSIC hydrodynamics~\cite{Schenke:2010rr}. The horizontal dash-dotted line starting at $\tau=15$\,fm represents lattice QCD predictions for the chiral crossover transition (as a weighted mean of the two available predictions~\cite{HotQCD:2018pds,Borsanyi:2020fev}).}
\label{fig:temp}
\end{figure}

In Fig.~\ref{fig:temp} we collect calculations of the temperature evolution for the central cell in 0-10\% central Pb-Pb collision at $\sqrtsNN = 5.02$ TeV for the expanding QCD medium using the approaches included in the present effort.  For the hydrodynamical models (Duke, Munich-KSU, Nantes, and Tsinghua), the results shown are for the central cell ($x=y=\eta_s=0$), which is usually the hottest region of the plasma; the 4 results agree quite well, in particular for the former three, while the latter one cools faster due to the EoS based on a massless parton gas (and a first-order transition) which is ``harder" (generating a faster transverse expansion) and has a smaller temperature at given entropy density, compared to the lQCD-based EoS used in the other 3 models. A marked difference in the hydro models is the assumption of the initial thermalization time, $\tau_0$, which controls the initial temperature, $T_0$, varying between close to 700\,MeV at $\tau_0$ = 0.2\,fm for Munich-KSU and $\sim$400\,MeV at $\tau_0$ $\simeq$ 1.5\,fm for Nantes; this may also be affected by the initial temperature profiles used in the calculations (\eg, two- vs.~three-dimensional).
For the Saclay (ideal Bjorken hydrodynamics), Santiago (comover), Tsinghua (2+1D ideal hydrodynamics), and the TAMU fireball (isentropically expanding blastwave with lQCD EoS) models, spatial averages lead to lower temperatures, and the primarily longitudinal expansion in the former two leads to a significantly slower cooling at later times (which affects the more weakly bound quarkonia). 
In the SHM, the MUSIC 3+1D hydrodynamic calculations~\cite{Schenke:2010nt} with IP-Glasma initial conditions are used in a blast-wave parametrization, although the pertinent quarkonium production is only evaluated at the pseudo-critical temperature, $T_{\rm pc}$ $\simeq$ 155\,MeV. 
For the Nantes approach additional information is provided in terms of the average temperature encountered at the position of a HQ pair propagating through the fireball, which is significantly smaller than the values in the central cell but agrees fairly well with the spatially uniform profile of the TAMU fireball evolution.
The temperature evolution extracted from the PHSD transport model has a rather steep decrease, starting at about 660\,MeV and dropping down to $\sim$150\,MeV after about 9~fm.

\subsection{Reaction Rates}
\label{ssec_rate}
We first discuss charmonia in Sec.~\ref{sssec_charmonia} followed by bottomonia in Sec.~\ref{sssec_bottomonia}

\subsubsection{Charmonia}
\label{sssec_charmonia}
\begin{figure}[thb]
\begin{center}
\includegraphics[width=0.9\linewidth]{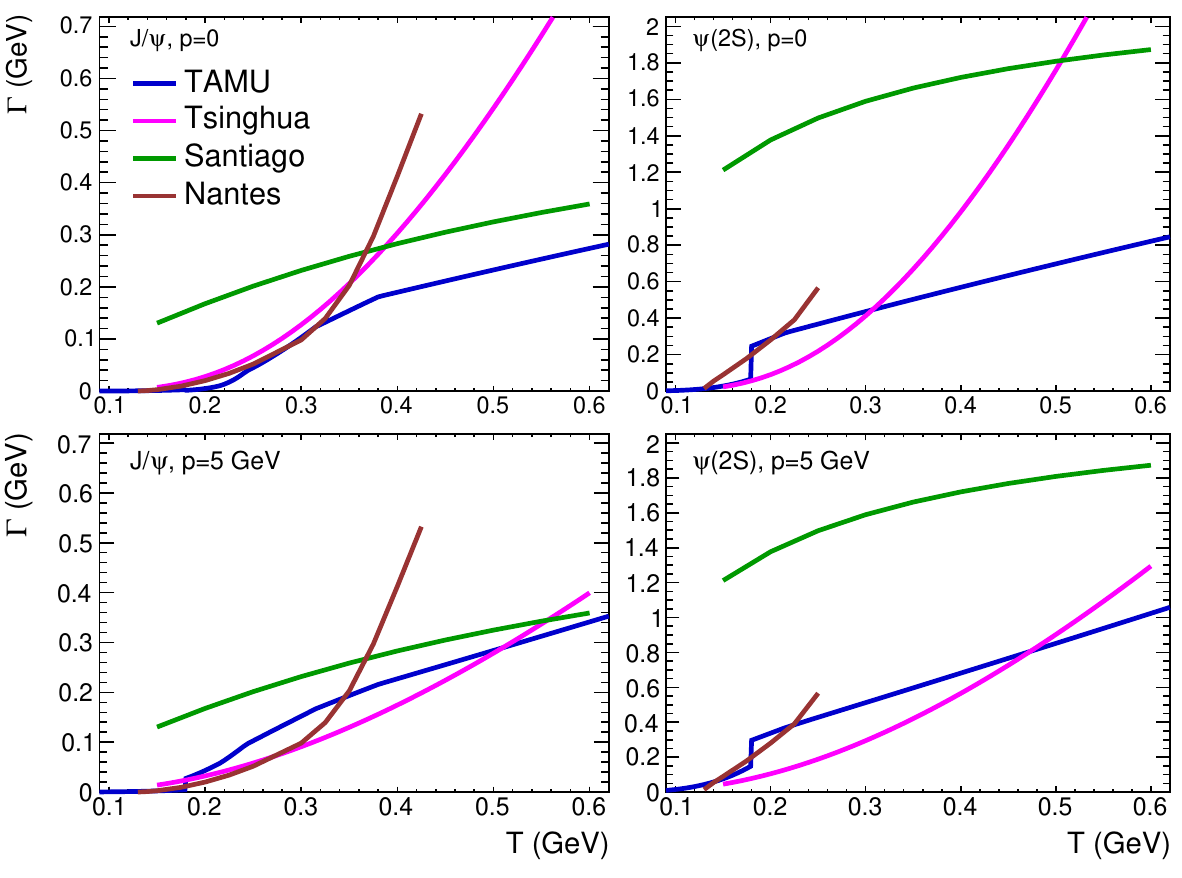}
\end{center}
\vspace{-7mm}
\caption{Temperature dependence of reaction rates for $\jpsi$ (left column) and $\psip$ (right column), and for 3-momentum fixed to 0 (5)\,GeV in the upper (lower) row.}
\label{fig:gam-t-psi}
\end{figure}
The available results for charmonium reaction rates, \ie, the 1S and 2S vector states, are compiled in Fig.~\ref{fig:gam-t-psi} as a function of temperature for 2 different momenta.
We note that most OQS approaches work with an EFT scale hierarchy which is not applicable to charmonia. In the Nantes approach, results are only shown for temperatures below the dissociation temperature, while for TAMU the rates above the dissociation temperature simply correspond to two times the HQ collision 
rate (as a means to characterize the decorrelation of a primordially produced HQ pair that would end up in a quarkonium state in a pp collision). 
The Tsinghua model assumes a constant $\jpsi$ binding energy and geometric scaling for the rates of excited states, while the comover model does not evaluate charmonium rates microscopically but via a constant dissociation cross section fit to data for each state.

At vanishing momentum, the $\jpsi$ rates from TAMU, Tsinghua and Nantes agree well at temperatures below $\sim$350\,MeV. This bodes well for the comparison between TAMU and Nantes, as both rates are computed with a similar mechanism (inelastic parton scattering closely related to the imaginary part of the potential) and a potential that leads to similar in-medium binding energies, cf.~Fig.~\ref{fig:Eb-psi}. 
For Tsinghua and Nantes the further increase beyond $T\simeq350$\,MeV is markedly stronger, but likely due to different mechanisms. For Tsinghua, this is rooted in the gluo-dissociation cross section with a constant in-medium binding energy which is peaked near the binding energy; at higher momenta this reduces the rate at higher $T$. 
For TAMU, once the binding energy vanishes for $T\gtrsim380$\,MeV, and for Santiago with constant comover cross section, the $T$ dependence becomes rather weak, close to linear or weaker. This can also be seen for the $\psip$. For the latter, the Nantes and TAMU rates are again comparable, albeit in this case with rather different input for the bound-state properties. The Santiago and Tsinghua $\psip$ rates are obtained from the $\jpsi$ rate by geometric scaling of the radii. Especially the Santiago rates are much larger than all other $\psi2S$ rates at low $T$ (we recall that they can explain the $\psip$ suppression in pA collisions), while the TAMU rates include an extra $K$-factor of $\sim$3 to mimic nonperturbative effects and enable a better description of pA results.

At finite charmonium momentum (taken as $p$=5\,GeV in the bottom panels of Fig.~\ref{fig:gam-t-psi}), the Nantes and comover results remain unchanged for both $\jpsi$ and $\psip$. 
On the other hand, the TAMU results increase (as consequence of a perturbative matrix element for the heavy-light coupling and a larger phase space available for dissociation) while the Tsinghua results decrease (as a consequence of the gluo-dissociation cross section being probed beyond its peak structure). 

\subsubsection{Bottomonia}
\label{sssec_bottomonia}
\begin{figure}[!thb]
\begin{center}
\includegraphics[width=0.9\linewidth]{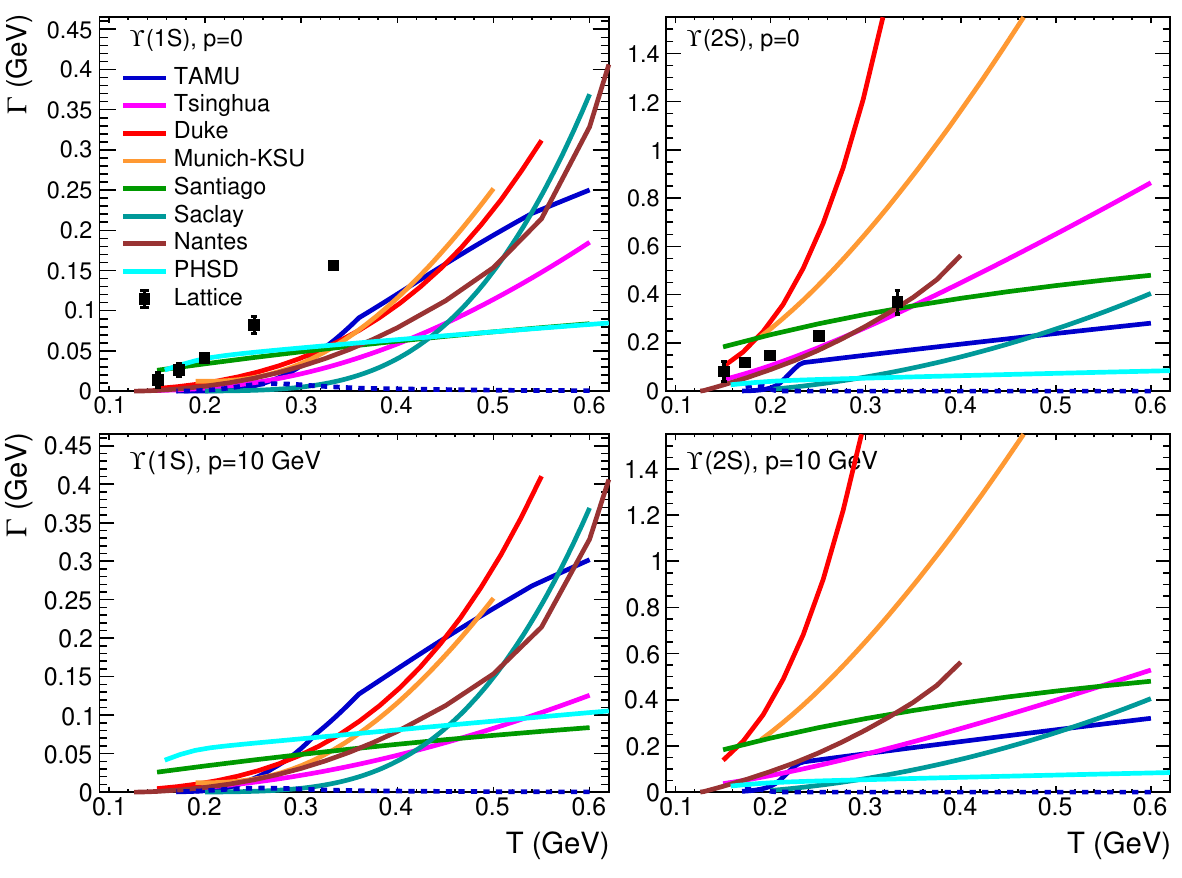}
\end{center}
\vspace{-7mm}
\caption{Temperature dependence of bottomonium reaction rates, for $\YiS$ (left column) and $\YiiS$ (right) at 3-momentum $p$=0 (upper row) and 10\,GeV (lower row). Dashed line is for the TAMU gluo-dissociation case. The lattice results are taken from Ref.~\cite{Larsen:2019zqv}.}
\label{fig:gam-t-ups}
\end{figure}
\begin{figure}[!thb]
\begin{center}
\includegraphics[width=1\linewidth]{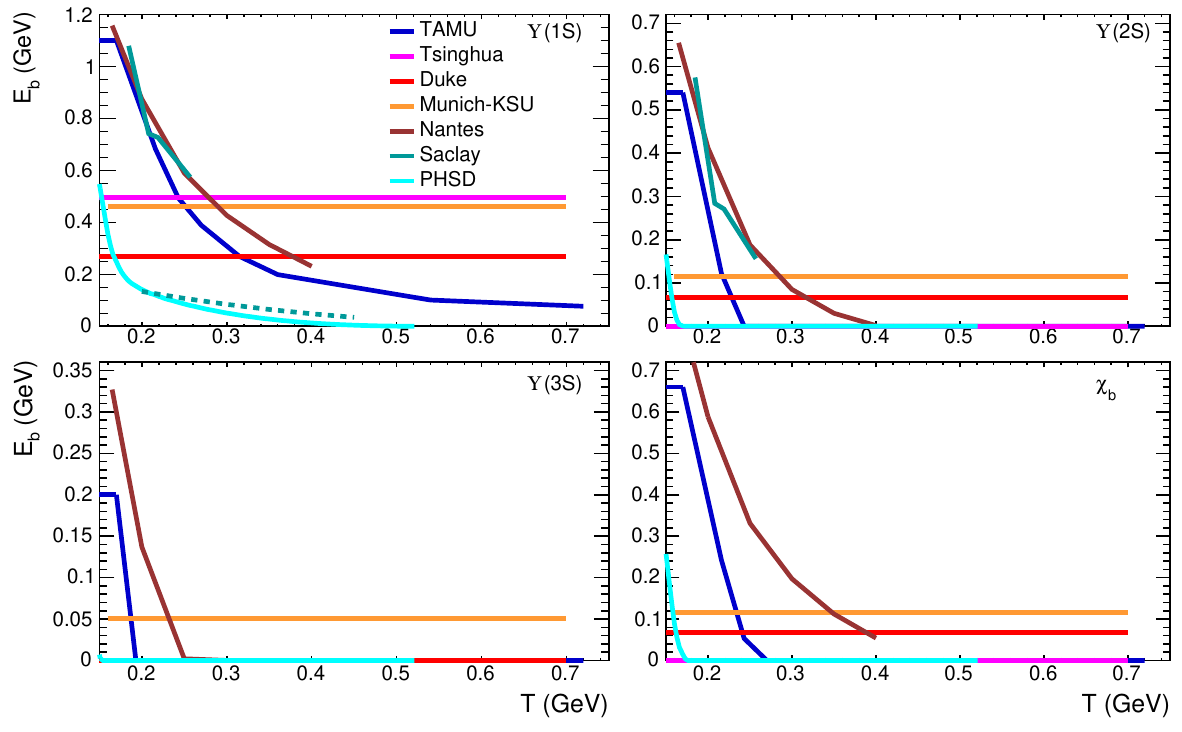}
\end{center}
\vspace{-7mm}
\caption{
Temperature dependence of bottomonium binding energies, \ie, for 
$\YiS$ (upper left), $\YiiS$ (upper right), $\YiiiS$ (lower left) and $\chi_b$ (lower right).
Dashed line: Saclay, perturbative; 
(the Santiago comover approach does not use the concept of binding energy).
}
\label{fig:Eb-ups}
\end{figure}

Next, we turn to the $T$ dependence of the inelastic bottomonium rates, compiled in Fig.~\ref{fig:gam-t-ups}, again for 2 different 3-momenta. 
The Duke, Munich-KSU, Nantes, TAMU, Tsinghua, PHSD and Santiago (comover) rates for $\YiS$ at $p$=0 approximately agree in the phenomenologically most relevant range of $T\simeq$~300-400\,MeV. While the underlying mechanism (inelastic parton scattering) and in-medium binding for Nantes and TAMU are similar (see Fig.~\ref{fig:Eb-ups}), the Tsinghua (gluo-dissociation) and Duke results utilize constant quark masses and binding energies, which are also quite different from each other being based on either a Cornell potential (large binding and large HQ mass) or color-Coulomb potential (small binding and small HQ mass), respectively. The $\YiS$ widths for Saclay (with a binding similar to Nantes and TAMU) are rather small.
Finally, the PHSD and comover results have the weakest $T$-dependence, and they are quite large at relatively low $T$. 
Moving on to $p$=10\,GeV, the Santiago, Munich-KSU, Nantes, PHSD and Saclay widths do not change significantly (or at all), while the Duke and TAMU widths increase and the Tsinghua width decreases. The overall spread in the results increases. 
For the $\YiiS$, the calculated widths generally increase substantially relative to the $\YiS$, especially at low temperatures, with the exception of PHSD (which utilize the same reaction rates for $\YiS$ and $\YiiS$).
The large spread (the agreement between Nantes and Tsinghua for $p$=0 must be considered a coincidence given the very different mechanisms), suggests that phenomenological constraints currently suffer from large uncertainty; \eg, even small contributions from regeneration can make a big difference in a phenomenological extraction based on a small $\raa$, whereas regeneration is expected to be relatively less important for the $\YiS$ (except for the SHM).
 
\subsection{Quarkonium binding energies and heavy-quark masses}
\label{ssec_Eb}
\begin{figure}[hbt]
\begin{center}
\includegraphics[width=1\linewidth]{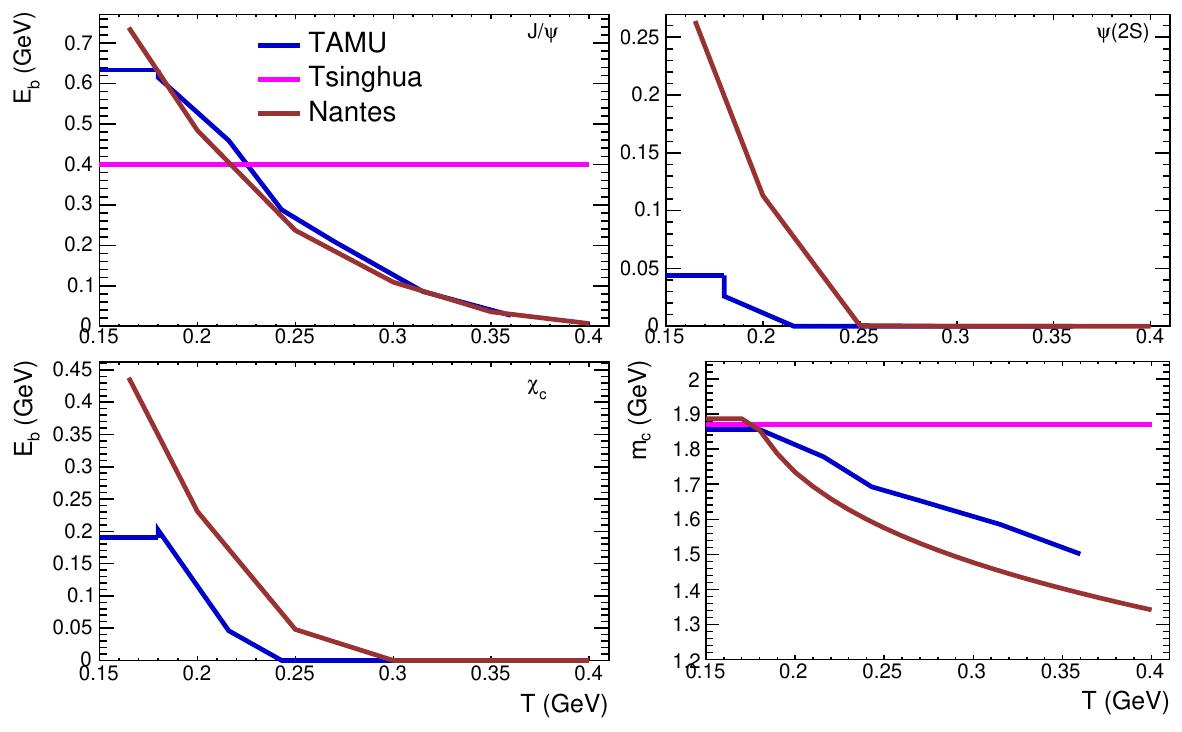}
\end{center}
\vspace{-7mm}
\caption{Temperature dependence of charmonium binding energies, \ie, for $\jpsi$ (upper left), $\psip$ (upper right) and $\chi_c$ (lower left), as well as the underlying charm-quark mass (lower right).}
\label{fig:Eb-psi}
\end{figure}
As already mentioned above, the in-medium binding energies of quarkonia are an important ingredient to compute their dissociation rates, as this determines the available phase space for inelastic reactions, and more compact states (at larger binding) are also subject to interference effects (that depend on the wave function) which further suppress the rate. 
We define the binding energy in the standard way as the difference between the bound-state mass and the $Q\overline{Q}$ threshold (defined as twice the in-medium HQ mass).

\begin{figure}[!bht]
\begin{center}
\includegraphics[width=0.5\linewidth]{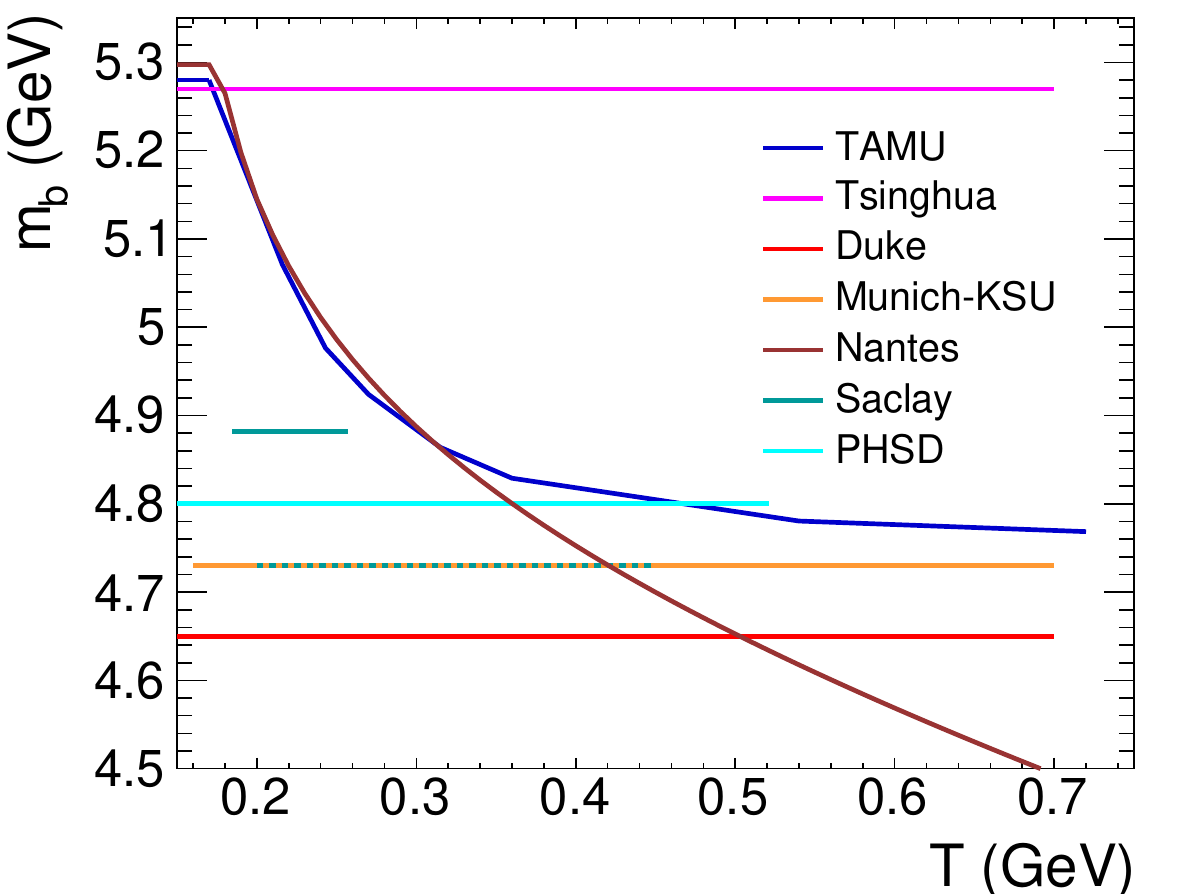}
\end{center}
\vspace{-5mm}
\caption{Temperature dependence of the bottom-quark mass. Dashed line: Saclay, perturbative.}
\label{fig:m_b}
\end{figure}
For charmonia, shown in Fig.~\ref{fig:Eb-psi}, we can reiterate what we discussed before, \ie, a rather good agreement between Nantes and TAMU, while Tsinghua uses a constant $E_b$, in part motivated by the restriction to gluo-dissociation which becomes increasingly inefficient for small $E_b$~\cite{Grandchamp:2001pf} but is extended to excited states by geometric-size scaling. For the excited states the Nantes binding energies are somewhat larger, even exceeding the commonly quoted vacuum values (relative to the open-charm threshold), see, however, Sec.~\ref{subsec:Nantes} for caveats on the model applicability at low $T$. 

For the bottomonium sector, more results are available. One may broadly classify them as constant vs.~$T$-dependent and being based on either a Cornell or a color-Coulomb only potential. Specifically, the Duke, Munich-KSU and Tsinghua groups employ constant $\YiS$ binding energies that are at about half of the vacuum value (similarly for $\YiiS$ and $\chi_b$ for Duke and Munich-KSU), while Nantes, TAMU, Saclay and PHSD have a strong $T$-dependence which, for the former three, starts out near the vacuum value around $T_{\rm pc}$. Furthermore, only Nantes and TAMU have an in-medium $b$-quark mass, which drops with $T$ and results in a near-constant $\YiS$ mass close to the vacuum value. Most other approaches have a significantly smaller $\YiS$ mass even close to $T_{\rm pc}$. Similar observations also hold for the excited states, where in particular the color-Coulomb based approaches have much smaller bound-state masses compared to the vacuum values, mostly due to the relatively small $b$-quark masses, cf.~Fig.~\ref{fig:m_b}.

\begin{figure}[!thb]
\includegraphics[width=.9\linewidth]{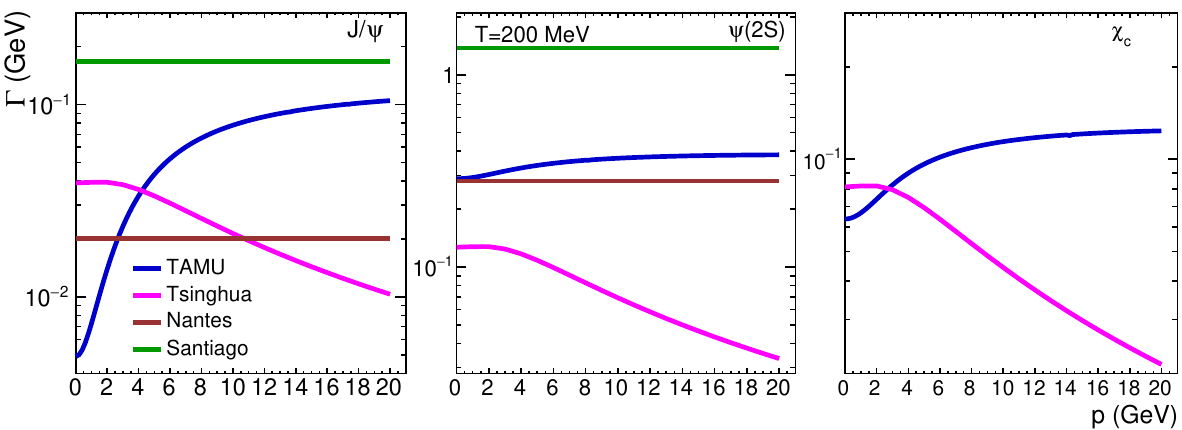}
\includegraphics[width=.9\linewidth]{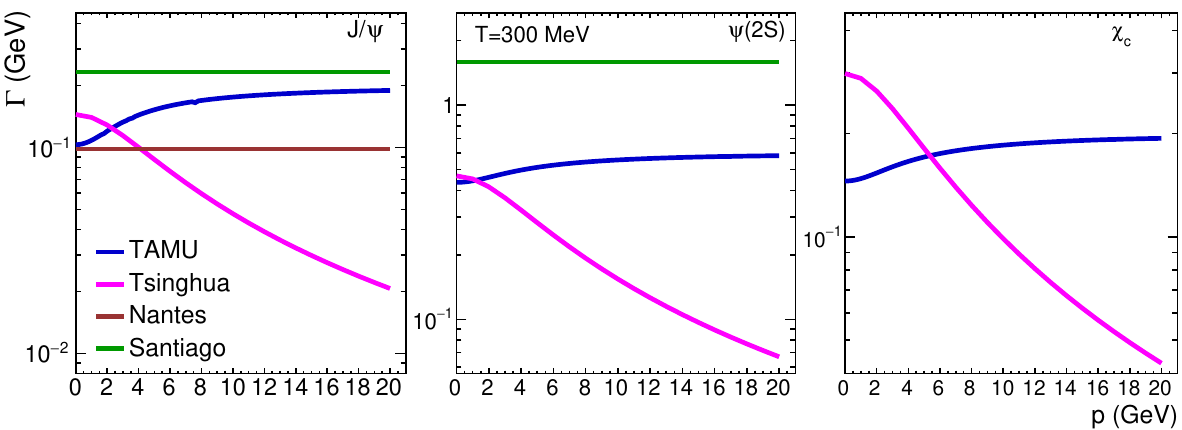}
\includegraphics[width=.9\linewidth]{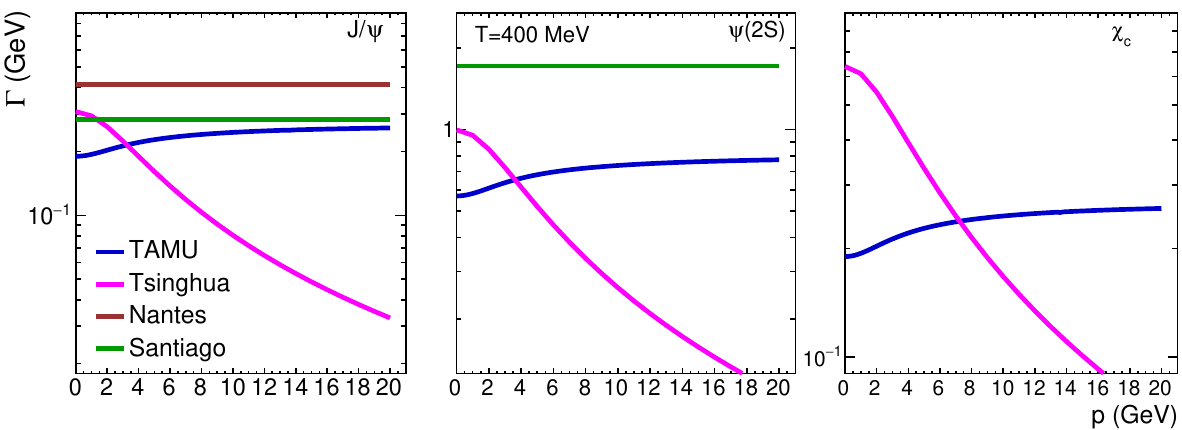}
\vspace{-3mm}
\caption{Momentum dependence of charmonium reaction rates for $T$=200 (upper row), 300 (middle row) and 400 (lower row) MeV.}
\label{fig:gam-p-psi}
\end{figure}

\subsection{Momentum Dependence of Reaction Rates}
\label{ssec_mom-dep}
Transverse-momentum spectra are a pivotal observable in heavy-ion collisions. In the context of quarkonium transport models one needs to calculate the 3-momentum dependence of the dissociation rate, which is directly reflected in the nuclear modification factor of the primordial-production component. However, because of detailed balance, the 3-momentum dependence also affects the regeneration of quarkonia, which is additionally influenced by the interplay with the phase space distribution of heavy quarks as they diffuse through the expanding fireball.  

We start again by inspecting available model results for charmonia, for 3 different temperatures, see Fig.~\ref{fig:gam-p-psi}. The Santiago (comover) and Nantes models are currently restricted to $p$=0 and therefore have no $p$-dependence. The Tsinghua results for the $\jpsi$ exhibit the typical momentum dependence generated by the peaked structure of the gluo-dissociation cross section, namely initially large values that quickly fall off once the center-of-mass energy exceeds the peak position. This effect is further accelerated with increasing temperature due to the increase in thermal-parton motion. As mentioned before, for the excited states, the rates are geometrically scaled by the bound-state size. In the TAMU model, the $\jpsi$ rate at $T$=200\,MeV exhibits a strong increase with momentum which is driven by steep increase in phase space due to a relatively large binding energy of about 550\,MeV, requiring a rather large threshold momentum of the thermal partons. For the $\psip$, where $E_b\simeq$\,10\,MeV at $T$~=~200~MeV (and vanishing at higher $T$), the phase-space suppression is absent and the remaining increase with $p$ of about 25\% is due to the perturbative-scattering matrix element, as can also be seen for the $\jpsi$ at $T$ = 400\,MeV.

\begin{figure}[!htb]
\begin{center}
\includegraphics[width=0.9\linewidth]{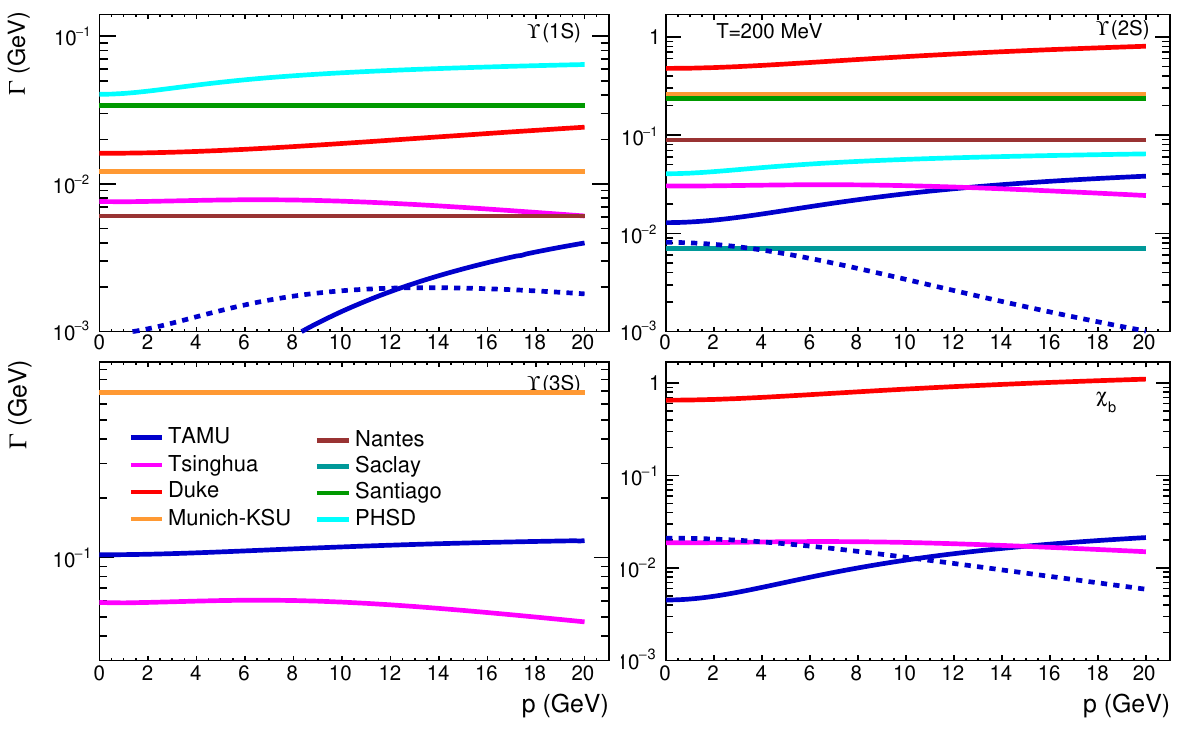}
\end{center}
\vspace{-7mm}
\caption{ 
Momentum dependence of bottomonium reaction rates for $T$ = 200 MeV, \ie, $\YiS$ (upper left), $\YiiS$ (upper right), $\YiiiS$ (lower left) and $\chi_b(1\mathrm{P})$ (lower right) (dashed lines: TAMU gluo-dissociation).}
\label{fig:gam-p200-ups}
\end{figure}
\begin{figure}[!tb]
\begin{center}
\includegraphics[width=0.9\linewidth]{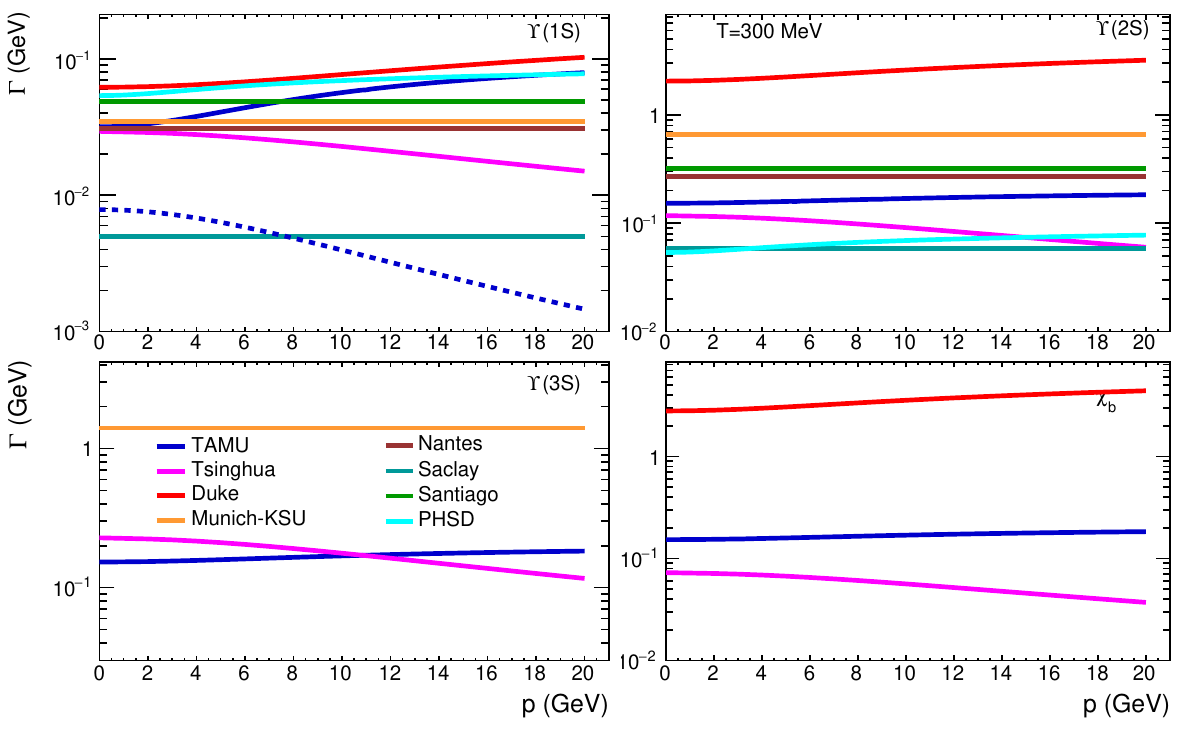}
\end{center}
\vspace{-7mm}
\caption{Same as Fig.~\ref{fig:gam-p200-ups}, but for $T$=300 MeV.
}
\label{fig:gam-p300-ups}
\end{figure}
\begin{figure}[htb]
\begin{center}
\includegraphics[width=0.9\linewidth]{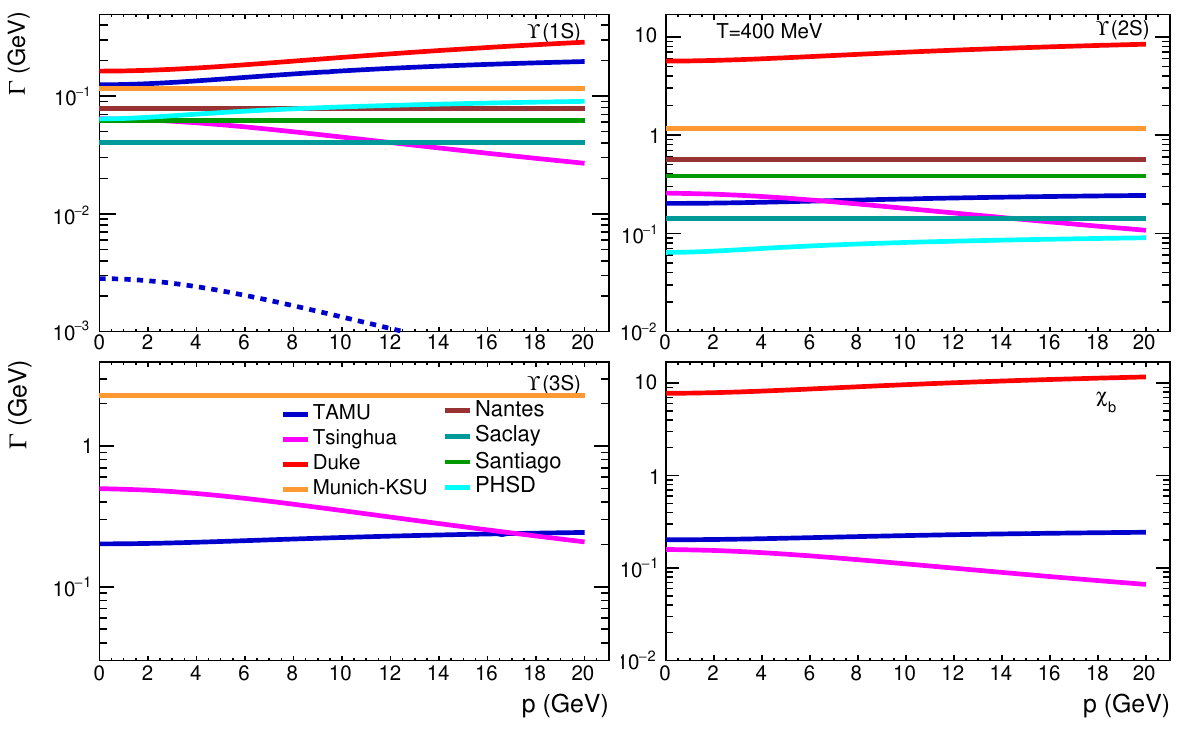}
\end{center}
\vspace{-7mm}
\caption{
Same as Fig.~\ref{fig:gam-p200-ups}, but for $T$=400 MeV. 
}
\label{fig:gam-p400-ups}
\end{figure}

The 3-momentum dependence of bottomonium reaction rates is compiled in Figs.~\ref{fig:gam-p200-ups}, \ref{fig:gam-p300-ups} and \ref{fig:gam-p400-ups} for temperatures of 200, 300 and 400\,MeV, respectively. At $T$ = 200\,MeV, the spread in the rates is substantial, spanning more than an order of magnitude for all states. For the $\YiS$, this spread is mitigated by the overall smallness of the rates (except for PHSD and the Santiago models, recall Fig.~\ref{fig:gam-t-ups}) which renders them phenomenologically of little relevance. Nevertheless, one finds a qualitative ordering according to the binding energy (recall Fig.~\ref{fig:Eb-ups}), with decreasing rates following the increasing binding energy from PHSD to Duke to Munich-KSU to Tsinghua to Nantes and TAMU. While the Santiago, Munich-KSU and Nantes models have currently no $p$-dependence, PHSD and Duke show an increasing trend which is even more pronounced in TAMU due to the large binding (suppressing the phase space at small $p$), while the gluo-dissociation of Tsinghua is decreasing 
(roughly comparable to the TAMU gluo-dissociation for the $\chi_b$ which has approximately the same binding at this temperature). For the $\YiiS$ the spread is also large (with little $p$-dependence in most models), which, as noted before, is likely due to current uncertainties in the phenomenological implementations. Similar remarks apply to $\YiiiS$ and $\chi_b(1\mathrm{P})$.

At higher temperatures, the model agreement for the $\YiS$ is better (as noted in the context of the $T$-dependence). The TAMU and Tsinghua rates for the excited states at $T\ge300$\,MeV are probably underestimated, since once the binding vanishes, one is basically dealing with heavy-light couplings, which are underestimated when using perturbative diagrams (recall, \eg, the $K$-factor for the $\psip$)~\cite{Du:2019tjf,He:2022ywp}. This is probably more realistic in models based on the dipole expansion where the rate is taken as $\Gamma\simeq r^2 \kappa$ (with the HQ momentum diffusion coupling, $\kappa$, usually taken from lattice-QCD results), which incorporates non-perturbative interaction strength (which is large~\cite{Rapp:2018qla}). However, this expansion breaks down for small $E_b$ where $r$ becomes large, and similarly for the quantum-optical limit in the Duke model (where, \eg, $\Gamma_{\chi_b}\simeq 10$\,GeV at $T$ = 400\,MeV).

\subsection{Imaginary Potential in Coordinate Space}
\label{ssec_ImV}
Next we turn to models that include an explicit ($r$-dependent) imaginary part in their heavy-quark potentials. This enters into transport models based on open quantum systems where the quarkonium wave function is coupled to a QGP medium consisting of light degrees of freedom~\footnote{An $r$-dependent dissociation rate can also be obtained by incorporating an interference factor to diagrammatic computations with subsequent use in semiclassical transport~\cite{Song:2007gm,Du:2017qkv}}.  Upon integrating out the light degrees of freedom, one can obtain an evolution for the heavy-quarkonium reduced density matrix. In the Markovian limit that emerges when the time scale for medium relaxation, $1/T$, is much faster than the time for internal transitions (parametrically given by $1/E_b$), the resulting evolution equation is of Lindblad form~\cite{Katz:2022fpb,Brambilla:2023hkw}. The real part of the potential can be combined with the transition rates, defined in terms of jump operators, into a complex effective Hamiltonian. The Lindblad equation is then solved by using either the method of quantum trajectories or quantum diffusion with a stochastic potential where the correlators for the stochastic terms in the potential are set by the imaginary part of the potential. This leads to a non-unitary dynamics of the quantum mechanical wave function via the complex effective Hamiltonian, which results in suppression of quarkonium states. 

\begin{figure}[!thb]
\begin{center}
\includegraphics[width=0.8\linewidth]{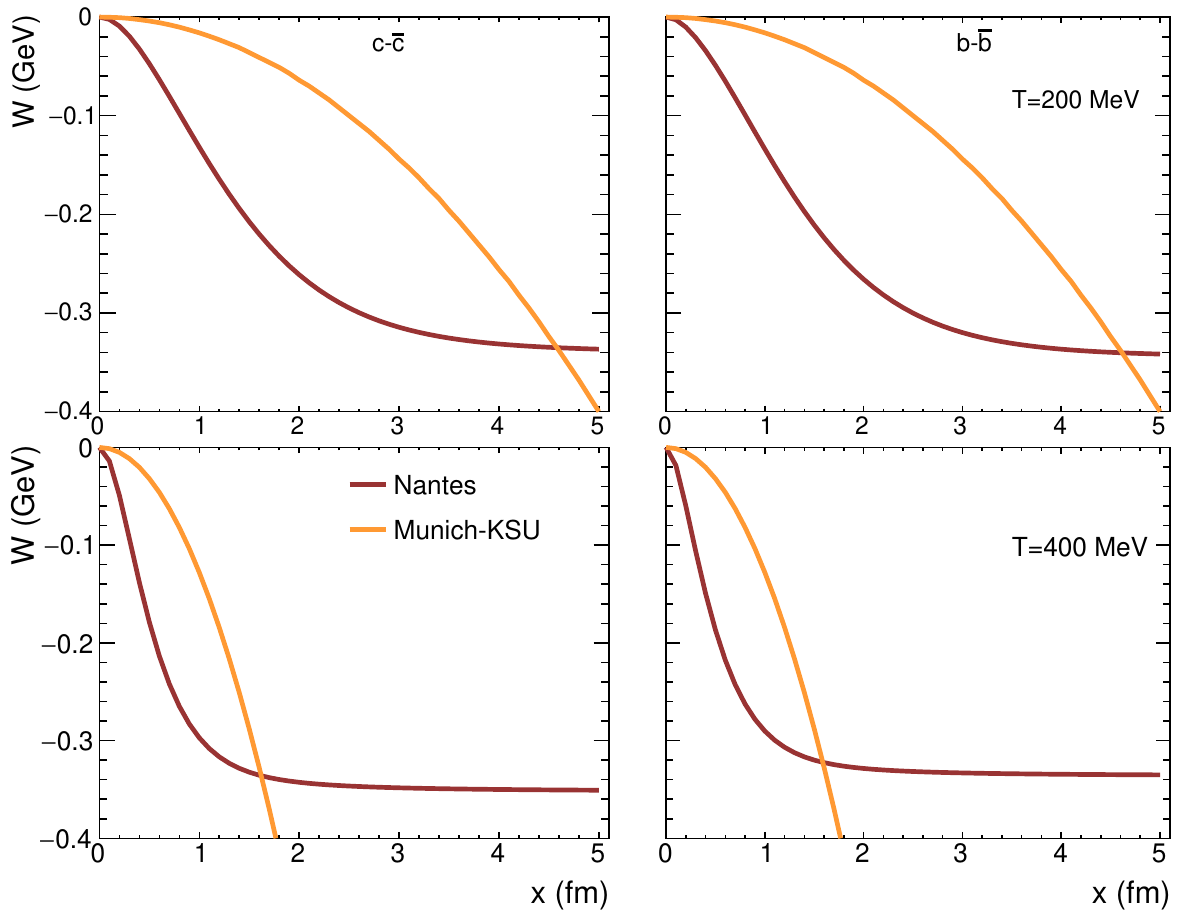}
\end{center}
\vspace{-7mm}
\caption{
The imaginary part of the effective potential for charmonium (left) and bottomonium (right). The top and bottom rows correspond to $T$ = 200 and 400 MeV, respectively.
}
\label{fig:homework5}
\end{figure}
In Fig.~\ref{fig:homework5} we compare the imaginary part, $W$, of the singlet effective potential, $V_{\rm singlet, eff} = \Re[V] + i W$, used in the Nantes and Munich-KSU OQS approaches. At low temperatures (upper panels), the Nantes and Munich-KSU results are rather different, while at higher temperatures (lower panels) they are closer, although the functional form remains qualitatively different. In the Munich-KSU approach, the imaginary part of the singlet effective potential is manifestly independent of the HQ mass and, at leading order in a $E_b/T$ expansion, given by $W = - i \hat\kappa r^2/2$, where $\hat\kappa = \kappa/T^3$ is the scaled heavy-quarkonium transport coefficient which can be defined via chromoelectric correlators; here, a temperature-dependent parameterization is used that was fit to lQCD data in Ref.~\cite{Brambilla:2020siz}.  We note that beyond leading order in $E_b/T$ the imaginary part of the effective Hamiltonian cannot be expressed as a local potential as there appear additional contributions proportional to the anti-commutator between the HQ relative momentum and their distance.  For purposes of comparison, both the Munich-KSU and Nantes groups did not include such momentum-dependent contributions to the width.

The dipole expansion in the Munich-KSU approach, which causes the $r^2$ behavior, limits the reliability of the imaginary part of the potential to relatively small distances. In practice, this problem is mitigated since the imaginary part of the effective potential causes the wave function to be rather localized near the origin when using the effective Hamiltonian evolution.  
From Fig.~\ref{fig:homework5} one sees that the imaginary part in the Nantes approach has a weak dependence on the HQ input mass. With increasing temperature its structure is compressed to smaller distances, a feature that is also seen in the Munich-KSU framework.

\subsection{Nuclear Modification Factor with a Common Reaction Rate}
\label{ssec_fix-rate}
\begin{figure}[htb]
\begin{tabular}{lr} \begin{minipage}{.49\textwidth}
\hspace{-.5cm}{\includegraphics[width=1.\textwidth]{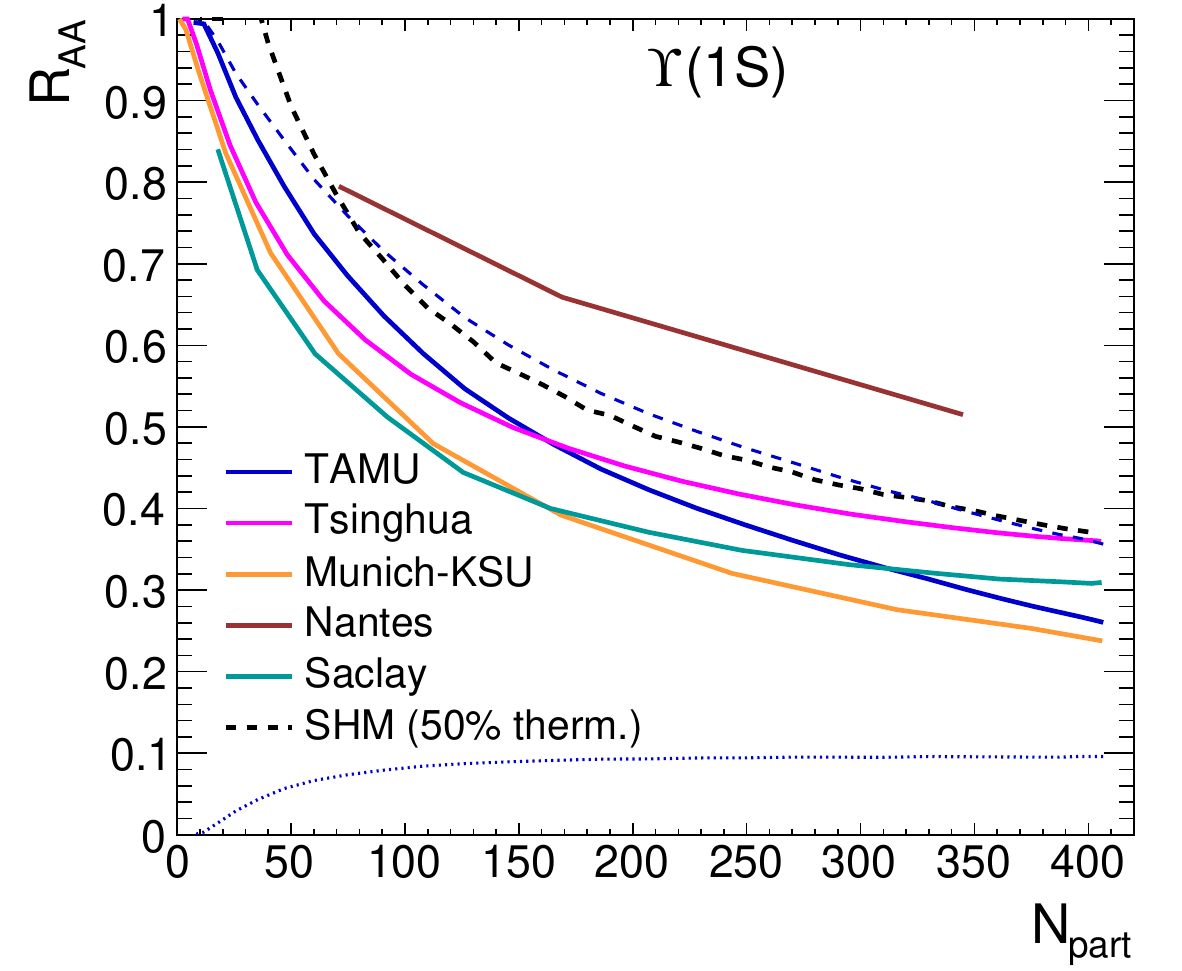}}
  \end{minipage} & \begin{minipage}{.49\textwidth}
    \hspace{-.8cm}{\includegraphics[width=1.\textwidth]{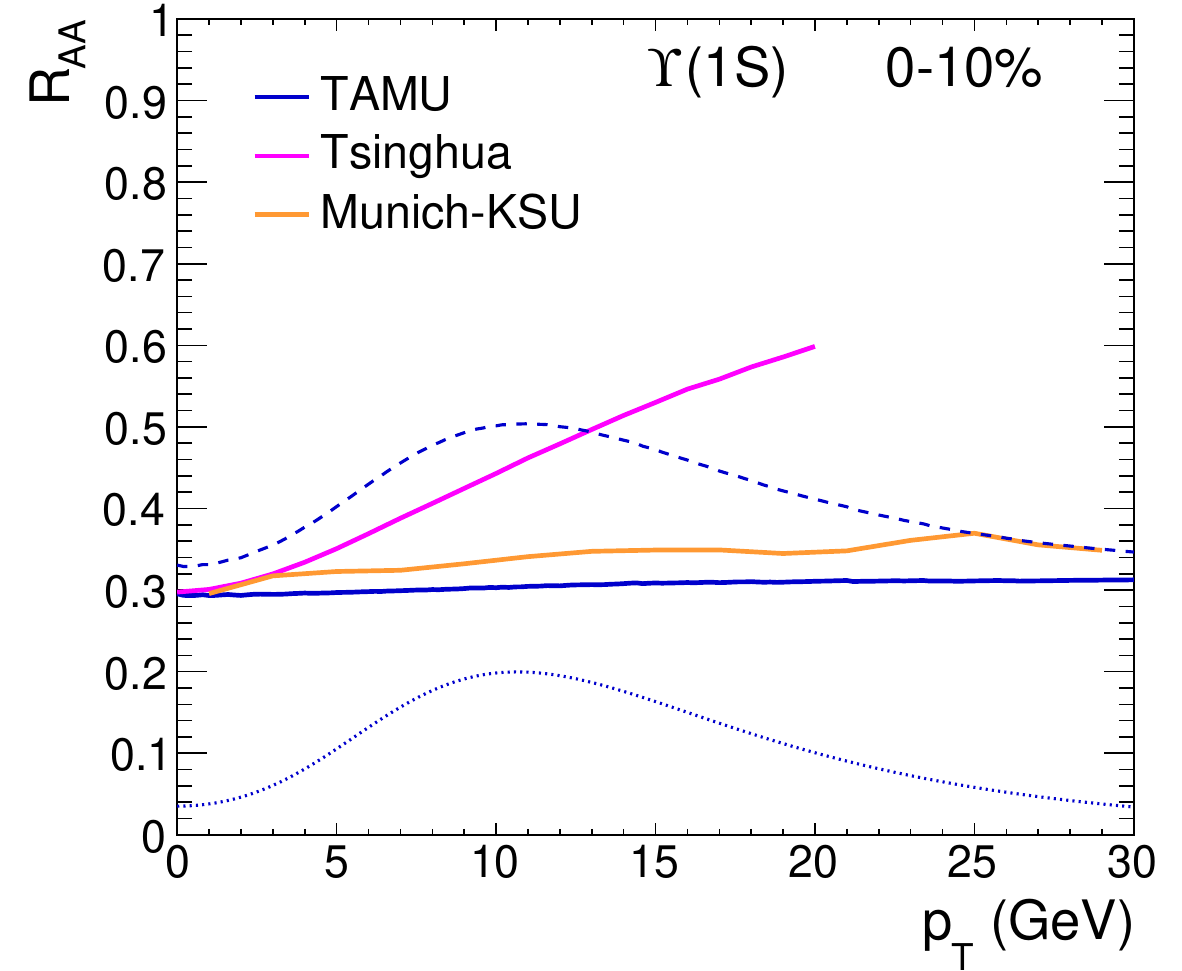}}
\end{minipage}\end{tabular}
\caption{
Left: $\raa$ as a function of $N_{\rm part}$ for $\YiS$, $\pT$-integrated; right: $\raa$ as a function of $\pT$ for 0-10\% centrality.  
 For the TAMU model the regeneration component is shown separately (dotted line) and also summed to the suppression component (dashed line). The results of the SHM, included in the left plot for the case of 50\% thermalized bottom quarks, constitute pure generation at $T_{\rm pc}$.}
\label{fig:homework3}
\end{figure}
Another comparison of the models was conducted for the $\raa$ observable for direct $\YiS$ production by imposing a simple but uniform parametrization for $\Gamma_\Upsilon$ (assumed to be constant vs. momentum), consisting of a linear increase from 0 at $T$=200\,MeV to 0.2 GeV at $T$=600\,MeV, which every group implemented in their respective evolution model\footnote{For models not directly relying on $\Gamma$ as an input for the dynamical evolution, a proper rescaling of the corresponding quantity -- like the imaginary potential $W$ -- has been applied.}. Initial formation time effects for the bound states, as well as CNM effects and feeddown, have been neglected. The results are shown in Fig.~\ref{fig:homework3} as a function of centrality and $\pT$. For the TAMU model the regeneration component is shown separately as well as its sum with the (suppressed) initial production. The results of the SHM, included for the case of 50\% thermalized bottom quarks, constitute pure generation at $T_{\rm pc}$.

At first sight, models vary significantly in the way they describe centrality dependence. 
However, closer inspection reveals that the Nantes results are anomalously high as the underlying EPOS4 background is assumed to thermalize only after $\approx 1.5\,{\rm fm}/c$ which leads to a rather low initial temperature, recall Fig.~\ref{fig:temp}.
The TAMU result is reasonably close to the 3 curves by Saclay, Munich-KSU and Tsinghua, with a shape close to Munich-KSU but with a higher yield in peripheral collisions which is presumably caused by larger medium thermalization times leading to smaller initial temperatures, cf.~Fig.~\ref{fig:temp}, while Tsinghua has a smaller initial temperature in central collisions but a larger one at "intermediate" centralities which could explain the stronger suppression in more peripheral collisions. The Saclay and the Munich-KSU results are closest to each other, which may be due to due to longer lifetime in the Bjorken model (Saclay) and escape effects in the Munich-KSU which render the latter's $\raa$ higher in peripheral but lower in central collisions where the initial temperature is higher.
Both models generate a stronger suppression than TAMU, which is largely consistent with the temperature evolution shown in Fig.~\ref{fig:temp}.

\subsection{Quarkonium Formation Time Effects}
\label{ssec_tauf}
 To study the impact of quarkonium formation times calculations were performed starting from a ``realistic" initial $Q\overline{Q}$ state (the one used in the respective dynamical model, usually reported as a "point-like initial state" in the OQS and the ground state in semiclassical approaches). This state was  evolved in a QGP at fixed temperature $T$=300 MeV, neglecting regeneration. The models provided  the "survival" probability as a function of  time to find this $Q\overline{Q}$ pair at $p$=0 in an eigenstate of the in-medium potential. 
 \begin{figure}[htb]
\begin{tabular}{lr} \begin{minipage}{.49\textwidth}
\hspace{-.5cm}{\includegraphics[width=1.03\textwidth]{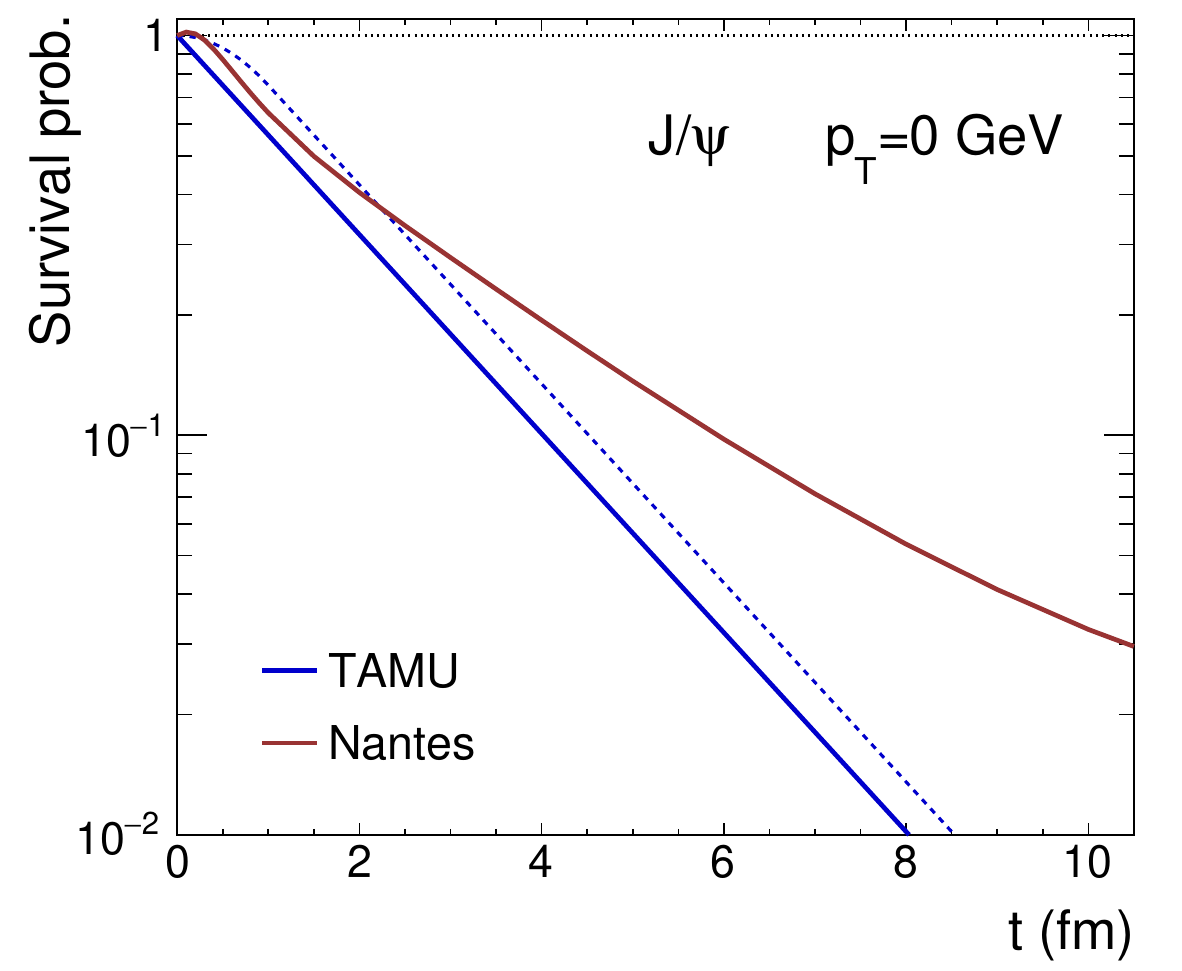}}
  \end{minipage} & \begin{minipage}{.49\textwidth}
    \hspace{-.8cm}{\includegraphics[width=1.03\textwidth]{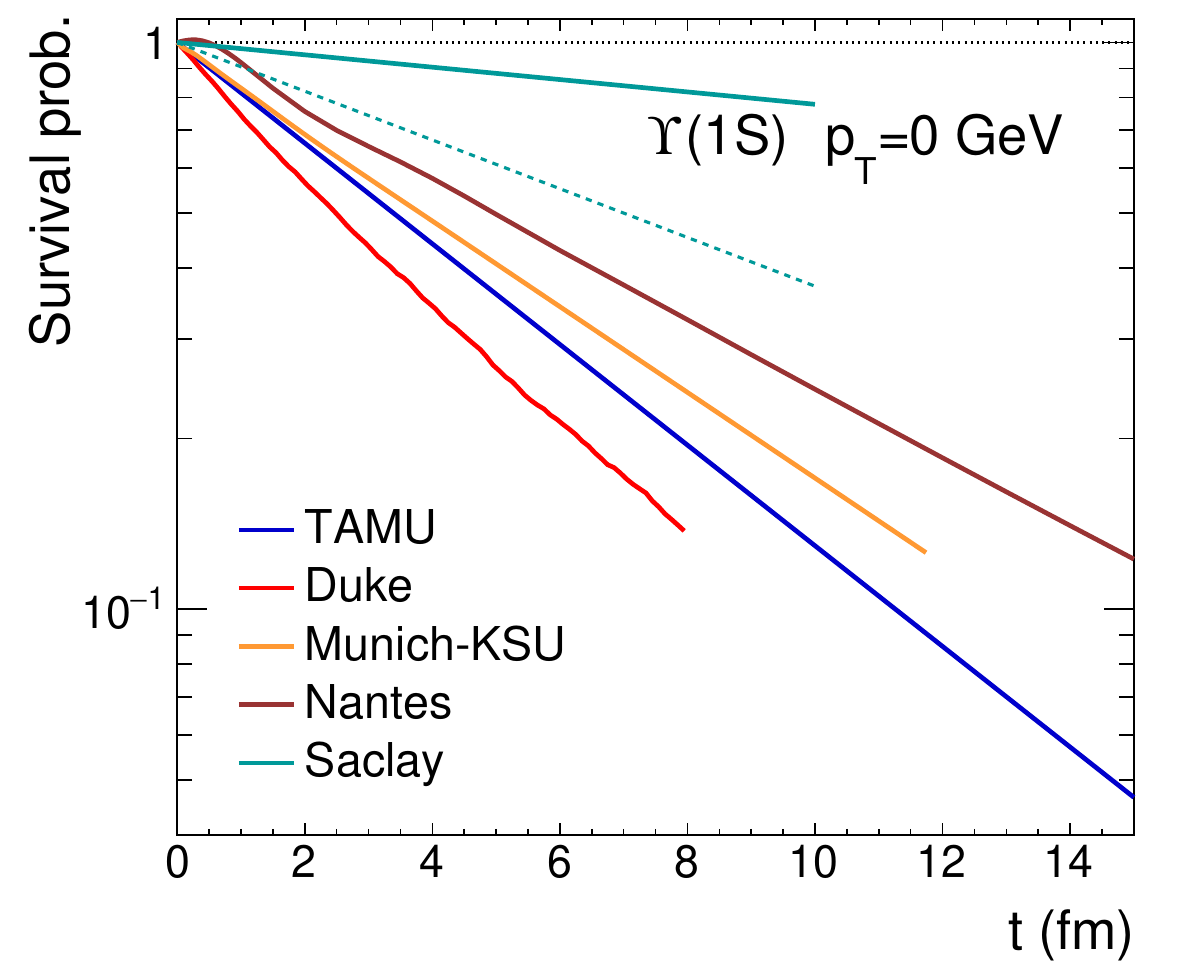}}
\end{minipage}\end{tabular}
\vspace{-3mm}
\caption{
The survival probability as a function of time for charmonium (left panel) and bottomonium (right panel) ground state at $\pT$=0 for an in-medium state. For the TAMU model, the dashed (solid) line in the left plot is for the case with (without) formation time. The dashed line in the right plot shows the perturbative case for the Saclay model (the default is lattice).  All calculations are for a constant temperature of 300 MeV. NB: Nantes and Duke contain regeneration.}
\label{fig:homework4}
\end{figure}

Figure~\ref{fig:homework4} illustrates how suppression mechanisms underlying the calculations of the decay rate are realized in a basic time evolution scenario at constant $T$. We focus the discussion on the bottomonium case which was addressed by most of the groups. In the TAMU approach, where the initial state is an in-medium $\YiS$ state and regeneration mechanisms were discarded for the purpose of this study, one finds as expected a survival probability $=\exp(-\Gamma t)$, where $\Gamma$ agrees with the reaction rate displayed in Fig.~\ref{fig:gam-t-ups} (including the gluo-dissociation mechanism); the inclusion of a formation time typically delays the evolution, with an offset of $\approx 0.05\,{\rm fm}/c$. The same exact agreement with the exponential decay law is obtained in the Saclay calculation as the regeneration was not considered in this implementation of the model. In the Duke approach the regeneration component was not removed, leading to a slight deviation with respect to the exponential decay initiated with a vacuum state of the $\YiS$\footnote{Note that "in-medium states" are not defined in the Duke approach owing to the $\Gamma \ll E_b$ hierarchy.}, of the order of 5\% after 8 fm/$c$. In the Munich-KSU calculation, the evolution starts from a compact state close to a Dirac $\delta$-function peak. 
While the evolution of the survival probability of the in-medium state decreases nearly exponentially, the associated decay rate is found to be twice the imaginary part of the 
eigenvalue  ($\approx 8.95 - 0.017i$ GeV) corresponding to the fundamental eigenstate of the non-hermitian effective Hamiltonian, which is an indication that this mode governs the evolution starting from very early time. In the Nantes approach, the initial condition is taken as a compact $b\bar{b}$ state as well. At early time (during typically the first 1 fm/$c$), the dynamics is impacted by quantum interference which leads to a non-exponential decay of the survival probability and could be interpreted as an effective formation time. After 1 fm/$c$, one observes an exponential decrease, with a decay rate close to the one evaluated in Fig.~\ref{fig:gam-t-ups} for an in-medium state.    

\section{Conclusions}
\label{sec_concl}
In this report we have summarized the common effort of a task force, composed of various theoretical groups, to scrutinize their models for the description of quarkonium production in high-energy heavy-ion collisions. To begin with, a synopsis of each model was given guided by 19 items to specify the inputs and spell out underlying assumptions. This revealed a large variety of the approaches, \eg, input potentials ranging from perturbative color-Coulomb to Cornell potentials with different degrees of screening, reaction rates based on gluo-dissociation and/or inelastic scattering, implemented into semiclassical vs.~quantum transport models, deviations in the treatment of regeneration (from none to diagonal to multiple independent $Q\bar{Q}$ recombinations), and different accounting for cold-nuclear-matter effects, to name a few. A set of calculations of derived quantities has been defined to study how these differences manifest themselves in key quantities of the transport approach, specifically in-medium binding energies and HQ masses, inelastic reaction rates in momentum space (and/or imaginary parts of the potential in coordinate space), and nuclear modification factors with controlled rates and medium conditions.    

Let us summarize the results in light of the 5 basic questions posed in the introduction. 
{\it First}, concerning the model consistency, the semiclassical approaches employ rather different inputs for the in-medium binding energy and the reaction rate (which is strongly affected by the binding energy), while all of them include regeneration in a way that accounts for multiple heavy-quark pairs in the system; in current quantum-transport approaches, which mostly focus on bottomonia, regeneration, if any, is only based on a single $b\bar{b}$ pair, sometimes further restricted to the configurations that in a pp collision would form an $\Upsilon$ state (which is a small subset). 
{\it Second}, concerning the equilibrium limits relevant for regeneration, they have not been explicitly compared, but from the model descriptions it appears that currently only the semiclassical treatments have control over them, although they are expected to be quantitatively different due to different HQ masses and binding energies. 
{\it Third}, concerning the significance of quantum effects, the comparisons of suppression factors appear to confirm that these are mostly relevant in the early stages; the long-time behavior of suppression can be characterized by exponential decays that correspond to the pertinent reaction rate in semiclassical approaches and the lowest eigenvalue in the quantum approaches. 
{\it Fourth}, concerning lQCD constraints, they have been implemented in a number of models, either in terms of directly computed quantities, \ie, transport coefficients (which, however, are restricted to vanishing 3-momentum), or more indirectly by computing lQCD quantities (\eg, free energies or Euclidean correlators) within a model approach to constrain its input quantities (like the potential or HQ masses); 
the latter variant usually offers broader phenomenological flexibility as well as microscopic insights.  
{\it Finally}, concerning the ultimate model uncertainties, it will be necessary to go beyond a diagnostic level of comparing the model calculations of specific quantities as conducted in this work. 
Clearly, a more systematic implementation of lQCD constraints on the input quantities (such as the in-medium potential) on an equal footing across model approaches is desirable. Then one could envision that a comparison of semiclassical to quantum transport approaches with the same microscopic input (such as in-medium potential and HQ masses, evaluated within the same process for the reaction rate, \eg, inelastic parton scattering) could reveal systematic uncertainties in the transport part. This might also identify in how far simplifications in certain components of the model calculations are justifiable. In doing all this, it will be important to account for the strongly coupled nature of the QGP, which manifests itself in both binding energies and reaction rates. While the former are a direct consequence of the in-medium QCD force (input potential), the latter are closely related to the single HQ transport coefficient, which has been experimentally established to be in the strong-coupling regime. Since the HQ interactions with the QGP which govern HQ transport can also be expected to be operative in the bound-state properties of quarkonia, a self-consistent treatment will provide the strongest constraints. 

\section*{Acknowledgements}

We thank the Extreme Matter Institute for support for this Rapid Reaction Task Force.  The work of XD was supported by the Deutsche
Forschungsgemeinschaft (DFG, German Research Foundation) through the CRC-TR 211 ’Strong-interaction matter under extreme conditions’– project number 315477589 – TRR 211; XD, MAE and EGF acknowledge support from the
Xunta de Galicia (Centro singular de investigacion de Galicia accreditation 2019-2022), the European Union ERDF, the “Maria de Maeztu” Units of Excellence program under projects CEX2020-001035-M and CEX2019-000918-M, the Spanish Research State Agency under projects PID2020-119632GB-I00 and PID2019-105614GB-C21, and the European Research Council under project ERC-2018-ADG-835105 YoctoLHC;
the work of MAE has also been supported by  the Generalitat de Catalunya under grant 2021-SGR-249.
MS was supported in part by U.S. Department of Energy award No.~DE-SC0013470; 
PP  was supported  by  the U.S.  Department  of  Energy,   Office  of  Science,   Office   of   Nuclear   Physics   through   Contract   No.   DE-SC0012704;
XY was supported in part by the U.S. Department of Energy, Office of Science, Office of Nuclear Physics, InQubator for Quantum Simulation (IQuS) under Award Number DE-SC0020970 via the program on Quantum Horizons: QIS Research and Innovation for Nuclear Science; SD is supported by the Centre national de la recherche scientifique (CNRS) and Région Pays de la Loire and acknowledges the support of Narodowe Centrum Nauki under grant no. 2019/34/E/ST2/00186.
PBG was supported by the European Union’s Horizon 2020 research and innovation program under grant agreement No 824093 (STRONG-2020). PVG was supported by the U.S. Department of Energy Award No. DE-SC0019095 and is grateful for the support and hospitality of the Fermilab theory group; RR and BW  have been supported by the U.S. National Science Foundation under grant nos. PHY-1913286 and PHY-2209335; PP, RR, MS, RV and BW have been supported by the U.S. Department of Energy, Office of Science, Office of Nuclear Physics through the Topical Collaboration in Nuclear Theory on Heavy-Flavor Theory (HEFTY) for QCD Matter under award no. DE-SC0023547

\setlength{\bibsep}{0.6ex}
\bibliography{rrtf}


\end{document}

%% file: models/dukemit.tex
\subsection{The Duke-MIT approach}

The Duke-MIT approach is based on a set of coupled Boltzmann equations that describe the in-medium evolution of both heavy quarks and quarkonia~\cite{Yao:2020xzw}. The Boltzmann equation for quarkonia describes their dissociation and recombination. It is derived by using the pNRQCD effective field theory in the hierarchy of energy scales
\begin{align}
M\gg Mv \gg Mv^2 \gtrsim T \,,
\end{align}
and the open quantum system approach in the quantum optical limit~\cite{Yao:2018nmy,Yao:2020eqy,Yao:2021lus}. The symbol $Mv^2\gtrsim T$ means that the temperature is on the order of or smaller than the binding energy; e.g., $T=1.5 Mv^2$ is still in the validity region of the framework. The dissociation and recombination processes are formulated in a factorized way such that their collision terms can be written as a convolution of a dipole transition term of the heavy quark-antiquark wave function and a generalized (chromoelectric) gluon distribution function of the QGP~\cite{Yao:2020eqy}. The generalized distribution function (GGD) is a chromoelectric field correlator defined in a gauge-invariant and non-perturbative way, and it naturally generalizes the gluo-induced and inelastic scattering processes to account for the situation where the medium partons inducing these processes can be off-shell and non-perturbative. When $T\gg \Lambda_{\rm QCD}$, the GGD has been calculated to next-to-leading order accuracy~\cite{Binder:2021otw}. When $T\sim \Lambda_{\rm QCD}$, the GGD has been calculated by holography~\cite{Nijs:2023dks,Nijs:2023dbc} and can be studied by Euclidean lattice QCD calculations~\cite{Scheihing-Hitschfeld:2023tuz}. The consistency between the dissociation and recombination implementations has been tested in a static QGP box~\cite{Yao:2017fuc}. The bulk dynamics of the QGP is described by a 2+1D viscous hydrodynamic equation that has been implemented in the ``VISHNU'' package and calibrated against the production of light particles~\cite{Moreland:2014oya}.
The Boltzmann equation for open heavy quarks describes their diffusion and energy loss in the QGP, which has been implemented in the ``Lido'' package and has been calibrated against open heavy-flavor production observables~\cite{Ke:2018tsh}. The effect of nPDFs is taken into account by using the EPPS16 parameterization~\cite{Eskola:2016oht} and the feed-down network is considered in the hadronic stage. The framework has been used in phenomenological studies of bottomonium production in heavy ion collisions~\cite{Yao:2020xzw}.


\ingredients{

    \ingredient{In-medium potential:} 
    An unscreened Coulomb potential is used as the in-medium potential. 

    \ingredient{Vacuum limit of potential/spectroscopy:}
    The Duke-MIT approach reproduces the $\Upsilon$(1S) mass in vacuum by construction but not the 2S and 3S excited states, which is a consequence of using a Coulomb potential.

    \ingredient{Reaction rates:}
    For dissociation, both gluo-dissociation and inelastic scattering contributions are included in a pNRQCD based calculation. The pNRQCD calculation takes into account the interference between the heavy quark scattering and the heavy antiquark scattering. Therefore, it goes beyond the quasi-free approximation. For recombination, similar processes (gluo-induced and inelastic) are included~\cite{Yao:2018sgn}. For unbound heavy quark-antiquark pairs, the diffusion and energy loss are included independently for each heavy quark.
    
    Three-momentum dependence comes in since the reaction rates are calculated in the rest frame of the heavy quark-antiquark pair for dissociation and recombination and boosted back to the laboratory frame~\cite{Yao:2019rcb}. The rest frame is a natural frame for the non-relativistic treatment in the effective field theory. However, for sufficiently large transverse momenta, the non-relativistic treatment breaks down in the rest frame since the partons from the medium get significantly boosted.

    \ingredient{Assumptions about the medium (degrees of freedom, etc.):}
    The calculations of the gluo-induced and inelastic processes mentioned above assume the medium consists of free quarks and gluons. However, the Boltzmann equation for quarkonium formulated in our framework can easily go beyond this. The nonperturbative nature of the medium partons relevant for quarkonium is encoded in terms of the chromoelectric field correlator mentioned above. It has been shown that, beyond leading-order in the strong coupling, this chromoelectric field correlator for quarkonium is different from the one used to define the heavy quark diffusion coefficient~\cite{Binder:2021otw,Scheihing-Hitschfeld:2022xqx}.

    \ingredient{Temperature dependence of heavy quark masses:}
    The heavy quark mass is treated as independent of temperature here. 

    \ingredient{Equilibrium limits in transport:}
    The heavy quark Boltzmann equation can drive the system of unbound heavy quarks to kinematic equilibrium. The quarkonium Boltzmann equation can drive the system of bound and unbound pairs into chemical equilibrium~\cite{Yao:2017fuc}.

    \ingredient{Constraints from lattice QCD:}
    At the moment, there is no constraint from lattice QCD calculations. However, the next systematic step is to apply lattice QCD techniques to calculate the chromoelectric field correlator mentioned above~\cite{Scheihing-Hitschfeld:2023tuz}.

    \ingredient{Range of applicability and how is this range established:}
    The range of applicability is established by the power counting of the effective field theory pNRQCD. When the temperature is much bigger than the inverse of a typical quarkonium size, quarkonium cannot exist as a well-defined bound state inside the medium, so the dynamics are effectively described in terms of an unbound but possibly correlated pair. When the temperature is much smaller than the inverse of the typical quarkonium size, the effective dynamics is described as quarkonium dissociation and recombination and these processes can be systematically calculated in pNRQCD. When the medium becomes nonperturbative, a nonperturbative determination is needed.  Due to the use of the optical limit there is also a upper limit on the temperature since the optical limit assumes that the binding energy of states is small compared to $\pi T$.
    
    \ingredient{Quantum features:}
    The Boltzmann equation for quarkonium is derived from the open quantum system in the quantum optical limit, with a systematic semi-classical (gradient) expansion. Quantum corrections to the recombination term have been worked out but not yet implemented. There is no quantum correction to the dissociation term~\cite{Yao:2020eqy,Yao:2021lus}.

    \ingredient{Regeneration:}
    The recombination is formulated in the open quantum system approach (quantum optical limit) and thus it is consistent with the description of dissociation~\cite{Yao:2018nmy,Yao:2020eqy,Yao:2021lus}. The recombination term in the Boltzmann equation is calculated in pNRQCD, which is valid when the quarkonium size is small, compared to the medium temperature. When the temperature is much bigger, the effective dynamics are described in terms of an unbound but possibly correlated heavy quark-antiquark pair, which is consistent with the quantum Brownian motion limit of the open quantum system framework.

    \ingredient{Coupling to open heavy-flavor sector:}
    The open heavy-flavor transport is an important part of the coupled Boltzmann equations. It provides the heavy quark-antiquark distribution that is needed as an input in the recombination term, which allows the study of both correlated and uncorrelated recombination processes. The open heavy-flavor transport is necessary for the system to reach kinetic equilibrium.

    \ingredient{Hadronic-Phase Transport:}
    No hadronic-phase transport except for feed down.

    \ingredient{Initial quark/quarkonium distributions:}
    Momentum distribution: Pythia with nPDF parametrized by EPPS16; spatial distribution: binary collision density obtained from Trento that has been calibrated against light particle production observables \cite{Moreland:2014oya,Bernhard:2016tnd}.

    \ingredient{Are cold nuclear matter effects, nPDF effects, etc. taken into account?:}
    nPDF parameterized by EPPS16.

    \ingredient{Constraints from pA and dA collisions:}
    EPPS16 uses some pA data to fit the nPDF. We also use p-Au data from STAR to fix the cold nuclear matter effects in 200 GeV Au-Au collisions.

    \ingredient{Medium evolution model:}
    2+1D viscous hydro (``VISHNU'') that has been calibrated against light particle production observables.
 
    \ingredient{Feed down implementation:}
    Feed down contributions from 2S, 3S, 1P, and 2P states for $\Upsilon$ states.

    \ingredient{Comparisons to experimental data:}
    Can describe $\Upsilon$(nS) data from both LHC and RHIC collisions, except for the recent $\Upsilon$(3S) data from CMS since recombination of $\Upsilon$(3S) has not been included.
    
    \ingredient{Phenomenological breadth:}
    We included $\Upsilon$(nS) and $\chi_b(nP)$ in the reaction network. We studied both RHIC and LHC collision energies.
}

%% file: models/munichksu.tex
\subsection{The Munich-KSU approach}

We utilize a set of coupled quantum evolution equations for the reduced density matrix derived in Refs.~\cite{Brambilla:2016wgg,Brambilla:2017zei,Brambilla:2022ynh} using the formalism of open quantum systems (OQS) and the effective field theory (EFT) potential nonrelativistic QCD (pNRQCD)
\cite{Brambilla:1999xf,Brambilla:2004jw}.
The resulting set of coupled evolution equations describes the time evolution of singlet and octet Coulombic states propagating in a thermal medium at temperature $T$ realizing the hierarchy of scales
\begin{equation}\label{eq:hierarchy_of_scales}
	M \gg 1/a_{0} \gg \pi T \sim gT, \Lambda_{\rm QCD}  \gg E \, ,
\end{equation}
where $M$ is the heavy quark mass, $a_{0}$ is the Bohr radius of the quarkonium state, $T$, is the temperature of the strongly-coupled QGP medium and $E$ is the Coulombic binding energy of the bound state.
In this regime, the evolution equations take the form of a Lindblad equation
\begin{equation}
	\frac{d\rho(t)}{dt} = -i \left[ H, \rho(t) \right] + \sum_{n=0}^1
	\left( C_{i}^{n} \rho(t) C^{n\dagger}_{i} - \frac{1}{2} \left\{ C^{n \dagger}_{i} C_{i}^{n}, \rho(t) \right\} \right),
	\label{eq:Lindblad}
\end{equation}
where $\rho(t)$ and $H$ are the quarkonium density matrix and in-medium Hamiltonian and the collapse operators $C_{i}$ encode interactions with the medium.
The anticommutator term is responsible for the in-medium width.
The first term in parentheses ensures that the evolution is trace preserving.
For a detailed discussion, see Eqs.~(2)-(5) and accompanying text of Ref.~\cite{Brambilla:2023hkw}.

We work in the bottom sector, which realizes the dilute limit described by the above Lindblad equation linear in the quarkonium density $\rho(t)$. 
We, furthermore, work up to order $\left( a_{0} \pi T \right)^{2}$ in the finite temperature pNRQCD multipole expansion and implement an additional expansion to linear order in $E/(\pi T)$.
The heavy quark mass is taken in the $1S$ scheme: 
$M = M_{b} = M_{\Upsilon(1S)} / 2 = 4.73$ GeV where $M_{\Upsilon(1S)}$ is taken from the Particle Data Group \cite{Zyla:2020zbs}.
The Bohr radius is computed by solving its self-defining relation
\begin{equation}
	a_{0} = \frac{2}{C_{F}\alpha_{\textrm s} ( 1 / a_{0}) M_{b}} \, ,
\end{equation}
where $C_{F} = (N_{c}^{2} - 1) / (2N_{c}) = 4/3$ is the quadratic Casimir of the fundamental representation of $\text{SU}(N_{c} = 3)$, $\alpha_{s}(1 / a_{0})$ is the strong coupling evaluated at the energy scale $1 / a_{0}$ using the 1-loop, 3-flavor running coupling and $\Lambda_{\overline{\text{MS}}}^{N_{f}=3} = 332$ MeV \cite{Zyla:2020zbs}.
This gives $a_{0} = 0.678\text{ GeV}^{-1}$.
In the OQS+pNRQCD formalism, interactions with the medium are calculated systematically and occur as in-medium modifications to the heavy quarkonium potential (the real and imaginary parts of which cause a thermal mass shift $\delta M$ and thermal width $\Gamma$, respectively) and color singlet to octet, octet to octet and octet to singlet transitions (the latter implementing recombination of the heavy quark-antiquark pairs).
For the hierarchy of scales realized in Eq.~(\ref{eq:hierarchy_of_scales}), the imaginary part of the in-medium potential is controlled by the heavy quarkonium momentum diffusion coefficient $\kappa(T)$ and the real part by $\gamma(T)$, which is the dispersive counterpart of $\kappa(T)$; these transport coefficients are fixed from independent, unquenched lattice measurements of $\Gamma$ and $\delta M$~\cite{Brambilla:2019tpt}.

\ingredients{

    \ingredient{In-medium potential:} 
    We use a vacuum Coulomb potential plus a quadratic in-medium modification through the transport coefficient $\hat\gamma$.
    The Coulombic binding energy $|E| = 1/(M a_{0}^{2})=460$ MeV sets the scale of the spacing of the energy levels.

    \ingredient{Vacuum limit of potential/spectroscopy:}
    In the kinematical situation described by Eq.~\eqref{eq:hierarchy_of_scales}, 
    the real part of the in-medium quarkonium potential is the sum of a Coulomb part, $-C_F\alpha_s/r$, and 
    a term proportional to $\gamma(T)\, r^2$~\cite{Brambilla:2008cx,Brambilla:2016wgg}.
    Taking the vacuum limit $T\to 0$, the in-medium potential reduces to the Coulomb potential plus a correction proportional to $\gamma(T=0)\, r^2 \sim   \Lambda_{\rm QCD}^3\, r^2$~\cite{Brambilla:1999xf}.
    We emphasize that to accurately reproduce the vacuum spectrum (in particular fine and hyperfine splittings), higher-order loop corrections in $\alpha_s$ to the Coulomb potential may be necessary to include (cf. Ref.~\cite{Brambilla:2004jw}).  
    In the adopted hierarchy of energy scales,
    they are parametrically subleading with respect to the considered in-medium quarkonium potential.
    The imaginary part of the in-medium quarkonium potential is proportional to $\kappa(T)\, r^2$~\cite{Brambilla:2008cx,Brambilla:2016wgg}.

    \ingredient{Reaction rates:}
    Using the OQS+pNRQCD formalism, we systematically calculate interactions with the medium at strong coupling: our self-energy contributions induce a thermal shift in the quarkonium energy, a thermal quarkonium width and singlet-octet, octet-octet and octet-singlet transitions that describe recombination and account for gluodissociation~\cite{Brambilla:2011sg} parton dissociation~\cite{Brambilla:2013dpa} and screening effects (in the short-distance limit).
    We note that the thermal decay width can be computed as either (1) an expansion in $E/(\pi T)$ with the state of the art expression extending up to order $E/(\pi T)$ or (2) using the eigenvalue of the non-Hermitian Lagrangian which emerges at a given order in the $E/(\pi T)$ expansion.  For a detailed discussion of method (1) and explicit expressions, see Sec.~4.1 of Ref.~\cite{Brambilla:2022ynh}.  For this report we use method (2) since, in practice, it provides a better description of the dynamical evolution than using widths based on the vacuum states.  We consider a quarkonium state comoving with the medium; the width is, therefore, independent of $p$.

    \ingredient{Assumptions about the medium (degrees of freedom, etc.):}
    We assume a strongly-coupled QCD plasma in local thermal equilibrium.  The degrees of freedom are singlet and octet quarkonium states.  Interactions with in-medium gluons are integrated out and described using the pNRQCD Lagrangian.
    In a medium realizing the hierarchy of scales of Eq.~(\ref{eq:hierarchy_of_scales}), the relevant transport properties of the medium are encoded in the heavy quarkonium momentum diffusion coefficient $\kappa (T)$ and its dispersive counterpart $\gamma (T)$.

    \ingredient{Temperature dependence of heavy quark masses:}
    The heavy quark mass is treated as independent of temperature. Similar effects are already included in the chromoelectric field correlator mentioned above~\cite{Yao:2020eqy}. We work in the $1S$ scheme for the heavy-quark mass in which $M = M_{b} = M_{\Upsilon(1S)} / 2$.  It is important to note that the mass of the bottom quark is not an observable quantity and is, furthermore, a scheme-dependent quantity.
    In our framework, the heavy-quark mass is a parameter entering the evolution equations; it receives no medium corrections.
    The bound-state mass, on the other hand, is an observable quantity, and its thermal correction up to order $a_{0}^{2}(\pi T)^{3}$ in the multipole expansion  of pNRQCD at finite temperature is given by
    $\delta M_{\YiS} = \frac{3}{2} \,a_{0}^{2} \,\gamma(T)$.
    \ingredient{Equilibrium limits in transport:}
    We do not enforce and have not observed (on time scales relevant to HICs) an equilibrium limit; this is a subject of ongoing investigation.

    \ingredient{Constraints from lattice QCD:}
    The transport coefficients $\kappa (T) $ and $\gamma (T)$, which fully encode the interaction of the quarkonium state with the medium, are calculated from lattice measurements of the thermal width $\Gamma$ \cite{Larsen:2019bwy,Bala:2021fkm} and the thermal mass shift $\delta M$ \cite{Bala:2021fkm} of the $\Upsilon(1S)$.
	For a presentation of the method (with older lattice data), see Ref.~\cite{Brambilla:2019tpt}.

    \ingredient{Range of applicability and how is this range established:}
    We do not extend the medium evolution to temperatures below $T_f=190$ MeV.
    As discussed in our executive summary, our evolution equations are derived assuming the hierarchy of scales $\pi T \gg E$; furthermore, our in-medium width operator is computed order-by-order as an expansion in $E/(\pi T)$.
    At zeroth-order in this expansion \cite{Brambilla:2016wgg,Brambilla:2017zei,Brambilla:2020qwo,Brambilla:2021wkt}, we do not extend our analysis below $T_f=250$ MeV. 
    Our state of the art results include terms up to and including order $E/(\pi T)$; a phenomenological comparison against the zeroth-order results (see discussion at the end of sec.~4.2 of Ref.~\cite{Brambilla:2022ynh}) leads us to select $T_f=190$ MeV as a lower bound when including these terms.
    Finally, to ensure that $\pi T < 1/a_0$ is fulfilled at all times, we do not extend our analysis above $T \sim 500$ MeV.

    \ingredient{Quantum features:}
    Our in-medium EFT description of the system is fully quantum.  In practice, we solve the Lindblad equation by computing the real-time evolution of the Schr\"odinger equation for a large ensemble using the quantum trajectories algorithm.  This provides a numerical solution of the evolution of the reduced density matrix that describes both singlet and octet states.

    \ingredient{Regeneration:}
    pNRQCD octet to singlet chromoelectric dipole transitions implement quantum regeneration.  The quantum mechanical wave function includes both bound and free $b\bar{b}$ pairs and transitions between these two types of states.

    \ingredient{Coupling to open heavy-flavor sector:}
    We work in the dilute limit and thus only in the bottom sector; open bottom states are implemented via the octet (unbound) contribution.

    \ingredient{Hadronic-Phase Transport:}
    When the local temperature of the medium falls below $T_{f}$, we terminate the coupling with the medium and evolve the state in the vacuum; in practice, this corresponds to termination of the evolution.
We apply our feed-down procedure to the resulting survival probabilities and do not implement hadronic-phase transport.	

    \ingredient{Initial quark/quarkonium distributions:}
    As the heavy quark mass $M$, which sets the scale of heavy quarkonium formation, is the largest scale of the combined system, our initial state is point-like with respect to the medium.
We thus utilize a peaked Gaussian as a numerically tractable approximation of a position-space delta function.   We sample the initial positions of these quantum mechanical wave-packets from the AA binary collision profile for a given impact parameter and sample their initial transverse momentum from a $\pT/(\pT^2 + \langle M\rangle^2)^2$ distribution where $\langle M\rangle$ is the average mass of a bottomonium state.

    \ingredient{Are cold nuclear matter effects, nPDF effects, etc. taken into account?:}
    We do not include cold nuclear matter or nPDF effects.

    \ingredient{Constraints from pA and dA collisions:}
    We do not include constraints from pA and dA collisions.

    \ingredient{Medium evolution model:}
    We make use of a 3+1D dissipative hydrodynamics code based on the relativistic quasiparticle anisotropic hydrodynamics (aHydroQP) framework~\cite{Alqahtani:2015qja,Alqahtani:2016rth,Alqahtani:2017mhy}.
	The aHydroQP framework has been tuned to reproduce a large set of experimental soft-hadronic observables, such as the total charged-hadron multiplicity, identified hadron spectra, integrated and identified hadron elliptic flow, and HBT radii at both RHIC and LHC nucleus-nucleus collision energies~\cite{Alqahtani:2017jwl,Alqahtani:2017tnq,Nopoush:2014pfa,Almaalol:2018gjh,Alqahtani:2020daq,Alqahtani:2020paa}.
	The evolution uses a realistic equation of state determined from lattice QCD measurements \cite{Bazavov:2013txa} and self-consistently computed second- and higher-order dissipative transport coefficients~\cite{Alqahtani:2017mhy}.
 
    \ingredient{Feed down implementation:}
    We implement a feed-down procedure on our results for the survival probability before comparing against experimental measurements.
	The feed-down procedure consists of multiplying a vector containing the survival probabilities by a feed-down matrix, the entries of which are determined from the branching fractions of the considered states obtained from the Particle Data Group listings.
	For a more detailed description, including the states considered and the precise values of the entries of the feed down matrix, see sec.~6.4 of Ref.~\cite{Brambilla:2020qwo}.

    \ingredient{Comparisons to experimental data:}
    We compare with experimental measurements of the nuclear modification factor $\raa$ of the $\YiS$, $\YiiS$ and $\YiiiS$ states as functions of the number of participants and of transverse momentum; double ratios of the aforementioned nuclear modification factors; and the elliptic flow $v_{2}$ of the aforementioned states from the ALICE \cite{ALICE:2020wwx}, ATLAS \cite{ATLAS:2022xso} and CMS \cite{CMS:2018zza,CMS-PAS-HIN-21-007} collaborations.

    \ingredient{Phenomenological breadth:}
    We work exclusively in the bottom sector and compare with experimental data for the $\YiS$, $\YiiS$ and $\YiiiS$ states obtained in Pb-Pb collisions at $\sqrt{s_{\textrm NN}} = 5.02$~TeV.
}

%% file: models/nantes.tex
\subsection{The Nantes model}
\label{subsec:Nantes}

The Nantes model is based on earlier work by Blaizot and Escobedo~\cite{Blaizot:2017ypk}. Using NRQCD and the open quantum systems framework, coupled quantum master equations are derived in the quantum Brownian motion regime, describing the dynamics of a single quark-antiquark pair in the QGP. Our work extended their equations in order to ensure that the positivity of the density operator is preserved~\cite{Delorme2021,Delorme:2023}. The static screening in the medium and the dynamical processes occurring between the $Q\bar{Q}$ pair and the medium constituents are described by a complex potential. The quantum master equations (QMEs) derived take into account both diffusive and dissipative effects in the medium. The dissociation and diagonal recombination are described by the singlet $\leftrightarrow$ octet color transitions of the pair. All together, this results in a quantum master equation of the Lindblad type expressed as
\begin{equation}
\frac{d\mathcal{D}_{Q\bar{Q}}}{dt}=\mathcal{L}_0[\mathcal{D}_{Q\bar{Q}}]+	\mathcal{L}_1[\mathcal{D}_{Q\bar{Q}}]+\mathcal{L}_2[\mathcal{D}_{Q\bar{Q}}]+\mathcal{L}_3[\mathcal{D}_{Q\bar{Q}}]+\mathcal{L}_4[\mathcal{D}_{Q\bar{Q}}],
\label{eq:QMENAntes}
\end{equation}
where $\mathcal{D}_{Q\bar{Q}}$ is the density operator containing a singlet and an octet component. $\mathcal{L}_0$ and $\mathcal{L}_1$ represent the kinetic energy and the (screened) real potential, while  $\mathcal{L}_2$,  $\mathcal{L}_3$, and $\mathcal{L}_4$ respectively account for the fluctuations, the dissipation and the preservation of positivity; they depend on the assumed imaginary part $W$ of the complex potential.  

The off-diagonal recombination is not treated as only one pair is considered in our modeling. The medium is assumed to have a uniform temperature, which is either fixed during the whole evolution or is dynamical. The dynamical medium follows an average temperature profile obtained from EPOS4, corresponding to the average temperature along the path of multiple pairs. The resolution of the quantum master equations is performed in one dimension. We have developed a new complex potential, specifically tailored to model the three-dimensional phenomenology \cite{Katz:2022fpb}. In addition, to obtain tractable equations for the evolution of a $Q\bar{Q}$ pair, the center of mass degrees of freedom were integrated out.  

\ingredients{
	
\ingredient{In-medium potential:} One-dimensional temperature-dependent complex potential parameterized \cite{Katz:2022fpb} to reproduce decay widths and mass spectra of the three-dimensional potential by Lafferty and Rothkopf \cite{Lafferty:2019jpr}.
	
\ingredient{Vacuum limit of potential/spectroscopy:} In the vacuum, the real part of the potential is linear with a saturation corresponding to twice the mass of the lightest D or B meson. The mass spectra of charmonium and bottomonium states are reproduced with a precision of 20 MeV.
	
\ingredient{Reaction rates:} The reaction rates are obtained from an imaginary potential, which is one-dimensional and fitted to a three-dimensional potential \cite{Lafferty:2019jpr}. Each term $\mathcal{L}_2$, $\mathcal{L}_3$, and $\mathcal{L}_4$ brings it own contribution, with increasing power of the relative velocity. The dominant one stemming from the $\mathcal{L}_2$ term simply leads to $\Gamma_2=2 \langle \psi| W |\psi \rangle$.   
	
\ingredient{Assumptions about the medium (degrees of freedom, etc.):} The medium constituents are light quarks and gluons. The medium is assumed to be in equilibrium.
	
\ingredient{Temperature dependence of heavy quark masses:} The temperature dependence of quark mass is taken as $m_0+V_\infty(T)/2$, where $m_0$ is the bare mass -- taken as 1.47 GeV for c quarks and 4.88 GeV for b quarks and $V_\infty$ is the asymptotic value of the 1D real potential.
	
\ingredient{Equilibrium limits in transport:} The QME (\ref{eq:QMENAntes}) admits a unique asymptotic state for each temperature and such a state is reached as the final outcome of any initial state. For large relative distance, the complex potential saturates and the QME for the $Q\bar{Q}$ pair splits into two independent QMEs for each quark. In the semi-classical limit, these QMEs are equivalent to Fokker-Planck equations (with a momentum diffusion coefficient expressed as $W''(0)$) and naturally lead to Boltzmann distributions for individual quarks. In the quantum regime, the HQ effective temperature is  10\%  to 20\% higher than the medium temperature. This is due to the UV regularization procedure adopted for the imaginary potential. When the evolution is performed in a box, the asymptotic spatial densities of both $Q$ and $\bar{Q}$ are finite and some $Q\bar{Q}$ spatial correlation emerges at small relative distance. The corresponding asymptotic distribution of quarkonia is presently under investigation.
	
\ingredient{Constraints from lattice QCD:} Lattice QCD is used to constrain the complex potential.
	
\ingredient{Range of applicability and how is this range established:} The model is developed in the quantum Brownian regime due to the assumed scale hierarchy.
	
\ingredient{Quantum features:} The treatment within the open quantum system framework is fully quantum (under the usual assumptions made in the Quantum Brownian regime).
	
\ingredient{Regeneration:} ``Diagonal'' regeneration is included but not ``off-diagonal'' regeneration as only one pair is considered.
	
\ingredient{Coupling to open heavy-flavor sector:} As explained previously, the $Q\bar{Q}$ dynamics naturally factorize when the distance between $Q$ and $\bar{Q}$ is large. This has however little influence on the quarkonia probability measured at finite time since only pairs which do not evolve too far apart during the evolution contribute.  
	
\ingredient{Hadronic-Phase Transport:} No transport in the hadronic phase.
	
\ingredient{Initial quark/quarkonium distributions:} 
The initial distribution of heavy quarks is taken from the EPOS4 framework. The ensuing transverse distribution of charmed hadrons is found in good agreement with the experimental distributions for the cases investigated up to now \cite{Zhao:2023ucp}. In AA collisions, the spatial distribution of heavy quarks and heavy quarks pairs follows a $N_{\rm coll}$ scaling. In our approach, there is no initial quarkonia per se but $Q\bar{Q}$ pairs. The initial internal degrees of freedom for these pairs are described as a pure-state density operator either in the singlet or in the octet representation. As for the relative distance dependence, compact Gaussian states are privileged, although it is also possible to initiate the evolution with any eigenstate from the vacuum or thermal potential.
	
\ingredient{Are cold nuclear matter effects, nPDF effects, etc. taken into account?:} No Cold Nuclear Matter effects are included.
	
\ingredient{Constraints from pA and dA collisions:} No constraints from pA/dA collisions.
	
\ingredient{Medium evolution model:} Two cases are considered: a static medium with fixed temperature and an evolving medium following an average temperature profile from EPOS4.
	
\ingredient{Feed down implementation:} No realistic feeddown is included.
	
\ingredient{Comparisons to experimental data:} No comparison to experimental data for the moment.
	
\ingredient{Phenomenological breadth:} Charmonium and bottomonium systems can be studied within our model. As our model is one-dimensional, we only consider even (S-like) and odd (P-like) eigenstates of the real part of the complex potential.}

%% file: models/phsd.tex
\subsection{Parton-hadron string dynamics}

Quarkonium production in the Parton-Hadron-String Dynamics (PHSD) is composed of the production of heavy quark pairs, the interaction and dynamics of heavy (anti)quarks in the medium and the Wigner projection of heavy quark pairs onto the quarkonium wave function based on the Remler formalism~\cite{Remler:1975re,Remler:1975fm,Remler:1981du,Gyulassy:1982pe,Villar:2022sbv,Song:2023ywt}.

First, the Wigner production is carried out in pp collisions without any nuclear matter effects. Heavy quarks and heavy antiquarks are separated from each other in space by Monte Carlo such that the average distance is on the order of the inverse mass and each momentum is provided by the PYTHIA event generator~\cite{Sjostrand:2006za} after rescaling the rapidity and transverse momentum of heavy quark to mimic the FONLL calculations~\cite{Cacciari:2012ny}. Then the Wigner projection finds out the most suitable radius of each quarkonium state from the comparison to the experimental data in pp collisions~\cite{Song:2017phm}.

In heavy-ion collisions the (anti)shadowing effects are implemented using the EPS09~\cite{Eskola:2009uj} which modifies heavy quark production and affects quarkonium production as well. Produced heavy quarks and antiquarks interact in the QGP within the Dynamical Quasi-Particle Model (DQPM)~\cite{Moreau:2019vhw}.
Quarkonium properties in the QGP, such as the dissociation temperature and radius are obtained by solving the Schr\"{o}dinger equation with the heavy quark potential from lattice QCD. Once the local temperature is lower than the dissociation temperature, quarkonium begins to form through the Wigner projection with a temperature-dependent radius.
Since each state of quarkonium has a different dissociation temperature, this projection takes place at different times and positions in heavy-ion collisions.
We take into account 1S, 1P, and 2S states for charmonia and 1S, 1P, 2S, 2P, and 3S states for bottomonia and also feed down from excited states to a lower state as well as $B$-hadron decay to charmonium.

Whenever a heavy (anti)quark scatters in QGP, a Wigner projection is carried out and the density of each quarkonium state is updated. If a heavy quark pair forms a bound state, its cross section with thermal partons will be smaller than the sum of the cross sections of heavy quarks and of heavy antiquarks due to the interference terms.
Therefore, we introduce an ad-hoc suppression factor to the heavy quark scattering such that only part of the scatterings affects quarkonium production and dissociation.
The suppression factor is roughly 10 \% for bottomonia in Pb+Pb collisions at \mbox{$\sqrt{s_{\rm NN}} =$ 5.02 TeV}, which produces results consistent with experimental data~\cite{Song:2023zma}.

\ingredients{

\ingredient{In-medium potential:} 
The present form of PHSD uses the lattice free energy as heavy quark potential. But we can switch the potential to the internal energy or a combination of both energies.
By solving the Schr\"{o}dinger equation with the heavy quark potential, one obtains the dissociation temperatures and the binding energies of charmonia ($J/\psi$,~$\chi_c$,~$\psi^\prime$) and of bottomonia ($\Upsilon(n{\rm S}$),~$\chi_b(m{\rm P})$).
If a dissociation temperature is lower than $T_c$, it is taken to be $T_c$.

\ingredient{Vacuum limit of potential/spectroscopy:}
The heavy quark mass in vacuum is adjusted such that the ground state of quarkonium has the physical mass in the Particle Data Book.
As a result, the charm quark mass is taken to be 1.26 GeV and the bottom quark mass 4.62 GeV.
These are only used to solve the Schr\"{o}dinger equation and the masses of open charm and bottom quarks in the PHSD are, respectively, 1.5 GeV and 4.8 GeV.

\ingredient{Reaction rates:}
The reaction rate of quarkonium is directly related to the reaction rate of heavy quarks, based on the Remler formalism~\cite{Remler:1975re,Remler:1975fm,Remler:1981du,Gyulassy:1982pe,Song:2023ywt}.
It is similar to a quasi-free approach. However, if the heavy quark and heavy anti-quark form a bound state, the reaction rate will be smaller than twice the reaction rate of heavy quarks due to interference terms of heavy quark scattering and heavy antiquark scattering.
So an adhoc suppression factor is introduced to the reaction rate of heavy quark, which is about 0.1 for bottomonia.

\ingredient{Assumptions about the medium (degrees of freedom, etc.):}
A QGP in PHSD is composed of off-shell massive partons whose pole mass and spectral width depend on temperature and baryon chemical potential and are fitted to the equation-of-state (EoS) from the lattice QCD calculations~\cite{Moreau:2019vhw}.
The scattering cross sections of heavy quarks and light partons are calculated at leading order in the strong coupling constant, but the propagator of massive offshell partons implements a resummation of the parton propagator and removes all singularities, making it unnecessary to introduce an explicit Debye screening mass~\cite{Berrehrah:2013mua}.

\ingredient{Temperature dependence of heavy quark masses:}
In principle heavy quark will also be off-shell in medium.
But that is not taken into account in the present form of the PHSD, since its effects on the observables are expected to be small. So the heavy quark mass is constant and does not depend on temperature in the PHSD simulations~\cite{Song:2015sfa}.
The charm quark mass is taken to be 1.5 GeV and bottom quark mass 4.8 GeV~\cite{Song:2016rzw}.

\ingredient{Equilibrium limits in transport:}
We have recently tested the Remler formalism which the PHSD adopts in thermalized and thermalizing boxes and found that the results are in good agreement with the statistical model~\cite{Song:2023ywt}. 

\ingredient{Constraints from lattice QCD:}
In the PHSD approach, the pole mass and spectral width of massive off-shell partons which interact with heavy quark are fitted to lattice EoS through the strong coupling that depends on temperature and baryon chemical potential~\cite{Moreau:2019vhw}.
And the heavy quark spatial diffusion coefficient from lattice calculations is also described in the DQPM with an adhoc multiplication factor of 2~\cite{Song:2016rzw,Song:2019cqz}.

\ingredient{Range of applicability and how is this range established:}
In principle there is no limitation for the application of the Remler formalism, because it is calculating the quarkonium density in quantum statistics.
In practice, the dynamics of heavy quarks are calculated resorting to semi-classical trajectories which are a good approximation to the exact quantum mechanical evolution when the local temperature is larger than the binding energies. At lower temperatures, quantum corrections should however be applied.

\ingredient{Quantum features:}
Once the local temperature is lower than the dissociation temperature of quarkonium in heavy-ion collisions, there is a possibility for $Q$ and $\bar{Q}$ which are close in phase space 
to form a bound state, based on the projection of the pair on the quarkonium wave function. Whenever a heavy (anti)quark scatters in the QGP, the projection is carried out and updated.

\ingredient{Regeneration:}
In the Remler formalism each scattering of a (anti)heavy quark brings about quarkonium dissociation and regeneration simultaneously, based on the change of 2-body Wigner density. 
Since the formalism provides only the change of quarkonium density with time, it is not simple to separate dissociation and regeneration.

	\ingredient{Coupling to open heavy-flavor sector:}
 Quarkonium production and dissociation are closely related to the dynamics of heavy (anti)quarks, since the quarkonium exists in QGP as a density probability, which is calculated from the distribution of heavy (anti)quarks in coordinate and momentum space.

	\ingredient{Hadronic-Phase Transport:}
The dissociation cross section of quarkonium into a $Q\bar{Q}$ pair by light pseudoscalar or vector mesons is obtained by assuming a constant transition amplitude, which is fitted to experimental data at SPS energy~\cite{Bratkovskaya:2003ux}. The transition amplitude squared is multiplied by 3, if there is a vector meson such as ${\rm D}^\star$ or ${\rm B}^\star$ in the final state due to spin degeneracy, and divided by 3, if the light meson has strangeness. The reverse reaction is realized by the detailed balance.

	\ingredient{Initial quark/quarkonium distributions:}
The initial spatial distribution of heavy quark pairs is given by the distribution of primary nucleon+nucleon scatterings in the Glauber model, and the initial momentum distribution is generated by the PYTHIA event generator~\cite{Sjostrand:2006za} and then tuned by rescaling rapidity and/or transverse momentum of heavy quark pairs such that the distribution becomes similar to that from the FONLL calculations~\cite{Cacciari:2005rk}. 
In pp collisions the produced heavy quark pair is projected on the quarkonium wave function after separating the heavy quark and antiquark by a distance which is inversely proportional to heavy quark mass on average. In heavy-ion collisions the projection is carried out when the local temperature is lower than the dissociation temperature of each quarkonium state.

\ingredient{Are cold nuclear matter effects, nPDF effects, etc. taken into account?:}
 In the PHSD approach (anti)shadowing effects on heavy quark production are included using EPS09~\cite{Eskola:2009uj,Song:2015ykw}.
This suppresses the production of heavy flavor at low $p_T$ and at mid-rapidity, depending on collision energy.
It also affects the production of quarkonium.

\ingredient{Constraints from pA and dA collisions:}
There is no additional constraint from pA and dA collisions except introducing EPS09 for the (anti)shadowing effects.
 
\ingredient{Medium evolution model:}
The production and time evolution of a nuclear medium in heavy-ion collisions are described using PHSD~\cite{Cassing:2008sv,Cassing:2008nn,Cassing:2009vt,Bratkovskaya:2011wp,Linnyk:2015rco} where initial partons are produced through the string fragmentation or string melting, and then evolve and interact in off-shell dynamics.
Around the critical temperature the off-shell partons hadronize to off-shell hadrons which eventually freeze out as on-shell hadrons.

\ingredient{Feed down implementation:}
We first realize the feed down in pp collisions, because experimental data are available there. The branching ratio of each excited state to a lower energy state is given by the Particle Data Book, and its contribution to the ground state is controlled by adjusting the radius of the excited states, because the yield of each state depends on its radius in the Wigner projection~\cite{Song:2017phm}. 
The same method is applied to heavy-ion collisions.
We also take into account B-hadron decay to charmonium which is not negligible at LHC energies.

\ingredient{Comparisons to experimental data:}
We have tested our model first in pp collisions and found good agreement with the experimental data on the rapidity distributions and $p_T$-spectra of $J/\psi$, $\Upsilon$ and their excited states~\cite{Song:2017phm,Song:2023zma}.
Based on these successful results in pp collisions, the Remler formalism is applied to bottomonia production in Pb-Pb collisions at $\sqrt{s_{\rm NN}} =$ 5.02 TeV and it produces the results consistent with experimental data~\cite{Song:2023zma}.

	\ingredient{Phenomenological breadth:}
Our model is applicable to any quarkonium state, if its wave function is given. This method does not depend on collision energy or system size.

}

%% file: models/saclay.tex
\def \als {\alpha_{\mathrm{s}}}
\def \bk {\mathbf{k}}
\def \nc {N_c}
\def \cf {C_F}
\def \br {\mathbf{r}}
\newcommand{\rme}{{\rm e}}
\def\x{{\boldsymbol x}}
\def\a{{\boldsymbol a}}
\def\y{{\boldsymbol y}}
\def\F{{\boldsymbol F}}
\def\v{{\boldsymbol v}}
\def\r{{\boldsymbol r}}
\def\u{{\boldsymbol u}}
\def\bmnabla{\mbox{\boldmath$\nabla$}}
\def\q{{\boldsymbol q}}
\def\b{{\boldsymbol b}}
\def\s{{\boldsymbol s}}
\def\p{{\boldsymbol p}}
\def\y{{\boldsymbol y}}
\def\J{{\boldsymbol J}}
\def\P{{\boldsymbol P}}\def\Q{{\boldsymbol Q}}\def\R{{\boldsymbol R}}
\def\X{{\boldsymbol X}}
\def\Y{{\boldsymbol Y}}
\def\nab{{\boldsymbol \nabla}}
\def\colr{\color{red}}
\def\bra#1{\langle#1\vert}
\def\ket#1{\vert#1\rangle}
\newcommand{\eq}{{\text{eq}}}
\newcommand{\rmd}{{\rm d}}
\newcommand{\del}{\partial}
\newcommand{\nn}{\nonumber\\ }

\subsection{The Saclay model}

The aim of the ``Saclay model'' is to highlight the importance of the finite energy gap between singlets and octets (which essentially corresponds to the binding energy). It is an elaboration of a model that was developed in \cite{Blaizot:2018oev} for illustration purposes.  We studied there the conditions that a master equation should fulfill in order to lead to thermal equilibrium. We found that taking into account the finite energy gap between energy levels was crucial,  and we derived a general set of equations (valid in the $E\gg\Gamma$ limit) that describe an evolution in which the free energy monotonically decreases.

As a simple model we consider equations that consist of a rate equation for the singlet combined with a Langevin equation for the octet component. The rate equation that governs the populations of singlets ($p_n^s)$ and octets ($ p_m^o$) is given by
\begin{equation}
\frac{\rmd p_n^s}{\rmd t}=4g^2C_F\sum_m\left(p_m^o-p_n^se^{-\frac{E_m^o-E_n^s}{T}}\right)\int_\q\Gamma^>(E_m^o-E_n^s,{\q})|\langle n^s|{\cal S}_{\q\cdot\r} |m^o\rangle|^2\,,
\label{eq:evops}
\end{equation}
where ${\cal S}_{\q\cdot\r}=\sin\left(\frac{\q\cdot\r}{2}\right)$, with $\r$ the relative coordinate, and $\Gamma^>$ is a correlator of  $A_0$ fields in the Coulomb gauge, evaluated in the hard-thermal-loop (HTL) approximation. The octet can also decay into a singlet. However, we checked in \cite{Blaizot:2018oev} that the probability for that to happen during the lifetime of the fireball is negligible, so we ignore it, which is achieved in practice by taking $p_m^o=0$ in eq.~(\ref{eq:evops}). We note that in this model, the decay width of the bound state depends both on its  binding energy and  its wave function.

In \cite{Blaizot:2021xqa} we made a phenomenological application of this model. We did that considering two different choices for the real part of the potential. In the first case, we considered a Yukawa potential with the running of the coupling constants that would correspond to a computation in which $1/r\sim m_D$. In the second scenario, we considered the real part of the potential that was fitted to the lattice data of \cite{Lafferty:2019jpr}. The results presented here extend some of the results reported in \cite{Blaizot:2021xqa}.

\ingredients{

\ingredient{In-medium potential:} 
As just mentioned,  two different scenarios are considered, which, for simplicity,  we name  the ``perturbative'' and the ``lattice-inspired'' scenarios \cite{Blaizot:2021xqa}. In both cases, we solve the Schr\"{o}dinger equation using the real part of the potential. In the perturbative case, we use a Yukawa potential in which $\alpha_s$ is evaluated at the Bohr radius of $\Upsilon(1S)$ ($1/a_0=1322\ {\rm MeV}$) and the $\alpha_s$ for the calculation of the Debye mass is evaluated at $\mu=2\pi T$. In the lattice inspired case, we use the real part of the potential fitted to the lattice data of \cite{Lafferty:2019jpr}.

\ingredient{Vacuum limit of potential/spectroscopy:}
In the perturbative case, we would get a Coulomb potential at zero temperature. In the lattice inspired scenario, we recover a Cornell potential. Note however that with the parameters used this Cornell potential (composed of a Coulomb plus a linear confining part) reproduces the spectroscopy only approximately.

	\ingredient{Reaction rates:}
We assume that the leading mechanism is inelastic scattering with medium particles. This is computed in the HTL one-gluon exchange approximation using the wave functions and binding energies obtained by solving the Schr\"{o}dinger equation with the real part of the potential. The octet wave function is modeled in the large $N_c$ limit. Our reaction rate can be encoded in a frequency-dependent imaginary potential (which reduces to the commonly used static imaginary potential when the frequency dependence is ignored). It was studied and computed in \cite{Blaizot:2018oev}.

	\ingredient{Assumptions about the medium (degrees of freedom, etc.):}
We assume the accuracy of the one-gluon exchange approximation and HTLs. The only exception is the real part of the potential in the lattice-inspired scenario, which is taken from a fit to lattice data.

	\ingredient{Temperature dependence of heavy quark masses:}
	Our heavy quark masses do not depend on temperature.
	
	\ingredient{Equilibrium limits in transport:}
Strictly speaking, the equations we use do not bring the system to equilibrium because we did not include regeneration. However, we checked in 
\cite{Blaizot:2018oev} that this was an accurate approximation for the typical time scales that we are interested in at the LHC. 

	\ingredient{Constraints from lattice QCD:}
In the lattice inspired scenario we use as input lattice data on the real part of the potential.

	\ingredient{Range of applicability and how is this range established:}
One condition for the model to be valid is that  the decay width be much smaller than the energy gap between singlets and octets. We start with a quantum non-Markovian master equation and only in this limit we arrive to a Markovian rate equation that we can solve.
 The validity of other approximations used (one-gluon exchange and validity of the HTL propagators) is difficult to establish.
 
	\ingredient{Quantum features:}
Quantum features are included in the rate (wave function and binding energy, as well as the gap between energy levels).

	\ingredient{Regeneration:}
We do not consider regeneration since we work in the dilute limit (valid for bottomonium) and we have checked that in this limit regeneration effects are not important for the relevant  LHC time-scales.

	\ingredient{Coupling to open heavy-flavor sector:}
 There is no direct relation to open heavy-flavor. In our model the octet evolves following a Langevin equation that uses the heavy quark diffusion coefficient as input. However, since we do not include regeneration, this has little impact on the results that we present.
 
	\ingredient{Hadronic-Phase Transport:}
We do not consider hadronic effects.

	\ingredient{Initial quark/quarkonium distributions:}
In the present simple approach, the survival probability coincides with $R_{AA}$. The spatial distribution of quarkonium at the initial time is proportional to the density of collisions (unintegrated number of collisions in the Glauber model). We do not consider the momentum distribution.

	\ingredient{Are cold nuclear matter effects, nPDF effects, etc. taken into account?:}
 No
 
	\ingredient{Constraints from pA and dA collisions:}
 No
 
	\ingredient{Medium evolution model:}
We assume a Bjorken expansion using the equation-of-state of a free gas. The initial energy density scales as the density of participants.

	\ingredient{Feed down implementation:}
We do not include feed down.

	\ingredient{Comparisons to experimental data:}
Our motivation was to underline the importance of the finite energy gap between singlets and octets. No attempt was made to quantitatively reproduce experimental data.

	\ingredient{Phenomenological breadth:}
Our main focus is on bottomonium since the dilute approximation and the condition $E\gg\Gamma$ are more likely to be fulfilled in that case.

}

%% file: models/comover.tex
\subsection{The Santiago comover interaction model}
\label{comovers}

The comover interaction model (CIM)~\cite{Capella:1996va,Armesto:1997sa,Armesto:1998rc,Capella:2000zp,Capella:2005cn} was originally proposed to explain the suppression of quarkonium states by final-state interactions with a hadronic or partonic medium. 
In this framework, the suppression arises from the scattering of the nascent quarkonia with comoving particles, {\it i.e.} particles with similar rapidities whose density is directly connected to the particle multiplicity measured at that rapidity for the corresponding colliding system. 
The comover interaction is governed by the Boltzmann gain and loss differential equations in a transport theory for a quarkonium state. 
The main parameter of the model is the comover-quarkonium cross section that results from the convolution of the comover momentum distribution in the transverse plane and the momentum-dependent comover-quarkonium cross section, proportional to the quarkonium geometrical cross section~\cite{Ferreiro:2018wbd}.
This model includes the initial-state nuclear effects through nuclear shadowing, {\it i.e.}, the nuclear modification of the parton distribution functions. This model takes into account not only the dissociation of quarkonium, but also the possibility of recombination of $Q\bar{Q}$ into secondary quarkonium states~\cite{Capella:2007jv,Ferreiro:2012rq}.
We recall two basic features of the comover approach. First, larger quarkonia are more affected by dissociation, due to larger interaction cross sections. As a consequence, excited states are more suppressed than the ground states. 
Second, the suppression increases with comover densities, which are proportional to particle multiplicities. Thus, the suppression increases with centrality in nucleus-nucleus collisions and is stronger in the nucleus direction for proton-nucleus collisions~\cite{Ferreiro:2014bia}.

\ingredients{
	\ingredient{Reaction rates:}
	The rates are based on gluon and pion dissociation; they depend on the comover momentum distribution in the transverse plane (assumed to be characterized by an effective temperature, $T_{\rm eff}$) and the momentum dependent comover-quarkonium cross section as $\Gamma^{\cal Q}(E^{\rm co},T) = \sigma^{\rm co-{\cal Q}}(E^{\rm co})\, \rho(E^{\rm co},T)$. The average over the comover energies is carried out using
	\beq
	\Gamma^{\cal Q}(T)=
	\int_{E^Q_{\rm thr}}^\infty dE^{\rm co }\, \sigma^{\cal Q}_{\rm geo}  \left(1-\frac{E^{\cal Q}_{\rm thr}}{E^{\rm co }}\right)^n\,   \frac{\rho^{\rm co }}
		{{e^{E^{\rm co }/T_{\rm eff}}-1}} \ ,
	\eeq
where $E^{\cal Q}_{\rm thr}=M_{\cal Q}+m_{\rm co}-2 M_{B\ \rm{or}\ D}$  corresponds to the threshold energy to break the quarkonium bound state (and as such is sensitive to the binding energies of the different states),  
$E^{\rm co}=\sqrt{p^2+m_{\rm co}^2}$ is the energy of the comover in the quarkonium rest frame (with $m_{\rm co}$=0 and 140\,MeV for gluons and pions, respectively), and $\rho^{\rm co }$ is the \textit{transverse} density of comovers, proportional to their multiplicities. The geometric cross section, $\sigma^{\cal Q}_{\rm geo}\simeq\pi r_{\cal{Q}}^2$, depends on the bound-state size; the phase space parameters $n \simeq$ 0.5-2 and the effective temperature, $T_{\rm eff}$$\simeq$ 200-300\,MeV control the energy dependence and are fitted to the CMS and ATLAS p-Pb data. 
\ingredient{Assumptions about the medium (degrees of freedom, etc.):}
	The medium is formed by the comoving particles, considered to be partons or hadrons, \ie, gluons or pions. The relevant degrees of freedom are mostly hadrons in proton-nucleus collisions, whereas the gluons become relevant in the hotter nucleus-nucleus environment. As for the momentum distribution of the comovers in the transverse plane, we take a Bose-Einstein distribution $1/(e^{E^{\rm co }/T_{\rm eff}}-1)$.
	\ingredient{Temperature dependence of heavy quark masses:}
	The heavy quark masses are independent of temperature. They do not appear explicitly in the model.
	\ingredient{Equilibrium limits in transport:}
	We use the Boltzmann equation to describe the dissociation of quarkonium, 
	\beq
	\label{eq:comovrateeq}
	\tau \frac{\mbox{d} \rho^{\cal Q}}{\mbox{d} \tau} \, \left( b,s,y \right)
	\;=\; -\sigma^{\rm co -{\cal Q}}\; \rho^{\rm co }(b,s,y)\; \rho^{\cal Q}(b,s,y) \;,
	\eeq
where $\sigma^{\rm co -{\cal Q}}$ is the the energy-averaged quarkonium-comover-interaction cross section of quarkonium dissociation due to interactions with the comoving medium characterized by transverse density~$\rho^{\rm co }(b,s,y)$ at $\tau_i$. By integrating this equation from $\tau_i$ to $\tau_f$, one obtains the survival probability 
$S^{\rm co }_{\cal Q}(b,s,y)  \;=\; \exp \Big\{-\sigma^{\rm co -{\cal Q}}\, \rho^{\rm co }(b,s,y)\, \ln\Big({\rho^{\rm co }(b,s,y)}/{\rho_{pp} (y)}\Big) \Big\}$ of a quarkonium  interacting with comovers, where the argument of the logarithm comes from $\tau_f/\tau_i$ converted to ratios of densities where we assumed that the interaction stops at $\tau_f$ when the densities have diluted, reaching the value of the proton-proton density at the same energy and rapidity, $\rho_{pp}$.
	\ingredient{Constraints from lattice QCD:}  No lattice constraints are used.
	\ingredient{Range of applicability and how is this range established:}
	Our approach can be used to describe quarkonium production in both proton-nucleus and nucleus-nucleus collisions at SPS, RHIC, and LHC energies without the need to invoke any other phenomena. It may also be at play in high-density proton-proton collisions.
	\ingredient{Quantum features:}  N/A
	\ingredient{Regeneration:}
	We consider the regeneration contribution for charmonium states. The regeneration through the uncorrelated charm and anticharm quark is represented by the gain term in the Boltzmann equation
	\beq
	\label{eq:comovrateeq-2}
	\tau \frac{\mbox{d} \rho^{\cal Q}}{\mbox{d} \tau} \, \left( b,s,y \right)
	\;=\; -\sigma^{\rm co -{\cal Q}}\; [\rho^{\rm co }(b,s,y)\; \rho^{\cal Q}(b,s,y) - \rho^q(b,s,y)\; \rho^{\bar{q}}(b,s,y)]\; .
	\eeq
	No regeneration is considered for bottomonium states.
	\ingredient{Coupling to open heavy-flavor sector:}  N/A
	\ingredient{Hadronic-Phase Transport:} The Boltzmann transport equations apply for both the partonic and hadronic phases.
	\ingredient{Initial quark/quarkonium distributions:} The initial quarkonium and heavy quark distributions are proportional to the number of binary collisions.
	\ingredient{Are cold nuclear matter effects, nPDF effects, etc. taken into account?:} We take into account nCTEQ15 or EPPS16 shadowing. Nuclear absorption is included at low (SPS) energies.
	\ingredient{Constraints from pA and dA collisions:}
 	The parameter $n$ that characterizes how quickly the cross section approaches
	the geometrical cross section,
	$\sigma^{\rm co-{\cal Q}}(E^{\rm co}) =\sigma^Q_{\rm geo}  (1-{E^Q_{\rm thr}}/{E^{\rm co }})^n$ is fixed through a fit to proton-nucleus data. Both ground- and excited-quarkonium states are taken into account.
	\ingredient{Medium evolution model:} 
    We assume a dilution in time of the comover densities due to longitudinal motion, which leads to a $\tau^{-1}$ 
    dependence on proper time, since Bjorken expansion is included. Transverse expansion is neglected.
	\ingredient{Feed down implementation:} Realistic feed-down contributions are included.
	\ingredient{Comparisons to experimental data and phenomenological breadth:} The model has been used to reproduce the nuclear modification factor of ground and excited charmonium and bottomonium states in proton-nucleus and nucleus-nucleus collisions at SPS, RHIC, and LHC energies.
}

%% file: models/shm.tex
\subsection{The statistical hadronization model}

The Statistical Hadronization Model (SHM) \cite{Braun-Munzinger:2000csl,Andronic:2006ky,Andronic:2021erx} assumes full dissociation of all quarkonium states in the QGP and exclusive generation at the QCD (crossover) phase boundary, in a concurrent hadronization process of all quark flavors. While the production of the heavy quarks occurs in hard parton scattering processes at initial stages of the collision, their thermalization, an essential condition for the applicability of SHM, is realized in the expansion of the deconfined medium.
In the SHM, the absence of chemical equilibrium for heavy quarks is accounted for by introducing a fugacity factor ($g_c$ in case of charm quarks). The fugacity is not a free parameter but is obtained from the balance equation \cite{Andronic:2021erx}
that accounts for the distribution of all initially produced heavy quarks into hadrons at the phase boundary, with a thermal weight constrained by exact charm conservation. With this approach, the knowledge of the heavy quark production cross section along with the thermal parameters obtained from the analysis of the yields of hadrons composed of light quarks \cite{Andronic:2017pug}, is sufficient to determine the total ($\pT$-integrated) yield of hadrons containing heavy quarks in ultra-relativistic nuclear collisions.

With the assumption of the kinetic freeze-out taking place also at the phase boundary and employing hydrodynamics to determine the collective expansion velocity (at the freeze-out hypersurface), the transverse momentum distribution can be calculated as well. A corona contribution is added (both for the total and $\pT$-differential yields), based on measurements in pp collisions.

\ingredients{

	\ingredient{In-medium potential:}  Full dissociation (screening) is assumed for all quarkonium states.

	\ingredient{Vacuum limit of potential/spectroscopy:} Not directly relevant, but quarkonium (vacuum) masses are essential inputs for the model.

	\ingredient{Reaction rates:} N/A

	\ingredient{Assumptions about the medium (degrees of freedom, etc.):} Not explicit, but full (partial) thermalization of charm (bottom) quarks in the quark-gluon phase is assumed.

	\ingredient{Temperature dependence of heavy quark masses:} N/A

	\ingredient{Equilibrium limits in transport:} N/A

	\ingredient{Constraints from lattice QCD:}
Not considered in an explicit way. Full dissociation of all quarkonium states is assumed.

	\ingredient{Range of applicability and how is this range established:}
Applicable for AA collisions. Given the assumption of (full) heavy-quark thermalization, the SHM is most justified at LHC energies, but it was applied for lower energies too, down to SPS energies \cite{Andronic:2007zu}.
 
	\ingredient{Quantum features:} Not explicit (quantum numbers are of course considered and very relevant for SHM).

	\ingredient{Regeneration:} Exclusive generation at the QCD crossover boundary.

	\ingredient{Coupling to open heavy-flavor sector:}
 Full open heavy-flavor chemistry is predicted and plays a major role in the model. In fact, the knowledge of the full spectrum of the heavy-quark resonances is crucial for the quarkonium production in SHM. While the PDG hadron spectrum is the default implementation in the model, a version with enhanced charm baryon resonances was also explored  \cite{Andronic:2021erx}. This leaves the charmonium predictions unchanged, under the assumption that the possible existence of such additional states (not yet measured) leads to a commensurately larger (by 1\%) total charm production cross section.

	\ingredient{Hadronic-Phase Transport:} None.

	\ingredient{Initial quark/quarkonium distributions:}
No explicit treatment of initial heavy-quark distributions.

	\ingredient{Are cold nuclear matter effects, nPDF effects, etc. taken into account?:}
 Yes, currently based on the ALICE measurements of D$^0$ mesons in (central) Pb-Pb collisions \cite{ALICE:2021rxa}. The charm production cross section in Pb-Pb collisions is the value describing in SHM the D$^0$ data in central Pb-Pb collisions. The equivalent shadowing value, 30\%, is used for the bottom sector as well.

	\ingredient{Constraints from pA and dA collisions:}
 Such constraints apply in SHM only if interpreted exclusively as shadowing on the total initial heavy-quark production. This was employed earlier, but is since recently superseded by the direct knowledge via the measurement of D$^0$-meson production in (central) Pb-Pb collisions.
 
	\ingredient{Medium evolution model:}
Hydrodynamics (MUSIC \cite{Schenke:2010rr} and Fluidum \cite{Floerchinger:2018pje}), either in an explicit manner \cite{Andronic:2023tui} or needed to extract the $\beta$ parameter at $T$=156.5 MeV \cite{Andronic:2021erx}, the temperature at which we assume kinetic freeze-out of quarkonium and open heavy flavor hadrons.

	\ingredient{Feed down implementation:}
Feed down from excited quarkonia is considered, based on statistical production, more recently also from bottom hadrons.

	\ingredient{Comparisons to experimental data:}
Full comparisons to the centrality dependence of integral yields ($R_{AA}$) and to the $\pT$ spectra and $\raa$ vs. $\pT$ in central and semi-central collisions.

	\ingredient{Phenomenological breadth:}
Yes, with the remark that $\Upsilon$ treatment is based on an ad-hoc thermalization fraction in the bottom sector \cite{Andronic:2022ucg}.

}

%% file: models/tamu.tex
\subsection{The Texas A\&M University (TAMU) model}

The starting point of the TAMU transport approach is a set of kinetic rate equations which describe the time evolution of quarkonium yields, $N_{\mathcal{Q}}$, according to~\cite{Grandchamp:2003uw,Zhao:2010nk}
\begin{equation}
   \frac{dN_{\mathcal{Q}}(\tau)}{d\tau}=-\Gamma_{\mathcal{Q}}(T(\tau))\left[N_{\mathcal{Q}}(\tau)-N_{\mathcal{Q}}^{\rm eq}(T(\tau))\right] \  .
\label{rate-eq}
\end{equation}
This provides a versatile tool that has been uniformly applied to a wide variety of quarkonia, $\mathcal{Q}$, including both ground- and excited-state charmonia~\cite{Zhao:2011cv,Du:2015wha} and bottomonia~\cite{Grandchamp:2005yw,Du:2017qkv}, as well as $B_c$ mesons~\cite{Wu:2023djn} and the exotic $X(3872)$~\cite{Wu:2020zbx}, over a large range of collision energies, from SPS via RHIC to the LHC~\cite{Rapp:2017chc}. The key transport parameter is the inelastic reaction rate, $\Gamma_{\mathcal{Q}}$. In the QGP it is calculated based on in-medium HQ masses and binding energies obtained from a thermodynamic $T$-matrix approach~\cite{Riek:2010fk,Riek:2010py} constrained by lattice QCD. The rate is dominated by ``quasifree'' dissociation processes~\cite{Grandchamp:2001pf,Zhao:2010nk,Du:2017qkv}, which are computed using perturbative diagrams with an effective but universal coupling constant $\alpha_s$ as the main parameter. In the hadronic phase, the reaction rate is obtained from effective interactions with a large set of hadronic states (currently restricted to charmonia).
The long-time equilibrium limit of each state,  $N_{\mathcal{Q}}^{\rm eq}$, is manifest in the regeneration term and computed from relative chemical equilibrium between the HF states in the system at given temperature. Transverse-momentum spectra have been computed from the Boltzmann equation for the suppressed primordial yields supplemented with a blast-wave approximation for the regenerated yields~\cite{Zhao:2007hh}. The latter has recently been improved by accounting for transported (off-equilibrium) HQ spectra obtained from relativistic Langevin simulations~\cite{He:2021zej,Du:2022uvj}.


\ingredients{

	\ingredient{In-medium potential:} 
The in-medium binding energies for charmonia~\cite{Zhao:2010nk} and bottomonia~\cite{Du:2017qkv} are taken from in-medium $T$-matrix calculations 
using the finite-temperature HQ internal energy from lattice QCD as the in-medium potential proxy~\cite{Riek:2010fk,Riek:2010py}.  

	\ingredient{Vacuum limit of potential/spectroscopy:}
The vacuum potential, as the zero-temperature limit of a screened Cornell potential, reproduces quarkonium spectroscopy (currently without spin-induced interactions). This provides an important benchmark for a smooth transition to realistic binding energies at moderate temperatures.

	\ingredient{Reaction rates:}
The inelastic reaction rates for quarkonia are evaluated from both leading-order gluo-dissociation, $g+ {\mathcal Q} \to Q + \bar{Q} $, and next-to-leading order (NLO) inelastic parton scattering, $p +{\cal Q} \to p+ Q + \bar{Q} $ where $p=q, \bar{q},g$. The NLO processes are evaluated in a quasi-free approximation, as a half-off-shell inelastic scattering off the heavy quark (plus antiquark) in the bound state, thereby accounting for the in-medium binding and conserving four-momentum. The $\psip$ rate in the QGP has been multiplied by an additional phenomenological $K$-factor of 3 extracted from d-Au collision data at RHIC (see the pertinent item below).
For bottomonia, interference effects between the scattering off the $b$ and $\bar{b}$ quarks are accounted for using an interference factor, $[(1-\exp(i\vec q\cdot\vec r)]$, that depends on the size, $r$, of the bound state via the 3-momentum transfer, $\vec{q}$. This leads to an $r$-dependent reduction of the width, also referred to as an imaginary part of the potential; in practice, it is mostly relevant for the $\YiS$~\cite{Du:2017qkv} and therefore not implemented for charmonia. Reaction rates in hadronic matter are included for charmonia as described below under {\it Hadronic-Phase Transport}.

	\ingredient{Assumptions about the medium (degrees of freedom, etc.):}
The thermal QGP medium is modeled with massive quasiparticles that are used to compute the inelastic dissociation reactions. The quasiparticle masses are taken as $m_{g,q} \propto gT$, to approximately describe the energy density of the QGP down to temperatures of  approximately 190\,MeV. A transition to a hadron resonance gas is performed using a mixed-phase construction at a temperature of $T_c=180$\,MeV (with charmonium reaction rates estimated from effective hadronic models). For bottomonia, hadronic dissociation is currently neglected. The QGP EoS has been updated with a lQCD parameterization~\cite{He:2011zx}, matched to a hadron resonance gas below a transition temperature of $T_c$=170\,MeV. The impact of this modification on bottomonium transport has been found to be small~\cite{Du:2017qkv}.

	\ingredient{Temperature dependence of heavy-quark masses:}
The HQ mass is composed of a bare mass and a temperature-dependent in-medium contribution determined from the infinite-distance limit of the potential. The bare mass is fitted to the quarkonium masses in vacuum where the potential saturates at a string-breaking distance of about 1\,fm. The temperature corrections to the HQ mass are constrained together with the in-medium potential by results from lattice QCD for the heavy-quark internal energy~\cite{Riek:2010fk}.

	\ingredient{Equilibrium limits in transport:}
The quarkonium equilibrium limits are calculated from the statistical model. In the QGP, the total abundance of heavy quarks (with their in-medium masses) is assumed to be conserved (given by their production in primordial $NN$ collisions), using fugacity factors, $\gamma_Q$, throughout the fireball evolution. The quarkonium equilibrium number then follows from the standard thermal density multiplied with a factor of $\gamma_Q^2$. In addition, a thermal-relaxation time correction is accounted for in the equilibrium limits to simulate the presence of non-equilibrated HQ distributions~\cite{Grandchamp:2002wp}, which imply a reduced phase space for quarkonium production and thus a smaller equilibrium limit~\cite{Song:2012at,Du:2022uvj}. 

	\ingredient{Constraints from lattice QCD:}
The quarkonium spectral functions in the thermodynamic $T$-matrix approach, from which the combination of in-medium quarkonium binding energies and HQ masses are taken, have been constrained by Euclidean correlation functions computed in lQCD~\cite{Riek:2010fk}, and cross-checked using the widths as employed in the rate equation~\cite{Zhao:2010nk}. Furthermore, the in-medium charm-quark masses have been checked against charm-quark susceptibilities in the QGP~\cite{Riek:2010py}, cf.~also Ref.~\cite{Liu:2021rjf} for more recent work.

	\ingredient{Range of applicability and how is this range established:}
 Both (LO) gluo-dissociation and (NLO) quasifree dissociation have been computed, with the former only relevant (albeit numerically small) in a small temperature window close to $T_c$ for $\jpsi$ and $\Upsilon$(1S,2S,1P); this follows the expected applicability for temperature ranges, $E_B\gg T$ and $E_B\lesssim T$, respectively, although with a rather large coefficient for the temperature, $T$, due to nonperturbative effects. A conceptual drawback is the current use of tree-level amplitudes to compute the quasifree reaction rates. A more realistic and consistent implementation directly using the nonperturbative in-medium $T$-matrix amplitudes is currently underway.

	\ingredient{Quantum features:}
The TAMU transport model is generally a semi-classical transport approach. A quantum feature is implemented in the early evolution after primordial production via bound-state formation times, $\tau_0^{\cal Q}$; they are different for the various quarkonia, scaled by their binding energies as $\tau_0^{\cal Q} \propto 1/E_B^{\cal q}$ (0.2-2$\,$fm$/c$), thus increasing for higher excited states. The build-up of wave packets for heavy quarkonia is assumed to reduce their dissociation rates by a factor  $\sim \tau/\tau_0^{\cal Q}$ for $\tau<\tau_0^{\cal}$, with the rates growing from zero to the equilibrium value linear in time, with an additional Lorentz-time dilation at finite $p_T$.

	\ingredient{Regeneration:}
Regeneration is included for all quarkonia, manifestly enforcing the statistical-equilibrium value in the long-time limit. For the $\pT$ spectra of the regeneration component thermal blast-wave spectra are employed following the fireball's flow profile at an average regeneration temperature which is smaller for more loosely bound states which emerge at lower temperatures but at higher flow velocities.

	\ingredient{Coupling to open heavy-flavor sector:}
  In the baseline implementation, the effect of the thermalization process of HQ spectra is implemented using a thermal-relaxation time approximation for the quarkonium equilibrium limits. Explicitly transported charm-quark spectra have recently been employed using the resonance recombination model (RRM) with space-momentum correlations (SMCs)~\cite{He:2021zej}, which extends the relevance of regeneration contributions by about a factor of two in $\pT$; this much improved the description of the $\jpsi$ $v_2$ at intermediate $\pT$ in Pb-Pb collisions at LHC energies.
 Furthermore, the Boltzmann equation for charmonia has recently been solved by implementing time-dependent (transported) charm-quark distributions from Langevin simulations into the regeneration term~\cite{Du:2022uvj}; these calculations have demonstrated the sensitivity of the ($\pT$-dependent) regeneration yield to the degree of charm-quark equilibration (corresponding to the thermal-relaxation time factor in the equilibrium limit, cf.~the item {\em Equilibrium limits in transport)}.
 
	\ingredient{Hadronic-Phase Transport:}
Transport in the hadronic phase is currently only implemented for charmonia. Starting from effective SU(4)-symmetric Lagrangians, reaction rates are calculated for charmonium dissociation induced by $\pi$- and $\rho$-mesons~\cite{Grandchamp:2002wp}. In Ref.~\cite{Du:2015wha} these calculations have been extended to include a large set of hadron resonance states through suitable changes in the available phase space. The $\jpsi$ rates in hadronic matter are generally small (up to $\sim 10$\,MeV), while those for the $\psip$ are larger (up to a few tens of MeV) and phenomenologically relevant.

	\ingredient{Initial quark/quarkonium distributions:}
The initial momentum distributions of primordial quarkonia in pp collisions are from fits to experimental $\pT$ spectra, while the initial spatial distributions follow a collision profile taken from the Glauber model.

	\ingredient{Are cold nuclear matter effects, nPDF effects, etc. taken into account?}
 Various CNM effects are included in the TAMU model:  Nuclear shadowing with both $N_{\rm part}$ and $\pT$ dependence; Cronin effect via a  Gaussian smearing to simulate nuclear $p_T$ broadening~\cite{Zhao:2007hh}, and nuclear absorption through effective (high-energy) ${\cal Q}$-$N$ cross sections extracted from pA data at SPS and RHIC energies~\cite{Zhao:2007hh}.

	\ingredient{Constraints from pA and dA collisions:}
Data from p/dA collisions at SPS and RHIC are primarily used to constrain the CNM effects. However, the medium assumed to be formed in d-Au reactions at RHIC has also been used to determine a nonperturbative correction to the QGP suppression rate of the $\psip$ in terms of a phenomenological $K$ factor of $\sim$2-3 multiplying its quasifree reaction rate~\cite{Du:2015wha}.

	\ingredient{Medium evolution model:}
The fireball model we used in the calculation is an isentropically 
and cylindrically expanding isotropic fireball~\cite{Grandchamp:2003uw,Zhao:2010nk,Zhao:2007hh,Grandchamp:2001pf}.
This expansion model reproduces the hadron spectra at thermal 
freezeout that are consistent with the empirically extracted light-hadron 
spectra ($\pi$, $K$, $p$) similar to hydro calculations. The entropy is 
estimated from the multiplicities of observed charged particles and 
assumed to be conserved during the adiabatic expansion. 

	\ingredient{Feed down implementation:}
Constant feeddown fractions are used for charmonia~\cite{Andronic:2015wma} and bottomonia~\cite{Du:2017qkv}. 

	\ingredient{Comparisons to experimental data:}
TAMU calculations have been used to compare to charmonium data at the SPS (Pb-Pb, S-U, In-In at 17 GeV),
RHIC (Au-Au at 39, 62 and 200 GeV and Cu-Cu at 200 GeV), and the LHC (Pb-Pb at 2.76 and 5.5 TeV), bottomonium data at RHIC (Au-Au at 200 GeV) and the LHC (Pb-Pb at 2.76 and 5.5 TeV), and $B_c$ data at the LHC (Pb-Pb at 5.5 TeV) (as well as in p/dA systems, see above). 


	\ingredient{Phenomenological breadth:}
 The TAMU model has been broadly applied to charmonium, bottomonium, $B_c$, and $X(3872)$ phenomenology in AA and p/dA collisions at SPS, RHIC, and LHC energies, 
 using the same input quantities within the same formalism.
Specifically: $\jpsi$, $\chi_c(1\mathrm{P})$ (spin averaged), $\psip$~\cite{Grandchamp:2003uw,Zhao:2010nk,Zhao:2011cv,Du:2015wha} and $X(3872)$~\cite{Wu:2020zbx} for charmonia,  $\YiS$, $\chi_b(1\mathrm{P})$, $\YiiS$, 
$\YiiiS$ and $\chi_b(1\mathrm{P})$ for bottomonia~\cite{Du:2017qkv}, as well as   $B_c(1\mathrm{S})$ and $B_c(1\mathrm{P})$ for charm-bottom mesons~\cite{Wu:2023djn}.

}

%% file: models/tsinghua.tex
\subsection{The Tsinghua model}

In the Tsinghua transport model, 
the evolution of quarkonium in hot and dense QCD matter is described by a relativistic Boltzmann transport equation~\cite{Liu:2010ej,Zhou:2014kka,Chen:2018kfo,Chen:2016dke,Zhao:2022ggw,Zhao:2020jqu}.
The phase-space distribution of quarkonium states, $f_\psi(p, x)$, is controlled by the relativistic Boltzmann transport equation (with Bjorken coordinates), 
\begin{eqnarray}
\left[ \cosh(y-\eta)\frac{\partial}{\partial \tau} +\frac{\sinh(y-\eta)}{\tau}\frac{\partial}{\partial \eta}+{\bf v}_T\cdot \nabla_T \right] f_{\psi} =-\alpha f_{\psi} +\beta \ ,
\label{eq.tsinghua}
\end{eqnarray}
where $\psi$ represents different states (e.g., $\psi=J/\psi, \chi_c, \psi^\prime$ for charmonia); $\eta$ and $y$ are the rapidities in coordinate and momentum space, respectively, and ${\bf v}_T={\bf p}_T/E_T$ is the quarkonium transverse velocity with the transverse energy $E_T=\sqrt{m_{\psi}^2+{\bf p}_T^2}$. 
The anomalous suppression and regeneration mechanisms are reflected in the loss term $\alpha$ and the gain term $\beta$. 
Hot medium effects such as color screening effect and gluo-dissociation are included in $\alpha$. The regeneration process, which is the inverse of the gluo-dissociation, is represented byn $\beta$,
\begin{eqnarray}
\alpha&=&\frac{1}{2E_T}\int \frac{d^3{\bf k}}{(2\pi)^32E_g}W_{g\psi}^{Q{\bar Q}}(s)f_g({p}_g,x), \\
\beta&=&\frac{1}{2E_T}\int \frac{d^3{\bf k} }{(2\pi)^32E_g}\frac{d^3{\bf p}_{\bar Q}}{(2\pi)^32E_{\bar Q}}\frac{d^3{\bf p}_{Q} }{(2\pi)^32E_{Q}}   W_{Q{\bar Q}}^{g\psi}(s)f_{\bar Q}({p}_{\bar Q},x)f_{Q}({p}_{Q},x) \nonumber\\
&\times &(2\pi)^4\delta^{(4)}(p+k-p_{\bar Q}-p_{Q})\theta(T-T_c) \ , \nonumber
\end{eqnarray}
where $f_g$ is the gluon distribution taken as a Bose distribution; $E_g(E_Q)$ is the energy of gluon (heavy quark) and ${\bf p}_g ({\bf p}_Q)$ is the momentum of gluon (heavy quark). The dissociation rate, $W_{g\psi}^{Q\bar Q}$, contains the in-medium binding energy which is reduced by the color screening effect, and also the gluo-dissociation cross section in the reaction $g+\psi\rightarrow Q+\bar Q$. In the $\beta$ term, the regeneration rate $W_{Q\bar Q}^{g\psi}$ for the inverse reaction of gluo-dissociation is connected with the dissociation rate via detailed balance. Quarkonium regeneration also depends on the 
densities of heavy quarks, $f_{Q,\bar Q}$. The HQ density in the expanding medium is controlled by the diffusion equation, since heavy quarks are strongly coupled with the QGP. With this collisional term, the transport equation can be solved analytically~\cite{Zhao:2020jqu}. 
 
In nucleus-nucleus collisions, the quarkonium initial distribution is treated as a superposition of quarkonium distribution in pp collisions. Cold-nuclear-matter effects, such as nuclear absorption, Cronin effect, and shadowing effect, are included by modifying the initial distribution extracted from pp collisions. 
For the non-prompt $J/\psi$ from $B$ decays, the Langevin equation is employed to simulate the energy loss of bottom quarks in the medium~\cite{Chen:2021akx}. The hadronization of bottom quarks into $B$ mesons is described with the instantaneous coalescence model~\cite{Chen:2021akx}. 

\ingredients{

\ingredient{In-medium potential:} 
Charmonia and bottomonia experience a color screening effect in the hot deconfined medium, which reduces the heavy-quark potential and the in-medium binding energies. The color screening effect increases with the temperature and distance. As quarkonium dissociation mainly happens in the early stage of the medium evolution with high temperatures, we employ an effective constant in-medium binding energy in the calculation of quarkonium dissociation~\cite{Chen:2018kfo,Chen:2016dke}.

	\ingredient{Vacuum limit of potential/spectroscopy:}
The vacuum limit of potential is Cornell potential, $V(r)=-\alpha_c/r+\sigma r$ with $\alpha_c=\pi/12$ and $\sigma=0.2~ \rm GeV^2$. The charm and bottom quark masses are taken as 1.5 GeV and 4.5 GeV, respectively. 

	\ingredient{Reaction rates:}
Gluo-dissociation, $g+\psi\to Q+\bar Q$, is considered as the dominant dissociation process in the QGP for tightly bound quarkonium. The cross-section, $\sigma_{g\psi}^{Q\bar Q}$, in vacuum can be derived through the operator product expansion method and was calculated firstly by Peskin and Bhanot~\cite{Peskin:1979va,Bhanot:1979vb}. The gluon density is taken as the Bose distribution $f_g=1/(e^{p\cdot u/T}-1)$, where $T$ and $u$ are the temperature and the four-velocity of the medium given by hydrodynamic models. At different temperatures, the channging density of thermal gluons gives the temperature dependence in the quarkonium dissociation rates. For excited quarkonium states, their dissociation rates are obtained via the geometric scaling of their radii over the ground-state one. 

	\ingredient{Assumptions about the medium (degrees of freedom, etc.):}
The hot deconfined medium generated in high-energy nuclear collisions is treated as an ideal gas consisting of massless $u$/$d$ quarks and gluons, and strange quarks with a mass of $m_s=150$ MeV. There is a first-order phase transition between QGP and the hadronic gas, where the critical temperature is $T_c=165$ MeV at the zero baryon chemical potential. 

	\ingredient{Temperature dependence of heavy quark masses:}
The heavy quark mass is independent of temperature.

	\ingredient{Equilibrium limits in transport:}
We use the Boltzmann equation to describe the dissociation and regeneration of quarkonium in relativistic heavy-ion collisions. The  
dissociation and regeneration are related to each other via the detailed balance. Thus, the equilibrium limit is naturally satisfied. 

	\ingredient{Constraints from lattice QCD:}
The in-medium heavy-quark potential is close to the internal energy extracted from the lattice QCD calculations. The quarkonium in-medium binding energy can be calculated with the two-body Schr\"odinger equation where the 
in-medium potential is taken as the internal energy extracted from the free energy evaluated in Ref.~\cite{Petreczky:2010yn}. An 
effectively constant binding energy is then extracted and used in the calculation of the quarkonium decay rate. 

	\ingredient{Range of applicability and how is this range established:}
As the formation time/decoherence of quarkonia was not considered, the Tsinghua model mostly applies in the quantum optical limit. The charm quark phase-space distribution is assumed to be a kinetically thermalized distribution as mentioned in ``initial quark/quarkonium distribution" section. This approximation is based on the observation of the large $v_2$ of $D$ mesons in experiment. It is assumed to be a good approximation for charm with low transverse momentum and/or in central collisions. In addition, regeneration is not considered for bottomonium due to the scarcity of bottom quark in the hot medium.

\ingredient{Quantum features:}
There are no quantum features in the Tsinghua model.

	\ingredient{Regeneration:}
Regeneration is considered for charmonium states, via  uncorrelated charm and anti-charm quark as represented by the gain term in the Boltzmann equation. 
The regeneration process is related to the dissociation process via detailed balance.

	\ingredient{Coupling to open heavy-flavor sector:}
 For prompt charmonium, the charm-quark is assumed to be kinetically thermalized as used in charmonium regeneration. For bottom quarks, energy loss in the medium is simulated with the Langevin equation. The final momentum distribution of $B$ mesons is used to calculate the production of non-prompt $J/\psi$. 
 
	\ingredient{Hadronic-Phase Transport:}
Charmonium experience additional suppression through scattering with $\pi$ and $\rho$ mesons in the hadron gas via 
$J/\psi+\pi \to D+\bar{D}^*, D^*+\bar D$, and $J/\psi+\rho \to D^*+\bar {D^*}, D+\bar D$, with inelastic cross-sections  taken from Ref.~\cite{Lin:1999ad};  $\pi$ and $\rho$ mesons are assumed to be thermalized in the hadronic phase. 

	\ingredient{Initial quark/quarkonium distributions:}
The quarkonium initial distribution in nucleus-nucleus collisions is treated as a superposition of quarkonium distributions in pp collisions. 
The quarkonium momentum distribution is obtained by fitting the 
experimental data of $J/\psi$ distribution in pp collisions, with a further modification from CNM effects. 
In the regeneration part, where the charm-quark momentum distribution is needed, it is taken as a kinetically thermalized distribution. In this case, the initial momentum distribution of charm quarks does not affect charmonium regeneration. The initial spatial distributions of charm quarks and quarkonium are proportional to 
the density of binary collisions, $n_{coll}({\bf x}_T)$. 

	\ingredient{Are cold nuclear matter effects, nPDF effects, etc. taken into account?:}
 Yes, CNM such as nuclear shadowing, the Cronin effect, and nuclear absorption are included. The survival probability is related to the nuclear absorption cross-section~\cite{Lin:1999ad}, $\sigma_{abs}^{J/\psi}$. The absorption cross sections for the excited states are obtained from $\sigma_{abs}^{J/\psi}$ through geometric scaling,
$\sigma_{abs}^\psi = \langle r^2_\psi \rangle/ \langle r^2_{J/\psi} \rangle \sigma_{abs}^{J/\psi}$,
where the mean-square-radius can be obtained by solving the two-body Schr\"odinger equation~\cite{Zhao:2020jqu}.
The Cronin effect is included in the initial distribution of quarkonium via the Gaussian smearing method~\cite{Huefner:2002tt}. 
The nuclear shadowing factor is calculated with the EPS09 package~\cite{Eskola:2009uj}.

	\ingredient{Constraints from pA and dA collisions:}
 The nuclear absorption cross-section, $\sigma_{abs}^{J/\psi}$, and Cronin momentum-broadening parameter, $a_{gN}$, are determined by fitting experimental data from pA and dA collisions.

	\ingredient{Medium evolution model:}
2+1D ideal hydrodynamics is mostly used to describe the medium evolution in Tsinghua model for many years~\cite{Zhuang:2003fu}), while a 3+1D viscous hydro (MUSIC package) has been used in past two years. In the present context paper, the results are obtained with the 2+1D ideal hydro. 
The initial condition (entropy density) of the hydrodynamics in the transverse plane is given by the two-component model~\cite{Kharzeev:2000ph}. The maximum entropy density is determined by the charged-hadron multiplicity observed in the experiment. The start time of hydrodynamics is $\tau_0=0.6\; \rm fm/c$ and the maximum temperature is $T_0=510 \; \rm MeV$ for central Pb-Pb collisions with $\sqrt{s_{\rm NN}}=5.02 \; \rm TeV$.

	\ingredient{Feed down implementation:}
For $\jpsi$, the feeddown contributions from $\chi_c$ and $\psip$ are considered. 
For $\Upsilon$, the feeddown contributions from higher states such as $\Upsilon(1\mathrm{P})$, $\YiiS$, $\Upsilon(2\mathrm{P})$, and $\YiiiS$ are considered. The feeddown branching ratios are taken fromthe  PDG~\cite{ParticleDataGroup:2018ovx}.

	\ingredient{Comparisons to experimental data:}
The Tsinghua model has been used to explain the nuclear modification factor, $\raa$, anisotropic flow, $v_n$, and mean transverse-momentum squared, $\langle \pT^2\rangle$, of charmonium and bottomonium states.

	\ingredient{Phenomenological breadth:}
The Tsinghua model has been applied in the studies of charmonium, bottomonium, and $B_c$ in both small (p-Pb) and large (Pb-Pb, Au-Au) collision systems, at the collision energies of SPS, RHIC, and LHC. 
}

%% file: models/vogt.tex
\subsection{Cold nuclear matter effects}

Cold nuclear matter effects generally refer to all modifications present in pA collisions, when there is a nuclear target, but not in pp collisions.  These effects are referred to as being ``cold" in the sense that it is assumed that no quark-gluon plasma is created.  Note that this assumption may be challenged in high-multiplicity pp and pA collisions.  These effects are also presented in AA collisions when quark-gluon plasma is also created.  The nuclear modifications of the parton distributions (nPDF effects) are typically included by a parameterization such as the EPPS16 \cite{Eskola:2016oht} parameterization. The centrality dependence of nPDF effects may also be taken into account, see Ref.~\cite{McGlinchey:2012bp}. Enhanced $k_T$ broadening in the nucleus relative to a proton, due to multiple scattering or a Cronin-type effect, is also sometimes included \cite{Vogt:2022glr}.  The $k_T$ broadening may or may not be related to energy loss in cold matter.  A number of energy loss models have been proposed, see for example Ref.~\cite{Arleo:2013zua}.  Such models can account for quarkonium suppression at high $x_F$.  Other models that include intrinsic charm without energy loss can also provide a good description of this suppression, see Ref.~\cite{Vogt:2022glr}.

Quarkonium absorption by nucleons has been studied by many different groups but has been suggested to be negligible at the LHC.  See Ref.~ \cite{Lourenco:2008sk} for a discussion of the energy dependence of absorption.  Dissociation of quarkonium by comovers has also long been suggested, see Sec.~\ref{comovers}.  The comovers can be considered to be partons or hadrons, see Refs.~\cite{Ftacnik:1988qv,Gavin:1988hs,Vogt:1988fj} for some early discussions of hadronic comover dissociation.  It was found that, in this description, the nuclear dependence of comover dissociation is similar to nuclear absorption \cite{Gavin:1990gm}.

In the traditional color evaporation model (CEM), the quarkonium 
production cross section is some fraction, $F_C$, of 
all $Q \overline Q$ pairs below the $H \overline H$ threshold where $H$ is
the lowest mass heavy-flavor hadron. The color of the
octet $Q \overline Q$ state is
`evaporated' through an unspecified process which does not significantly change the momentum.
The quarkonium yield may be only a small fraction of the total $Q\overline 
Q$ cross section below $2m_H$.
Schematically, the production cross section of quarkonium state $C$ in
a pp collision is
\begin{eqnarray}
\sigma_C^{\rm CEM}(s_{_{NN}})  =  F_C \sum_{i,j} 
\int_{4m^2}^{4m_H^2} d\hat{s}
\int dx_1 \, dx_2~ f_i^p(x_1,\mu_F^2)~ f_j^p(x_2,\mu_F^2)~ {\cal J}(\hat{s})
\hat\sigma_{ij}(\hat{s},\mu_F^2, \mu_R^2) \, 
\, , \label{sigtil}
\end{eqnarray} 
where $ij = q \overline q$ or $gg$ and $\hat\sigma_{ij}(\hat s)$ is the
$ij\rightarrow Q\overline Q$ cross section and $\mu_F$ and $\mu_R$ are the factorization and renormalization scales, respectively.  In pA collisions, the cold nuclear matter effects on quarkonium production are
\begin{eqnarray}
\sigma_{\rm CEM}(pA) = S_A^{\rm abs} F_C \sum_{i,j} 
\int_{4m^2}^{4m_H^2} d\hat{s}
\int dx_1 \, dx_2~ F_i^p(x_1,\mu_F^2,k_T)~ F_j^A(x_2,\mu_F^2,k_T)~ 
\hat\sigma_{ij}(\hat{s},\mu_F^2, \mu_R^2) \, \, , 
\label{sigCEM_pA}
\end{eqnarray}
where
\begin{eqnarray}
F_j^A(x_2,\mu_F^2,k_T) & = & R_j(x_2,\mu_F^2,A) f_j(x_2,\mu_F^2) G_A(k_T) \, \, \\
F_i^p(x_1,\mu_F^2,k_T) & = & f_i(x_1,\mu_F^2) G_p(k_T) \, \, ,
\end{eqnarray}
where $S_A$ is the survival probability for nucleon absorption, $R_j$ is the nuclear modification of the parton distributions (nPDF), and $G_p(k_T)$ and $G_A(k_T)$ account for transverse momentum broadening in the proton and nucleus, respectively \cite{Vogt:2022glr}.

The CEM has been improved (ICEM) to better account for feed down and the quarkonium mass \cite{Ma:2016exq}.  The unpolarized direct quarkonium production cross section in pp collisions in the ICEM is 
\begin{eqnarray}
\label{ch6-icem-cross-section}
\sigma = F_C^\prime \sum_{i,j}  \int^{2m_Q}_{M_C} dM \, dx_i \, dx_j \, f_i(x_i,\mu_F)f_j(x_j,\mu_F) \hat{\sigma}_{ij}(p_{c\bar{c}},\mu_R) |_{p_{c\bar{c}} = \frac{M}{M_C p_C}} \, \, .
\end{eqnarray}
Note that the change in integration range changes $F_C$ to $F_C^\prime$ while the change in the momentum range modifies the $\pT$ distribution of the quarkonium states relative to each other.

\ingredients{

	\ingredient{In-medium potential:}  N/A

	\ingredient{Vacuum limit of potential/spectroscopy:} N/A

	\ingredient{Reaction rates:} N/A

	\ingredient{Assumptions about the medium (degrees of freedom, etc.):} N/A

	\ingredient{Temperature dependence of heavy quark masses:} N/A

	\ingredient{Equilibrium limits in transport:} N/A  

	\ingredient{Constraints from lattice QCD:} N/A at the moment. If lattice could provide constraints on feed down, then it would be useful.  Also, for NRQCD-type formulations, calculating the LDMEs could be useful.

	\ingredient{Range of applicability and how is this range established:}
The calculation of initial production is applicable over all center of mass energies.

	\ingredient{Quantum features:} N/A

	\ingredient{Regeneration:} N/A at the moment, however, in the past the HVQMNR code
was used by Thews and Mangano to calculate regeneration by ``non-diagonal quarkonium production'' \cite{Thews:2005vj}.

	\ingredient{Coupling to open heavy-flavor sector:} N/A
 
	\ingredient{Hadronic-Phase Transport:} N/A

	\ingredient{Initial quark/quarkonium distributions:}
The charm quark mass and scale parameters used to calculate the $J/\psi$ distributions are \cite{Nelson:2012bc} $(m,\mu_F/m, \mu_R/m) = (1.27 \pm 0.09 \, {\rm GeV}, 2.1^{+2.55}_{-0.85},
1.6^{+0.11}_{-0.12})$.  In the case of $\Upsilon$ production, 
$(m,\mu_F/m, \mu_R/m) = (4.65 \pm 0.09 \, {\rm GeV}, 1.4^{+0.77}_{-0.49},
1.1^{+0.22}_{-0.20})$.  
The values of $F_C$ are fixed for the central parameter set in each case and all calculations employing other masses and scales use the same value of $F_C$ to
obtain the extent of the $J/\psi$ and $\Upsilon$ mass and scale uncertainty 
bands.
The normalization factors for the CEM are $F_{J/\psi} = 0.020393$ for the central result
with $(m,\mu_F/m, \mu_R/m)$ $ = (1.27 \, {\rm GeV}, 2.1,1.6)$ and $F_{\Upsilon} = 0.022$ with 
$(m,\mu_F/m, \mu_R/m) = (4.65 \, {\rm GeV}, 1.4,1.1)$.
The calculations use the CT10 proton parton distributions \cite{Vogt:2015uba}.  The quark mass makes a larger contribution to the cross section uncertainty than does the scale choice \cite{Vogt:2015uba}.

	\ingredient{Are cold nuclear matter effects, nPDF effects, etc. taken into account?:}
 The EPPS16 \cite{Eskola:2016oht} parameterization is employed for the nPDF calculation.  The intrinsic $k_T$ broadening employed in $p+p$ production is augmented in nuclei according to multiple scattering in the nucleus \cite{Vogt:2022glr}.  While absorption \cite{Lourenco:2008sk} is included at lower energies, it is considered negligible at LHC energies.
 
	\ingredient{Constraints from pA and dA collisions:}
 There are very few constraints from pA and dA collisions in the basic model.  There are constraints from these collisions in the global analyses of the nPDFs used in the calculations.  For example, EPS09 \cite{Eskola:2009uj} used $\pi^0$ data from RHIC to constrain the gluon distribution and EPPS16 \cite{Eskola:2016oht} used LHC p-Pb data on $W^\pm$ and $Z^0$ production to separate the antiquark distributions in the sea at high $Q^2$ and on dijet data to further constrain the gluon distribution.  The centrality dependence of $R_{\rm dAu}(\pT)$ was used to study the centrality dependence of shadowing \cite{McGlinchey:2012bp}.  Fixed-target pA data were used to constrain the $J/\psi$ absorption cross section \cite{Lourenco:2008sk}.  However, absorption is considered negligible at the LHC collider energies and is not included in the calculation \cite{Vogt:2022glr}.  The intrinsic $k_T$ employed in pp collisions was obtained from comparison to data but the broadening in pA collisions is based on a model and not a fit within these calculations \cite{Vogt:2022glr}.
 
	\ingredient{Medium evolution model:} N/A

	\ingredient{Feed down implementation:} 
The traditional CEM does not distinguish between the states, all of the distributions are assumed to be the same.  Thus, the feed down distributions are all identical to that of the ground state modulo emission of soft particles, photons or pions, in the decays.  In the improved CEM, the feed down distributions depend on the specific mass and momentum of the particular quarkonium state and are thus realistic in terms of the model \cite{Ma:2016exq,Cheung:2018tvq,Cheung:2018upe,Cheung:2021epq}. For example, the model reproduces the $\pT$ dependence of the $\psi(2{\rm S})/\psi$ ratio \cite{Ma:2016exq}.

	\ingredient{Comparisons to experimental data:}  
The model agrees well with experimental data, both for the traditional CEM and the improved CEM \cite{Cheung:2021epq}.

	\ingredient{Phenomenological breadth:} 
Both the traditional CEM and the improved CEM can be applied to charmonium and bottomonium production, including both S and P states, $J/\psi$, $\psi$(2S), and $\Upsilon$($n$S) as well as $\chi_c$ and $\chi_b$ states respectively \cite{Vogt:2015uba,Cheung:2018tvq,Cheung:2018upe,Cheung:2021epq}.  Charmonium production in particular has been studied from $p_{\rm lab} = 40$~GeV to $\sqrt{s} = 13$~TeV \cite{Vogt:2022glr}. 

The improved CEM has been developed to also calculate quarkonium polarization \cite{Cheung:2017loo,Cheung:2017osx,Cheung:2018tvq,Cheung:2018upe,Cheung:2021epq}.  The calculation was also extended to Pb-Pb collisions including only cold nuclear matter effects \cite{Cheung:2022nnq} and was shown to agree with the data.  Note that because the polarization calculation involves a ratio of cross sections with a certain spin, such effects generally cancel in the ratios \cite{Cheung:2022nnq}.  Indeed, only the charm quark mass generally affects the polarization but is not a strong effect \cite{Cheung:2018tvq}.

The ability of the CEM to consistently cover the entire energy range, from near production threshold to the highest available energies can be contrasted with the NRQCD approach which typically requires a $p_T$ cut to fit the LDMEs \cite{Brambilla:2014jmp} to the data and has only been matched to the low $p_T$ part of the spectrum by including a color-glass condensate contribution, matching to the high $p_T$ part \cite{Ma:2014mri}.  Such an approach cannot work at lower energies, such as fixed-target energies.  Additionally, the NRQCD LDMEs that have been fit to total cross section quarkonium data \cite{Maltoni:2006yp} also cannot describe $p_T$ distributions \cite{Feng:2015cba}.

}